\documentclass[aps,prd,groupedaddress,nofootinbib]{revtex4-2}

\usepackage{amsmath,amssymb}
\usepackage{graphicx}
\usepackage{slashed}
\usepackage{verbatim}
\usepackage{dsfont,latexsym} 
\usepackage{xcolor} 
\usepackage{hyperref} 
\usepackage{amsmath} 
\usepackage{bm}      

\hypersetup{
    colorlinks,
    citecolor=black,
    filecolor=black,
    linkcolor=black,
    urlcolor=black
}

\newcommand{\be}{\begin{equation}}
\newcommand{\ee}{\end{equation}}
\newcommand{\ba}{\begin{eqnarray}}
\newcommand{\ea}{\end{eqnarray}}
\newcommand{\ban}{\begin{eqnarray*}} 
\newcommand{\ean}{\end{eqnarray*}}

\def\b{{\mathbf b}}
\def\p{{\mathbf p}}
\def\q{{\mathbf q}}
\def\k{{\mathbf k}}
\def\x{{\mathbf x}}  
\def\y{{\mathbf y}}
\def\r{{\mathbf r}}
\def\z{{\mathbf z}}
\def\w{{\mathbf w}}
\def\v{{\mathbf v}}

\def\K{{\mathbf K}}
\def\P{{\mathbf P}}

\def\Qb{{\bar Q}}
\def\dd{{\rm d}}
\def\ubar{{\bar u}}
\def\T{{\mathcal{T}}}
\def\S{{\mathcal{S}}}
\def\C{{\mathcal{C}}}
\def\rmx{{\rm x}}

\begin{document}

\title{Twist-three quark TMDs from back-to-back dijet production in next-to-eikonal DIS}

\author{Tolga Altinoluk}
\affiliation{Theoretical Physics Division, National Centre for Nuclear Research, Pasteura 7, Warsaw, 02-093, Poland}
 
\author{Guillaume Beuf}
\affiliation{Theoretical Physics Division, National Centre for Nuclear Research, Pasteura 7, Warsaw, 02-093, Poland}

\author{Alina Czajka}
\affiliation{Theoretical Physics Division, National Centre for Nuclear Research, Pasteura 7, Warsaw, 02-093, Poland}

\date{\today}

\begin{abstract}
We study back-to-back dijet production in deep inelastic scattering (DIS) at next-to-eikonal accuracy within the Color Glass Condensate framework, focusing on the quark-background contributions. We compute the quark–gluon and quark–antiquark production cross sections through next-to-leading power in the correlation limit for longitudinally and transversely polarized photons. In the quark–gluon channel, we derive both the kinematic corrections to the hard parts accompanying the twist-two quark distribution and the genuine twist-three contributions involving three-body quark–gluon–quark correlators. In the quark–antiquark channel, the interference of the eikonal amplitude with the double $t$-channel quark-exchange amplitudes probes three-body correlators with a different transverse arrangement of the fields. We express the results in terms of transverse momentum dependent (TMD) correlators and provide the corresponding hard coefficients. Our analysis extends the operator-level connection between the next-to-eikonal CGC framework and quark TMDs beyond leading power in the correlation limit.
\end{abstract}

\maketitle

\tableofcontents

\section{Introduction}
\label{sec-intro}

The high-energy regime of Quantum Chromodynamics (QCD), where gluon densities inside hadronic wave functions become large, is conveniently described within the Color Glass Condensate (CGC) effective theory ~\cite{Gelis:2010nm,Albacete:2014fwa,Blaizot:2016qgz}. In this framework, the small-$x$ degrees of freedom of the target are treated as strong classical color fields generated by static color sources~\cite{McLerran:1993ni,McLerran:1993ka,McLerran:1994vd} carrying larger longitudinal momentum fractions, while their evolution towards smaller values of Bjorken-$x$ is governed by the non-linear  Balitsky-Kovchegov / Jalilian-Marian-Iancu-McLerran-Weigert-Leonidov-Kovner (BK/JIMWLK) evolution equations~\cite{Balitsky:1995ub,Kovchegov:1999yj,Kovchegov:1999ua,Jalilian-Marian:1996mkd,Jalilian-Marian:1997qno,Jalilian-Marian:1997jhx,Jalilian-Marian:1997ubg,Kovner:2000pt,Weigert:2000gi,Iancu:2000hn,Iancu:2001ad,Ferreiro:2001qy}. Owing to its systematic treatment of multiple scattering and gluon saturation effects, the CGC has become a standard theoretical framework for describing a broad class of high-energy scattering processes.

Among the processes that can be studied within the CGC framework, dijet production in deep inelastic scattering (DIS) has emerged as one of the cleanest probes of the transverse momentum structure of hadronic matter at small $x$. In the back-to-back correlation limit, where the transverse momentum imbalance of the dijet system is much smaller than the individual jet transverse momenta, the cross section factorizes into perturbatively calculable hard coefficients and nonperturbative transverse momentum dependent parton distribution functions (TMDs), thereby separating the perturbative short-distance dynamics from the nonperturbative information encoded in the TMDs. 
The correspondence between the CGC and TMD frameworks has by now been established for a broad class of observables (see~\cite{Petreska:2018cbf,Boussarie:2023izj} for recent reviews). Besides providing an operator-level connection between the two approaches, it has considerably improved our understanding of the small-$x$ dynamics underlying high-energy scattering~\cite{Dominguez:2011wm,Marquet:2016cgx,Marquet:2017xwy,Altinoluk:2018uax,Altinoluk:2018byz,Altinoluk:2020qet}.
The applicability of the back-to-back dynamics has subsequently been extended beyond the strict correlation limit through the systematic inclusion of kinematic power corrections, leading to the improved TMD framework at small-$x$~\cite{Kotko:2015ura,vanHameren:2016ftb,Altinoluk:2019wyu,Altinoluk:2019fui,Bury:2020ndc,Fujii:2020bkl,Boussarie:2020vzf,Altinoluk:2021ygv,Boussarie:2021ybe}.

Nevertheless, TMD factorization was first proposed away from the small-$x$ regime,  for back-to-back dihadron production in $e^+e^-$ collisions~\cite{Collins:1981uk,Collins:1981uw}, transverse momentum distribution in Drell-Yan production~\cite{Collins:1984kg} and semi-inclusive DIS (SIDIS)~\cite{Collins:1992kk,Ji:2004wu}, and proven to all orders in perturbation theory for these processes~\cite{Collins:2011zzd}, at leading power (LP) in the ratio of small scales over the hard scale (see~\cite{Boussarie:2023izj} for a recent review). Since then, TMD factorization (at moderate $x$) has been studied for various other processes, in particular for the back-to-back dijet production in DIS for either light or heavy quarks~\cite{Boer:2010zf,Pisano:2013cya,Zhang:2017uiz,delCastillo:2020omr,delCastillo:2021znl,Kang:2020xgk}, also considered in the present study but at small-$x$.    
Beyond LP, pioneering works~\cite{MuldersTangerman:1996Twist3,MuldersRodrigues:2001Gluon,BacchettaEtAl:2007SIDIS,LorcePasquini:2013Correlators} have focused on next-to-leading power (NLP) corrections at tree level to observables as well on the classification of TMD correlators appearing at NLP. More recently, the systematic study of TMD factorization at NLP, including one-loop evolution effects,
 has become a very active topic~\cite{Ebert:2018gsn,VladimirovMoosScimemi:2022NLP,Rodini:2022wki,EbertGaoStewart:2022NLP,GambergEtAl:2022NLP,RodiniVladimirov:2024SIDIS,RodiniAlvaroPasquini:2023Matching,Vladimirov:2023Kinematic,delCastillo:2023rng,Balitsky:2024ozy,Arroyo-Castro:2025slx,Jaarsma:2025ksf,Piloneta:2025jjb,Balitsky:2026nux}.

A complementary direction towards improving the theoretical description of the high-energy or low-$x$ regime consists in systematically going beyond the eikonal approximation that underlies the interaction of energetic partons with the target color field. Corrections to the eikonal approximation arise from retaining subleading power terms in the high-energy expansion and are systematically organized within the next-to-eikonal (NEik) framework. 
Besides providing systematic corrections to scattering amplitudes beyond the strict high-energy limit, the NEik expansion reveals a richer operator structure than that encountered at eikonal accuracy. During the past several years, a systematic formalism has been developed for deriving NEik corrections to propagators, scattering amplitudes, particle production processes, and helicity evolution within the CGC framework~\cite{Altinoluk:2014oxa,Altinoluk:2015gia,Altinoluk:2015xuy,Agostini:2019avp,Agostini:2019hkj,Altinoluk:2020oyd,Altinoluk:2021lvu,Agostini:2022ctk,Agostini:2022oge,Altinoluk:2022jkk,Altinoluk:2023qfr,Agostini:2023cvc,Agostini:2024xqs,Altinoluk:2024zom,Altinoluk:2024dba,Altinoluk:2024tyx,Altinoluk:2025ang,Altinoluk:2025ivn,Agostini:2025vvx,Altinoluk:2026ngv,Altinoluk:2026ymh,Kovchegov:2015pbl,Kovchegov:2016zex,Kovchegov:2016weo,Kovchegov:2017jxc,Kovchegov:2017lsr,Kovchegov:2018znm,Kovchegov:2018zeq,Cougoulic:2019aja,Kovchegov:2020kxg,Kovchegov:2020hgb,Cougoulic:2020tbc,Adamiak:2021ppq,Kovchegov:2021lvz,Kovchegov:2021iyc,Cougoulic:2022gbk,Kovchegov:2022kyy,Borden:2023ugd,Kovchegov:2024aus,Borden:2024bxa,Kovchegov:2024wjs,Kovchegov:2025gcg,Balitsky:2015qba,Balitsky:2016dgz,Chirilli:2018kkw,Chirilli:2021lif,Chirilli:2026vij,Chirilli:2026pkv,Jalilian-Marian:2017ttv,Jalilian-Marian:2018iui,Jalilian-Marian:2019kaf,Boussarie:2020fpb,Boussarie:2021wkn,Boussarie:2023xun,Li:2023tlw,Li:2024fdb,Li:2024xra,Li:2026azt,Mukherjee:2026cte,Mukherjee:2026six,Armesto:2026qwh}. Collectively, these developments have established a systematic framework for investigating subleading contributions in the high-energy expansion across a broad class of high-energy scattering processes.

The NEik formalism has also been applied to dijet production in deep inelastic scattering. The NEik amplitudes and cross sections for quark-gluon and quark-antiquark dijet production were derived in Refs.~\cite{Altinoluk:2022jkk,Altinoluk:2023qfr}, providing the necessary ingredients for investigating the correlation limit of these processes beyond the eikonal approximation. At the first nontrivial order in the expansion in the dipole size, the quark-background contribution to the quark-gluon channel was shown to reduce to the quark bilinear operator associated with the corresponding quark TMD~\cite{Altinoluk:2023qfr}. Going beyond this order naturally probes three-body quark operators involving field-strength insertions, while the corresponding double $t$-channel quark-exchange contribution to quark-antiquark dijet production gives access to analogous operator structures. These observations naturally motivate a systematic investigation of the higher-order operator content probed in the correlation limit of back-to-back DIS dijet production at NEik accuracy.


In this work, we study the quark-background contributions to back-to-back quark-gluon and quark-antiquark dijet production in DIS at NEik accuracy, extending the correlation-limit analysis through NLP. For the quark-gluon channel, we compute the kinematic corrections accompanying the quark bilinear operator and derive the three-body operator structures that give rise to genuine twist-three contributions. We establish their connection to the twist-three quark TMD correlators studied in Ref.~\cite{Rodini:2022wki}. For the quark-antiquark channel, the double $t$-channel quark-exchange contribution probes three-body correlators with a different transverse arrangement. In our case,  the quark and antiquark fields share the same transverse position, while the gluon field is transversely separated from them. These correlators are distinct from those considered in Ref.~\cite{Rodini:2022wki}, and we introduce their scalar parametrizations following an analogous construction. While this work was being completed, related three-body quark operator structures were independently obtained within an arbitrary-$x$ background-field formalism~\cite{Mukherjee:2026six}. Our analysis derives the operator structures directly within the NEik CGC framework and identifies the correlators probed by each dijet channel.

Beyond the derivation of the corresponding factorized cross sections, the operator structures identified in this work establish a direct connection between the NEik CGC framework and the twist-three quark TMD formalism. They enable a systematic future phenomenological studies of twist-three quark TMD effects in back-to-back DIS dijet production, particularly in view of the physics program of the future Electron-Ion Collider.

The remainder of this paper is organized as follows. In Sec.~\ref{sec:setup}, we introduce the kinematics and the momentum and coordinate variables used to describe the correlation limit. In Sec.~\ref{sec:quark-gluon}, we compute the quark-gluon dijet production cross section at NLP accuracy and establish its connection to quark TMD correlators. Section~\ref{sec:qbarq_production} presents the corresponding analysis for quark-antiquark dijet production, including the distinct three-body correlators probed in this channel. We summarize our findings and discuss future directions in Sec.~\ref{sec:summary}. Appendices~\ref{sec:app_1} and \ref{sec:app_2} provide the details of the Dirac algebra for the two production channels.

\section{Kinematics and the correlation limit}
\label{sec:setup}

In this work, we study dijet production in DIS at NEik accuracy, focusing on the quark background contributions. Although we first derive the scattering amplitudes in general kinematics, our main aim is to study the quark TMD contributions up to NLP in the back-to-back correlation limit For this purpose, it is convenient to introduce a new set of momentum and coordinate variables.

In our setup, the incoming virtual photon has virtuality $Q^2$, while the two produced jets carry transverse momenta $\k_1$ and $\k_2$ and longitudinal momenta $k_1^+$ and $k_2^+$. We define the longitudinal momentum fractions as
\begin{equation}
\label{eq:long_mom_frac}
z=\frac{k_1^+}{k_1^++k_2^+},
\qquad
1-z=\frac{k_2^+}{k_1^++k_2^+},
\end{equation}
and introduce the transverse momentum variables
\begin{equation}
\k=\k_1+\k_2,
\qquad
\P=(1-z)\k_1-z\k_2.
\end{equation}
Here, $\P$ characterizes the hard transverse momentum of the dijet system, while $\k$ denotes its transverse momentum imbalance. In terms of these variables, the individual jet momenta read
\begin{equation}
\k_1=\P+z\k,
\qquad
\k_2=-\P+(1-z)\k.
\label{cov_mom}
\end{equation}
A combination that will frequently appear in the scattering amplitudes is
\begin{equation}
\k_1-\frac{k_1^+}{k_2^+}\k_2
=
\frac{\P}{1-z}.
\end{equation}

In coordinate space, we introduce the transverse separation $\r$ and the longitudinal-momentum-weighted transverse position $\b$,
\begin{equation}
\r=\v-\z,
\qquad
\b=z\v+(1-z)\z,
\label{cov_cor}
\end{equation}
where $\v$ and $\z$ denote the transverse coordinates of the two outgoing partons. Equivalently,
\begin{equation}
\v=\b+(1-z)\r,
\qquad
\z=\b-z\r.
\end{equation}
With these definitions, the phase factor appearing in the scattering amplitudes becomes
\begin{equation}
e^{-i\v\cdot\k_1-i\z\cdot\k_2+i(\v-\z)\cdot\K}
=e^{-i\b\cdot\k-i\r\cdot(\P-\K)},
\label{phase_factors}
\end{equation}
where $\K$ denotes the transverse momentum flowing through the hard splitting kernel.

The back-to-back correlation limit is defined by the hierarchy
\begin{equation}
|\k|\ll|\P|.
\end{equation}
In coordinate space, the correlation limit motivates an expansion of the operator structures around vanishing transverse separation, $\r=0$. In the following sections, we perform this expansion for the quark background contributions to quark–gluon and quark–antiquark dijet production at NEik accuracy. For the quark-gluon channel, we compute both the kinematic corrections to the hard parts accompanying the two-body operators and the genuine twist-three contributions involving three-body quark operators with field-strength insertions. For the quark-antiquark channel, the quark-background contribution starts at twist-three accuracy and involves three-body quark operators with field-strength insertions.

\section{Quark-gluon dijet production}
\label{sec:quark-gluon}
Quark–gluon dijet production in DIS at NEik accuracy was studied in Ref.~\cite{Altinoluk:2023qfr}, where the scattering amplitudes were derived in general kinematics and the cross section was computed at LP in the correlation limit, corresponding to leading kinematic and genuine twist. In this section, we extend the correlation-limit calculation to NLP, including both the kinematic twist-three corrections to the hard parts accompanying the two-body operators and the genuine twist-three contributions involving three-body quark operators with field-strength insertions.

\subsection{Scattering amplitudes in general kinematics}
\label{sec:amplitude}

\begin{figure}[t]
\centering
\includegraphics[width=0.48\textwidth]{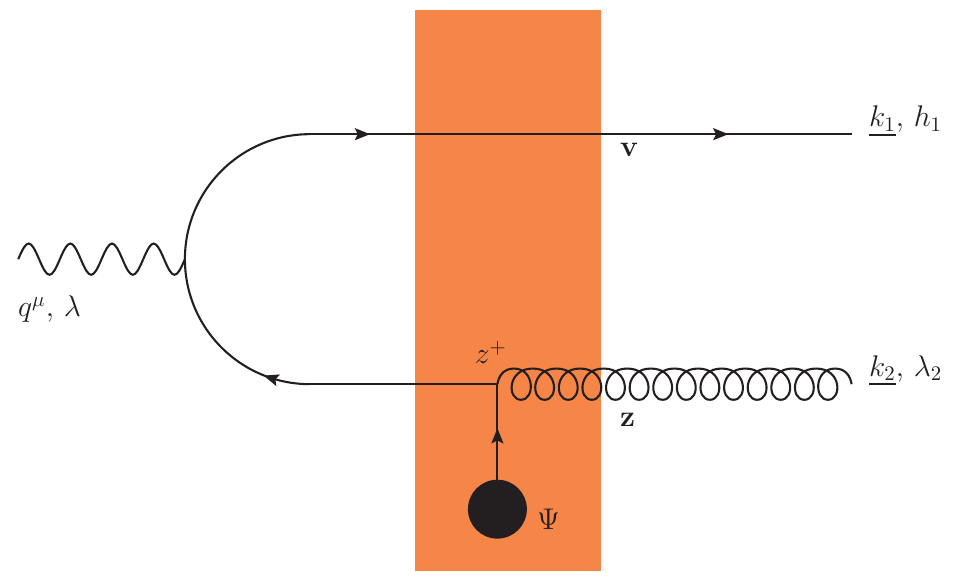}\hfill
\includegraphics[width=0.48\textwidth]{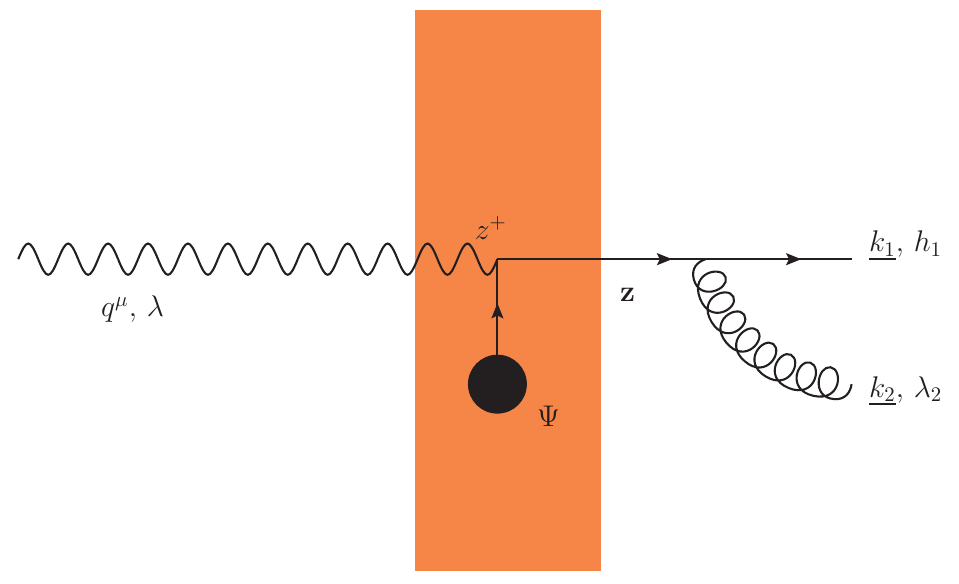}
\caption{Quark-gluon dijet production through the quark background, with photon splitting before the target (left) and inside the target (right).}
\label{fig:qg_bef_in}
\end{figure}

The scattering amplitude for quark–gluon dijet production receives two contributions, corresponding to the splitting of the virtual photon before or inside the target (see Fig.\ref{fig:qg_bef_in}). The expressions for these two contributions were derived in Ref.~\cite{Altinoluk:2023qfr}. In general kinematics, for the production of quark with momentum ${\underline k}_1\equiv(k_1^+,\k_1)$ and gluon ${\underline k}_2\equiv(k_2^+,\k_2)$, the corresponding scattering amplitudes, respectively read\footnote{For convenience, we use the shorthand notation $\int_{\x}\equiv\int\dd^2\x$ and $\int_{x^+}\equiv\int\dd x^+$ for transverse and longitudinal coordinate integrals, respectively.} 

\ba
i{\cal M}^{\rm bef}_{\gamma_{T,L}\to q_1g_2}&=&\frac{i e e_fg}{2q^+}  \int_{\v,\z}e^{-i\v\cdot\k_1-i\z\cdot\k_2}\int\frac{d^2\K}{(2\pi)^2}\frac{e^{i(\v-\z)\cdot\K}}{\left[ \K^2+m^2+\frac{k_1^+k^+_2}{(q^+)^2}Q^2\right]}\nonumber\\
&&\times 
{\bar u}(1)\frac{\gamma^+\gamma^-}{2} \Gamma^{\rm bef}_{T,L}
\int_{z^+} U_A\left(+\infty,z^+;\z\right)_{a_2b}
U_F(\v)U_F^\dagger\left(z^+,-\infty;\z\right)t^b
\Psi
(z^+,\z)\, ,
\label{bef_S_gen}
\\
i{\cal M}^{\rm in}_{\gamma_T\to q_1g_2}&=& \frac{i e e_fg}{2q^+}\, 
\frac{1}{ \left[ \left( \k_1-\frac{k_1^+}{k_2^+}\k_2\right)^2+m^2\right]}
\int_{\z}e^{-i\z\cdot(\k_1+\k_2)}
{\bar u}(1)\frac{\gamma^+\gamma^-}{2} \Gamma^{\rm in}_T
\int_{z^+} 
t^{a_2}U_F\left(+\infty,z^+;\z\right)
\Psi(z^+,\z)
\label{in_S_gen},
\ea
with 
\be
\label{gamma_bef}
\Gamma^{\rm bef}_L=2\frac{k_1^+k_2^+}{(q^+)^2}Q{\varepsilon^{j*}_{\lambda_2}}\gamma^j
\, , \hspace{0.5cm}
\Gamma^{\rm bef}_T=\varepsilon^i_\lambda \, {\varepsilon^{j*}_{\lambda_2}}
\left\{\K^l\left[\left(\frac{k_1^+-k_2^+}{q^+}\right)\delta^{il}-\frac{[\gamma^i,\gamma^l]}{2}\right]+m\gamma^i\right\} \gamma^j
\, ,
\ee

\be
\label{gamma_in}
\Gamma^{\rm in}_L=0 \, , \hspace{0.5cm}
\Gamma^{\rm in}_T=
{\varepsilon^{l*}_{\lambda_2}} \varepsilon^j_\lambda
\left\{ \left[\k_1^i-\frac{k_1^+}{k_2^+}\k_2^i\right]\left[-\left(\frac{2k_1^++k_2^+}{k_2^+}\right)\delta^{il}+\frac{[\gamma^i,\gamma^l]}{2}\right] +m\gamma^l\right\}
\gamma^j
\, .
\ee
Here, we have used $q^+=k_1^++k_2^+$. The indices $\lambda$ and $\lambda_2$ denote the polarizations of the incoming photon and the outgoing gluon, respectively, while $\bar u(1)$ is the outgoing-quark spinor. The vectors $\varepsilon_\lambda$ and $\varepsilon_{\lambda_2}$ denote the transverse polarization vectors of the photon and gluon, respectively. $\Psi$ denotes the target quark background field. The Wilson lines $U_F$ and $U_A$ account for the multiple interactions of the propagating parton with the target gluon background in the fundamental and adjoint representations, respectively. We use the shorthand notation
\begin{equation}
U_F(\x)\equiv U_F(+\infty,-\infty;\x).
\end{equation}
Both amplitudes contain a single insertion of the quark background field, together with the gluon background interactions encoded in the Wilson lines.

To study the back-to-back limit of the dijet production cross section, it is convenient to change variables from $(\k_1,\k_2)$ to $(\P,\k)$ and from $(\v,\z)$ to $(\r,\b)$, as defined in Eqs.~\eqref{cov_mom} and \eqref{cov_cor}, respectively. In terms of these variables, the ``before'' and ``inside'' amplitudes given in Eqs.~\eqref{bef_S_gen} and \eqref{in_S_gen} read
\ba
i{\cal M}^{\rm bef}_{\gamma^*_{T,L}\to q_1g_2}
&=&\frac{i e e_fg}{2q^+}  
\int_{\b,\r}
\int\frac{d^2\K}{(2\pi)^2}
\frac{e^{-i\k \cdot \b - i \r \cdot (\P-\K)}}{\left[ \K^2+\bar Q^2\right]} {\bar u}(1)\frac{\gamma^+\gamma^-}{2} \Gamma^{\rm bef}_{T,L}
\nonumber\\
&&
\times 
\int_{z^+} U_A\left(+\infty,z^+;\b-z\r\right)_{a_2b}
U_F(\b+(1-z)\r)U_F^\dagger\left(z^+,-\infty;\b-z\r\right)t^b
\Psi(z^+,\b-z\r)\, ,
\label{bef_S_new_var}\\
i{\cal M}^{\rm in}_{\gamma^*_T\to q_1g_2}
&=& 
\frac{i e e_fg}{2q^+}\, 
\frac{1}{ \left[ \frac{\P^2}{(1-z)^2}+m^2\right]}
\int_{\b}e^{-i\b\cdot \k}
{\bar u}(1)\frac{\gamma^+\gamma^-}{2} \Gamma^{\rm in}_T
\int_{z^+} 
t^{a_2}U_F\left(+\infty,z^+;\b  \right)
\Psi(z^+,\b)
\label{in_S_new_var},
\ea
where
\begin{equation}
\bar Q^2=m^2+z(1-z)Q^2,
\end{equation}
with $m$ being the quark mass. After the change of variables introduced above, and using the longitudinal momentum fraction defined in Eq.~\eqref{eq:long_mom_frac}, the Dirac hard factors $\Gamma_{L/T}^{\rm bef/in}$ take the form
\be
\label{gamma_bef_new}
\Gamma^{\rm bef}_L=2 z(1-z)\, Q{\varepsilon^{j*}_{\lambda_2}} \, \gamma^j
\, , \hspace{0.5cm}
\Gamma^{\rm bef}_T=\varepsilon^i_\lambda \, {\varepsilon^{j*}_{\lambda_2}}
\left\{\K^l\left[(2z-1)\delta^{il}-\frac{[\gamma^i,\gamma^l]}{2}\right]+m\gamma^i\right\} \gamma^j
\, ,
\ee
\be
\label{gamma_in_new}
\Gamma^{\rm in}_L=0 \, , \hspace{0.5cm}
\Gamma^{\rm in}_T=
{\varepsilon^{l*}_{\lambda_2}} \varepsilon^j_\lambda
\left\{ \frac{\P^i}{(1-z)}\left[- \frac{(1+z)}{(1-z)}\delta^{il}+\frac{[\gamma^i,\gamma^l]}{2}\right] +m\gamma^l\right\}
\gamma^j
\, .
\ee
The amplitudes in Eqs.~\eqref{bef_S_new_var} and \eqref{in_S_new_var} are still valid in general kinematics. In the next subsection, we perform the back-to-back correlation-limit expansion up to NLP accuracy.
%

\subsection{Scattering amplitudes in the back-to-back correlation limit}
\label{sec:ampl_corr_limit}

We are now ready to study the scattering amplitudes in the correlation limit defined in Section~II. The transverse-coordinate dependence of the ``before'' contribution in Eq.~(\ref{bef_S_new_var}) involves both $\b$ and $\r$, whereas the ``inside'' contribution in Eq.~(\ref{in_S_new_var}) depends only on $\b$. We therefore expand the operator structure in the former around $\r=0$, keeping the terms up to linear order in $\r$.

For this purpose, we introduce a generic expression containing the color structure of the ``before'' amplitude,  
\begin{equation}
\mathcal{O}_{qg}^{\rm bef}= \int_{\b, \r}e^{-i\b\cdot \k}  e^{-i\r\cdot \P} \, f(\r)\,  U_F\big(\b+(1-z)\r\big)\,  \mathcal{U}_{\rm bef}(\b-z\r)
\label{eq:operator_bef}
\end{equation}   
where $f(\r)$ represents the transverse separation dependence of the hard part  in the ``before'' amplitude and the effective $\mathcal{U}_{\rm bef}$ is defined as
\begin{equation}
\mathcal{U}_{\rm bef}(\z)\equiv \int_{z^+} U_A(+\infty, z^+;\z)_{{a_2} b}\, U_F^{\dagger}(z^+, -\infty;\z) t^b \Psi(z^+,\z)
\end{equation}
Expanding both factors in Eq. \eqref{eq:operator_bef} around $\b$ to linear order in $\r$ and integrating by parts in $\b$ to remove the derivative acting on $\mathcal{U}_{\rm bef}$, at NLP we obtain 
\begin{align}
O^{\rm bef}_{qg} \Big|_{\rm NLP}
&= \int_{\b,\r} e^{-i\b\cdot \k} \, e^{-i\r\cdot \P} \, f(\r)
\Big\{ \left[1 - iz\, \r\cdot \k\right] U_F(\b)\,\mathcal{U}_{\rm bef}(\b)
+ \r^i\left(\partial_{\b^i} U_F(\b)\right)\mathcal{U}_{\rm bef}(\b)
+ O\!\left(\frac{\k^2}{\P^2}\right)\Big\},
\label{eq:color_op_bef_NLP_1}
\end{align}
The term proportional to $\r\cdot\k$ arises from the derivative acting on the Fourier phase. When evaluated at the same transverse position, the product $U_F(\b)\mathcal{U}_{\rm bef}(\b)$ simplifies to 
\begin{align}
U_F(\b)\, {\mathcal U}_{\rm bef}(\b)&=\int_{z^+} U_A(+\infty,z^+;\b)_{a_2 b} \, U_F(\b)\, U_F^\dagger(z^+,-\infty;\b)\,  t^b \, \Psi(z^+,\b)
\nonumber\\
&= \int_{z^+} U_A(+\infty,z^+;\b)_{a_2 b} \, U_F(+\infty,z^+;\b) \, t^b\, \Psi(z^+,\b)
= \int_{z^+}t^{a_2}\, U_F(+\infty,z^+;\b)\Psi(z^+,\b)
\label{eq:UFUbef}
\end{align} 
where we have used the composition property of the fundamental Wilson lines and the relation between the fundamental and adjoint representations.
Moreover, following the treatment in Ref.~\cite{Altinoluk:2024zom}, the transverse derivative of the Wilson line can be expressed through a field-strength insertion,
\begin{equation}
\partial_{\b^i} U_F(\b) =
\int_ {z'^+} U_F(+\infty,z'^+;\b)\, \left[-ig F_i^{\; -}(z'^+,\b)\right]\,  U_F(z'^+,-\infty;\b),
\end{equation}
with $F_i^{\;-}=t^cF_{i,c}^{\;-}$. Using the same Wilson-line identities, the derivative term in Eq.~\eqref{eq:color_op_bef_NLP_1} becomes
\begin{equation}
\big( \partial_{\b^i}U_F(\b)\big) \mathcal{U}_{\rm bef}(\b)= \int_{z'^+, z^+} 
(-i) \Big[ g {F_{i}^{\; -}}_c(z'^+,\b)\Big] \, U_A(+\infty, z'^+;\b)_{dc}\, t^dt^{a_2}\, U_F(+\infty, z^+;\b)\, \Psi(z^+,\b)
\label{eq:derUFUbef}
\end{equation}
Substituting Eqs.~\eqref{eq:UFUbef} and \eqref{eq:derUFUbef} into Eq.~\eqref{eq:color_op_bef_NLP_1}, we obtain
\begin{align}
O^{\rm bef}_{qg}\Big|_{\rm NLP} &= \int_ {\b,\r} e^{-i\b\cdot \k} \,  e^{-i\r\cdot \P} \,  f(\r)
\bigg\{ \left[1 - iz\,\r\cdot \k\right]
\int _{z^+} t^{a_2}\, U_F(+\infty,z^+;\b)\,\Psi(z^+,\b)
\notag\\
&\quad + (-i) \r^i \int_{z^+, z'^+}
\left[U_A(+\infty,z'^+;\b)\right]_{dc}\, \Big[ g {F_{i}^{\; -}}_c(z'^+,\b)\Big]\,
t^d t^{a_2}\, U_F(+\infty,z^+;\b)\,\Psi(z^+,\b)
+ O\!\left(\frac{\k^2}{\P^2}\right) \bigg\}.
\label{eq:Obef_NLP_fin}
\end{align}
Substituting the generic expression of the operator expanded at NLP accuracy, Eq.\eqref{eq:Obef_NLP_fin}, into the ``before'' contribution to the scattering amplitude given in Eq. \eqref{bef_S_new_var}, we obtain 
\begin{align}
i\mathcal{M}^{\rm bef}_{\gamma^*_{T,L} \to q_1 g_2}
&= \frac{i \, e\, e_f\, g}{(2q^+)}
\int_{\b} e^{-i\k\cdot \b}
\int \frac{ {\rm d}\K}{(2\pi)^2}\,\frac{1}{\left[\K^2 + \Qb^2\right]}
\int_{\r} e^{-i\r\cdot(\P - \K)}
{\bar u}(1)\frac{\gamma^+\gamma^-}{2}\Gamma^{\rm bef}_{T,L}
\nonumber\\
&\hspace{-0.5cm} \times 
\left\{ \left[1-iz\,\r\cdot \k\right]\int_{z^+} t^{a_2}\,U_F(+\infty,z^+;\b)\Psi(z^+,\b)
\right.\nonumber\\
&\hspace{-0.5cm}\left.\quad + (-i) \r^i \int_{z^+,z'^+} 
\Big[U_A(+\infty,z'^+;\b)_{dc} \, g \, {F_{i}^{\; -}}_c(z'^+,\b)\Big]
\Big[t^d t^{a_2}U_F(+\infty,z^+;\b)\Psi(z^+,\b)\Big]
+ O\!\left(\frac{\k^2}{\P^2}\right) \right\}.
\end{align}
The integration over $\r$ gives a delta function at LP and its derivative at NLP:
\begin{equation}
\int_{\r}e^{-i\r\cdot(\P-\K)}=(2\pi)^2\, \delta^{(2)}(\P-\K)\; , 
\qquad
\int_{\r} \r^i\, e^{-i\r\cdot(\P-\K)}=i\partial_{\P^i}\Big[ (2\pi)^2\, \delta^{(2)}(\P-\K) \Big]
\end{equation}
Since $\Gamma_{T,L}^{\rm bef}$ may depend on  $\K$ but is independent of $\P$,  the $\P$ dependence originally only appears through the phase $e^{-i\r\cdot\P}$. Therefore, ``before" contribution to the scattering amplitude can be written as 
\begin{align} 
\label{M_bef_exp_1}
i\mathcal{M}^{\rm bef}_{\gamma_{T,L}^*\to q_1 g_2}
&= \frac{i\, e \, e_f\, g}{(2q^+)}
\int_{\b} e^{-i\k\cdot \b}
\bigg\{\left[1 + z\,\k^i \partial_{\P^i}\right]
\int_{z^+}\, t^{a_2}\, U_F(+\infty,z^+;\b)\, \Psi(z^+,\b)
\notag\\
&\hspace{1cm}+\;  \partial_{\P^i}\int_{z^+,z'^+}
\Big[U_A(+\infty,z'^+;\b)_{dc}\, g{F_{i}^{\; -}}_c(z'^+,\b)\Big]
\Big[t^dt^{a_2}U_F(+\infty,z^+;\b)\Psi(z^+,\b)\Big]
+O\!\left(\frac{\k^2}{\P^2}\right)\bigg\}
\notag\\
&\hspace{1cm} \times \;  \frac{1}{[\P^2 + \Qb^2]}\,\ubar(1)\frac{\gamma^+\gamma^-}{2}\Gamma^{\rm bef}_{T,L}(\P)\,\Psi(z^+,\b).
\end{align}
For a longitudinally polarized photon, the function $\Gamma_L^{\rm bef}$ given in Eq.~\eqref{gamma_bef_new} is independent of both $\K$ and $\P$. The derivative in Eq.~\eqref{M_bef_exp_1} therefore acts only on the denominator, yielding 
\begin{align}
\label{eq:ML_bef_NLP_fin}
i\mathcal{M}^{\rm bef}_{\gamma^*_L\to q_1 g_2}
&= \frac{i\, e\, e_f\, g}{(2q^+)}\,  2z(1-z)Q\,\varepsilon_{\lambda_2}^{j*}\,\ubar(1)\frac{\gamma^+\gamma^-}{2}\gamma^j
\int_{\b}  e^{-i\k\cdot \b}
\notag\\
& \hspace{-0.5cm}\times
\left\{\left(\frac{1}{[\P^2+\Qb^2]}
- \frac{2z(\P\cdot \k)}{[\P^2+\Qb^2]^2}\right)
\int_{z^+} t^{a_2}U_F(+\infty,z^+;\b)\Psi(z^+,\b)
\right.\notag\\
& \hspace{-0.5cm} \left. - \frac{2\P^i}{[\P^2+\Qb^2]^2}
\int_{z^+,z'^+} 
\Big[U_A(+\infty,z'^+;\b)_{dc}\, g{F_{i}^{\; -}}_c(z'^+,\b)\Big]
\Big[t^dt^{a_2}U_F(+\infty,z^+;\b)\Psi(z^+,\b)\Big]
+ O\!\left(\frac{\k^2}{\P^2}\right)\right\}.
\end{align}
For a transversely polarized photon, $\Gamma_T^{\rm bef}(\K)$ depends explicitly on $\K$ and becomes $\Gamma_T^{\rm bef}(\P)$ after the transverse integrations. The derivative therefore acts on both the denominator and the function $\Gamma_T^{\rm bef}(\P)$ itself:
\begin{equation}
\label{eq:der_on_T_hard/denom}     
\partial_{\P^i}\!
\bigg(\frac{\Gamma_T^{\rm bef}(\P)}{[\P^2+\Qb^2]}\bigg)
= -\frac{2\P^i}{[\P^2+\Qb^2]^2}\,\Gamma_T^{\rm bef}(\P)
+ \frac{1}{[\P^2+\Qb^2]}\Big(\partial_{\P^i}\Gamma_T^{\rm bef}(\P)\Big),
\end{equation}
where 
\begin{equation}
\label{eq:der_on_T_hard}
\partial_{\P^i}\Gamma_T^{\rm bef}(\P)
= \varepsilon_\lambda^m\varepsilon_{\lambda_2}^{j*}\delta_i^\ell
\left[(2z-1)\delta^{m\ell} - \frac{[\gamma^m,\gamma^\ell]}{2}\right]\gamma^j
= \varepsilon_\lambda^m\varepsilon_{\lambda_2}^{j*}
\left[(2z-1)\delta^{mi} - \frac{[\gamma^m,\gamma^i]}{2}\right]\gamma^j,
\end{equation}
Using Eq. \eqref{eq:der_on_T_hard/denom} together with Eq. \eqref{gamma_bef_new} for the LP hard factor and  Eq.\eqref{eq:der_on_T_hard} for its derivative, the ``before'' scattering amplitude for the transversely polarized photon can be written at NLP as 
\begin{align}
\label{eq:MT_bef_NLP_fin}
i\mathcal{M}^{\rm bef}_{\gamma_T^*\to q_1 g_2}
&= \frac{i\, e\, e_f\, g}{(2q^+)}\,\ubar(1)\frac{\gamma^+\gamma^-}{2}
\int_{\b} e^{-i\k\cdot \b}
\notag\\
&\quad\times
\Bigg\{
\bigg[\frac{\Gamma_T^{\rm bef}(\P)}{[\P^2+\Qb^2]}
- \frac{2z(\P\cdot \k)\,\Gamma_T^{\rm bef}(\P)}{[\P^2+\Qb^2]^2}
+ \varepsilon_\lambda^m\varepsilon_{\lambda_2}^{j*}
\frac{\k^i}{[\P^2+\Qb^2]}
\left((2z-1)\delta^{mi}-\frac{[\gamma^m,\gamma^i]}{2}\right)\gamma^j\bigg]
\notag\\
&\qquad\times \int_{z^+} t^{a_2}U_F(+\infty,z^+;\b)\Psi(z^+,\b)
\notag\\
&\quad + 
\left[-\frac{2\P^i}{[\P^2+\Qb^2]^2}\Gamma_T^{\rm bef}(\P)
+ \varepsilon_\lambda^m\varepsilon_{\lambda_2}^{j*}\frac{1}{[\P^2+\Qb^2]}
\left((2z-1)\delta^{mi}-\frac{[\gamma^m,\gamma^i]}{2}\right)\gamma^j \right]
\notag\\
&\qquad\times \int_{z^+,z'^+} 
\Big[U_A(+\infty,z'^+;\b)_{dc}\, g{F_{i}^{\; -}}_c(z'^+,\b)\Big]
\Big[t^dt^{a_2}U_F(+\infty,z^+;\b)\Psi(z^+,\b)\Big]\bigg\rgroup
+ O\!\left(\frac{\k^2}{\P^2}\right)
\Bigg\}
\end{align}
The ``before'' amplitudes for longitudinally and transversely polarized photons given in 
Eqs. \eqref{eq:ML_bef_NLP_fin} and \eqref{eq:MT_bef_NLP_fin} can be conveniently written in a common form as 
\begin{eqnarray}
\label{eq:M_bef_gen_fin}
i\mathcal{M}^{\rm bef}_{\gamma_{T,L}^*\to q_1 g_2}
& = &\frac{i\, e\, e_f\, g}{(2q^+)}\,\ubar(1)\frac{\gamma^+\gamma^-}{2}
\int_{\b}\,e^{-i\k\cdot \b}
\Bigg\{\frac{\Gamma_{T,L}^{\rm bef}(\P)}{[\P^2+\Qb^2]}
\int_{z^+}t^{a_2}U_F(+\infty,z^+;\b)\Psi(z^+,\b) 
\nonumber \\
& + & \; \Gamma_{T,L}^{\rm bef, NLP,i}
\bigg\lgroup z\k^i\int_{z^+}t^{a_2}U_F(+\infty, z^+;\b)\Psi(z^+,\b) 
\nonumber\\
&+& 
\int_{z^+,z'^+}
\Big[U_A(+\infty,z'^+;\b)_{dc}g{F_{i}^{\; -}}_c(z'^+,\b)\Big]
\Big[t^dt^{a_2}U_F(+\infty,z^+;\b)\Psi(z^+,\b)\Big]\bigg\rgroup
+ O\!\left(\frac{\k^2}{\P^2}\right)
\Bigg\}
\end{eqnarray}\
with the associated NLP hard factors are defined as 
\begin{align}
\label{eq:bef_NLP_i_L_new}
\Gamma_L^{{\rm bef, NLP},i}
&= 2z(1-z)Q\,\varepsilon_{\lambda_2}^{j*}\gamma^j\cdot\frac{(-2)\P^i}{[\P^2+\Qb^2]^2}, \\
\Gamma_T^{{\rm bef, NLP},i}
&= -\frac{2\P^i}{[\P^2+\Qb^2]^2}\Gamma_T^{\rm bef}(\P)
+ \frac{\varepsilon_\lambda^\ell \varepsilon_{\lambda_2}^{j*}}{[\P^2+\Qb^2]}
\left[(2z-1)\delta^{i\ell} - \frac{[\gamma^\ell,\gamma^i]}{2}\right]\gamma^j \nonumber\\
& = \varepsilon_\lambda^\ell \varepsilon_{\lambda_2}^{j*}
\Bigg\{\bigg[\frac{-2\, \P^i \P^k}{[\P^2+\Qb^2]^2}
+ \frac{\delta^{ik}}{[\P^2+\Qb^2]}\bigg]
\left[(2z-1)\delta^{k\ell} - \frac{[\gamma^\ell,\gamma^k]}{2}\right]
- \frac{2\P^i}{[\P^2+\Qb^2]^2}m\gamma^\ell \Bigg\}\gamma^j.
\label{eq:bef_NLP_i_T_new}
\end{align}
In the second line of Eq.~\eqref{eq:bef_NLP_i_T_new}, we have used the explicit expression for $\Gamma_T^{\rm bef}$ in Eq.~\eqref{gamma_bef_new}, evaluated at $\K=\P$.

Finally, the ``inside'' amplitude is given in Eq.~\eqref{in_S_new_var}, with its LP hard factor specified in Eq.~\eqref{gamma_in_new}. Since this contribution is independent of $\r$, no transverse-separation expansion is required.
\subsection{NEik quark-gluon dijet production cross section in DIS at NLP}
\label{sec:cross_sec}

Using the scattering amplitudes derived in the previous subsection, we now compute the quark–gluon dijet production cross section NLP accuracy. We treat the longitudinal and transverse polarizations of the incoming virtual photon separately, starting with the longitudinal case. In the most general case, the corresponding cross section can be written as
\begin{equation}
\label{eq:L_qg_Xsec_gen}
(2\pi)^6 \, (2k_1^+)\, (2k_2^+)\frac{\dd\sigma^{\gamma^*_L\to q_1g_2}}{\dd^2\k_1\, \dd k_1^+ \, \dd^2\k_2\, \dd k_2^+}
= (2q^+)\, (2\pi)\, \delta(k_1^++k_2^+-q^+)\, 
\sum_{h_1,\lambda_2}\sum_{\rm col.} \Big\langle\big| \mathcal{M}^{\rm bef}_{\gamma_{L}^*\to q_1 g_2}\big|^2\Big\rangle
\end{equation}
where we have used the fact that ``inside'' contribution to the scattering amplitude for a longitudinally polarized photon vanishes since $\Gamma_L^{\rm in}=0$ as shown in Eq. \eqref{gamma_in}. Here,  $\langle \cdots \rangle$ corresponds to the CGC averaging. The delta distribution on the right hand side of Eq. \eqref{eq:L_qg_Xsec_gen} arises from the static approximation which is valid at NEik accuracy for all the contributions containing quark background field.  Hence, any plus momentum transfer from the target is beyond the accuracy of our study. Therefore, it is natural to focus on the cross section which is integrated over this plus momentum of the dijet system. As a result, in the back-to-back variables defined in Sec.\ref{sec:setup}, the cross section reads 
\begin{equation}
(2\pi)^5 2z(1-z)\frac{\dd\sigma^{\gamma^*_L\to q_1g_2}}{\dd^2\k\, \dd z\, \dd^2\P}
=\sum_{h_1,\lambda_2}\sum_{\rm col.} \Big\langle\big| \mathcal{M}^{\rm bef}_{\gamma_{L}^*\to q_1 g_2}\big|^2\Big\rangle
\end{equation}
Using the explicit expression of the ``before'' amplitude for longitudinally polarized photon given in Eq. \eqref{eq:M_bef_gen_fin}, the cross section can be written as 
\begin{align}
(2\pi)^5 2z(1-z)\frac{\dd\sigma^{\gamma^*_L\to q_1g_2}}{\dd^2\k\, \dd z\, \dd^2\P} 
& = 
\frac{e^2\, e^2_f\, g^2}{(2q^+)^2}\int_{\b\b'}e^{i\k\cdot(\b'-\b)}\sum_{h_1,\lambda_2}\sum_{\rm col.}
\bigg\langle \bigg\{ 
\int_{z^+,z'^+}\bar{\Psi}(z'^+,\b')U^\dagger_F(+\infty,z'^+;\b') t^{a_2}
\nonumber\\
& \times 
\gamma^0
\bigg[ \frac{\Gamma^{\rm bef}_L}{\big[ \P^2+\Qb^2\big]} +z\k^i\Gamma_L^{{\rm bef, NLP},i}
\bigg]^\dagger \gamma^0 \frac{\gamma^-\gamma^+}{2} 
u(1)\bar{u}(1)
\frac{\gamma^+\gamma^-}{2}
\bigg[ \frac{\Gamma^{\rm bef}_L}{\big[ \P^2+\Qb^2\big]} +z\k^{i'}\Gamma_L^{{\rm bef, NLP},i'}
\bigg]
\nonumber\\
& \times t^{a_2}U_F(+\infty, z^+;\b)\Psi(z^+,\b)
\nonumber \\
& +
\int_{z^+,z'^+,z''^+}\!\!\!\! \bar{\Psi}(z''^+,\b')\, U^\dagger_F(+\infty, z''^+;\b')\,  t^{a_2}\gamma^0 \frac{\big(\Gamma_L^{\rm bef}\big)^\dagger}{\big[ \P^2+\Qb^2\big]}\gamma^0\, 
\frac{\gamma^-\gamma^+}{2}\, u(1)\bar{u}(1)\, \frac{\gamma^+\gamma^-}{2}\, 
\Gamma_{L}^{{\rm bef, NLP},i}
\nonumber\\
& \times
\Big[U_A(+\infty,z'^+;\b)_{dc}\, g{F_{i}^{\; -}}_c(z'^+,\b)\Big]
\Big[t^dt^{a_2}U_F(+\infty,z^+;\b)\Psi(z^+,\b)\Big] 
\nonumber\\
& +
\int_{z^+,z'^+,z''^+}\!\!\!\! \bar{\Psi}(z''^+,\b')\, U^\dagger_F(+\infty, z''^+;\b')\,   \, 
\Big[ t^{a_2}t^d \, g{F_{i}^{\; -}}_c(z'^+,\b')\, U^\dagger_A(+\infty,z'^+;\b')_{cd}\Big] 
\nonumber \\
& \times 
\gamma^0\Big( \Gamma_L^{{\rm bef, NLP}, i}\Big)^\dagger\gamma^0
\, \frac{\gamma^-\gamma^+}{2}\, u(1)\bar{u}(1)\, \frac{\gamma^+\gamma^-}{2}\, \frac{\Gamma_L^{\rm bef}}{\big[ \P^2+\Qb^2\big]} \,  t^{a_2}U_F(+\infty, z^+;\b)\Psi(z^+,\b) 
\nonumber \\
& + 
\mathcal{O}({\rm NNLP})\bigg\}\bigg\rangle
\label{eq:qg_Xsec_NLP_1}
\end{align}
This cross section can be simplified by performing the summmation over the colors
\begin{equation}
\label{eq:color_sum_bef}
\sum_{\rm col.} t^{a_2}t^{a_2}=C_F\, \bm{1}_F \qquad \text{and} \qquad
\sum_{\rm col.} t^{a_2}t^dt^{a_2}= \Big(C_F-\frac{C_A}{2}\Big) t^d
\end{equation}
with the quadratic Casimir operators in fundamental and adjoint representations are given as 
\begin{equation}
C_F=\frac{N_c^2-1}{2N_c}\qquad \text{and} \qquad C_A=N_c\, .
\end{equation}
Moreover, one can also simplify the Dirac structures appearing in Eq. \eqref{eq:qg_Xsec_NLP_1}. 
First of all, by using the explicit expressions for the hard factor at LP and NLP given in Eqs. \eqref{gamma_bef_new} and \eqref{eq:bef_NLP_i_L_new}, the first Dirac structure can be written as  
\begin{eqnarray}x
&& \hspace{-1.5cm}
\sum_{h_1,\lambda_2} \gamma^0\bigg[ \frac{\Gamma_L^{\rm bef}}{\big[ \P^2+\Qb^2\big]}+z\k^i\Gamma_L^{{\rm bef, NLP},i}\bigg]^\dagger \gamma^0 \, \frac{\gamma^-\gamma^+}{2} 
u(1)\bar{u}(1)
\frac{\gamma^+\gamma^-}{2}
\bigg[ \frac{\Gamma^{\rm bef}_L}{\big[ \P^2+\Qb^2\big]} +z\k^{i'}\Gamma_L^{{\rm bef, NLP},i'}
\bigg]
\nonumber \\
& & 
\hspace{2cm}
= 4z^2(1-z)^2Q^2\bigg[ \frac{1}{[\P^2+\Qb^2]^2}-4z\frac{\P\cdot\k}{[\P^2+\Qb^2]^3}\bigg]
\sum_{h_1,\lambda_2}\varepsilon^{j*}_{\lambda_2}\varepsilon^{j\prime}_{\lambda_2} 
\gamma^{j\prime}\, \frac{\gamma^-\gamma^+}{2} u(1)\bar{u}(1)\frac{\gamma^+\gamma^-}{2} \, \gamma^j
\label{eq:dirac_bef_NLP_1}
\end{eqnarray}
Noting that 
\begin{equation}
\label{eq:sum_h1_lambda_1_v1}
\sum_{h_1,\lambda_2}\varepsilon^{j*}_{\lambda_2}\varepsilon^{j\prime}_{\lambda_2} 
\gamma^{j\prime}\, \frac{\gamma^-\gamma^+}{2} u(1)\bar{u}(1)\frac{\gamma^+\gamma^-}{2} \, \gamma^j
= (2q^+)\, z\, \delta^{jj'\, }\Big(-\frac{\gamma^-}{2}\Big) \frac{\{\gamma^{j\prime},\gamma^j\}}{2}
= (2q^+)\, z\, \gamma^-
\end{equation}
where we have used 
\begin{equation}
\sum_{h_1}\frac{\gamma^-\gamma^+}{2}u(1){\bar u}(1)\frac{\gamma^+\gamma^-}{2}= (2q^+)\, z\, \frac{\gamma^-}{2}
\end{equation}
Substituting Eq. \eqref{eq:sum_h1_lambda_1_v1} into Eq. \eqref{eq:dirac_bef_NLP_1}, we obtain 
\begin{eqnarray}
&& \hspace{-1.5cm}
\sum_{h_1,\lambda_2} \gamma^0\bigg[ \frac{\Gamma_L^{\rm bef}}{\big[ \P^2+\Qb^2\big]}+z\k^i\Gamma_L^{{\rm bef, NLP},i}\bigg]^\dagger \gamma^0 \, \frac{\gamma^-\gamma^+}{2} 
u(1)\bar{u}(1)
\frac{\gamma^+\gamma^-}{2}
\bigg[ \frac{\Gamma^{\rm bef}_L}{\big[ \P^2+\Qb^2\big]} +z\k^{i'}\Gamma_L^{{\rm bef, NLP},i'}
\bigg]
\nonumber \\
& & \hspace{2cm} =
4\, z^2\, (1-z)^2\, Q^2\, (2q^+) \, z\, \gamma^-\Bigg[ \frac{1}{\big[ \P^2+\Qb^2\big]^2}-4z\frac{\P\cdot\k}{\big[\P^2+\Qb^2\big]^3}\Bigg]+ \mathcal{O}({\rm NNLP})
\label{eq:dirac_bef_NLP_2}
\end{eqnarray}
The computation of  the interference terms of $\Gamma_L^{\rm bef}$ and $\Gamma_L^{{\rm bef, NLP},i}$ appearing in Eq. \eqref{eq:bef_NLP_i_L_new} can be performed in a similar way and the result reads 
\begin{align}
\label{eq:dirac_bef_NLP_3}
\sum_{h_1,\lambda_2}
\gamma^0\frac{\big( \Gamma_L^{\rm bef}\big)^\dagger}{\big[ \P^2+\Qb^2\big]}\gamma^0
\, \frac{\gamma^-\gamma^+}{2}u(1){\bar u}(1)\frac{\gamma^+\gamma^-}{2}\, 
\Gamma_L^{{\rm bef, NLP},i}
&=
\sum_{h_1,\lambda_2}
\gamma^0 \Big( \Gamma_L^{{\rm bef,NLP}, i}\Big)^\dagger\gamma^0 \, 
\frac{\gamma^-\gamma^+}{2}u(1){\bar u}(1)\frac{\gamma^+\gamma^-}{2} \, 
\frac{\Gamma_L^{\rm bef}}{\big[ \P^2+\Qb^2\big]}
\nonumber\\
&
\hspace{-1cm}
=
4\, z^2\, (1-z)^2\, Q^2\, (2q^+)\, z\, \gamma^-(-2)\frac{\P^i}{\big[\P^2+\Qb^2\big]^3} 
+\mathcal{O}({\rm NNLP})
\end{align}
Finally, using Eqs. \eqref{eq:color_sum_bef}, \eqref{eq:dirac_bef_NLP_2} and \eqref{eq:dirac_bef_NLP_3} in Eq. \eqref{eq:qg_Xsec_NLP_1}, one can write the NEik quark-gluon dijet production cross section at NLP as 
\begin{align}
\label{eq:qg_Xsec_NLP_2}
&
(2\pi)^5 2z(1-z)\frac{\dd\sigma^{\gamma^*_L\to q_1g_2}}{\dd^2\k\, \dd z\, \dd^2\P} 
 = 
\frac{e^2\, e^2_f\, g^2}{(2q^+)} \, 4\, z^3 (1-z)^2\, Q^2\int_{\b\b'}e^{i\k\cdot(\b'-\b)}
\nonumber\\
& \times 
\Big\langle\Big\{ 
C_F\bigg[ \frac{1}{[\P^2+\Qb^2]^2}-4z\frac{\P\cdot\k}{[\P^2+\Qb^2]^3}\bigg]
\int_{z^+,z'^+}\bar{\Psi}(z'^+,\b')\gamma^-U^\dagger_F(+\infty,z'^+;\b')U_F(+\infty, z^+;\b)\Psi(z^+,\b)
\nonumber\\
& \hspace{0.5cm}-2\frac{\P^i}{[\P^2+\Qb^2]^3}\bigg(C_F-\frac{C_A}{2}\bigg)\int_{z^+,z'^+,z''^+}
\Big[ 
\bar{\Psi}(z''^+,\b')\gamma^-U^\dagger_F(+\infty, z''^+;\b') t^d 
U_A(+\infty, z'^+;\b)_{dc}\, g{F_{i}^{\; -}}_c(z'^+,\b) 
\nonumber\\
& \hspace{7cm} 
\times 
U_F(+\infty, z^+;\b)\Psi(z^+,\b)
\nonumber\\
& \hspace{6.3cm}
+ \bar{\Psi}(z''^+,\b')\gamma^- U_F^\dagger(+\infty,z''^+;\b')\, g{F_{i}^{\; -}}_c(z'^+,\b') 
U_A^\dagger(+\infty, z'^+;\b')_{cd} \, t^d 
\nonumber \\
& \hspace{7cm} 
\times 
U_F(+\infty,z^+;\b)\Psi(z^+,\b)\Big]+ \mathcal{O}({\rm NNLP})\Big\}\Big\rangle
\end{align}
Let us now define the following target-averaged color operators:  
\begin{align}
\label{eq:unshifted_11}
\T_{11}(\k) &\equiv \int_{\b\b'}\,e^{i\k\cdot(\b'-\b)}
\int_{z^+,z'^+}
\big\langle \, 
\bar\Psi(z'^+,\b')\gamma^-U^\dagger_F(+\infty,z'^+;\b') U_F(+\infty,z^+;\b)\Psi(z^+,\b) \, 
\big\rangle,
\\
\label{eq:unshifted_i12}
\T_{i,12}(\k) &\equiv \int_{\b\b'}e^{i\k\cdot(\b'-\b)}
\int_{z^+,z'^+,z''^+}
\big\langle \, 
\bar\Psi(z''^+,\b')\gamma^-U^\dagger_F(+\infty,z''^+;\b') \, t^d \, U_A(+\infty,z'^+;\b)_{dc}
\notag\\
&\hspace{9cm}\times 
\,g{F_i^{\; -}}_c(z'^+,\b)\,U_F(+\infty,z^+;\b)\Psi(z^+,\b)\big\rangle,
\\
\label{eq:unshifted_i21}
\T_{i,21}(\k) &\equiv \int_{\b\b'}e^{i\k\cdot(\b'-\b)}
\int_{z^+,z'^+,z''^+}
\big\langle
\bar\Psi(z''^+,\b')\gamma^-U^\dagger_F(+\infty,z''^+;\b')\,  t^d\, U_A(+\infty,z'^+;\b')_{dc}
\notag\\
& \hspace{9cm}\times 
g{F_i^{\; -}}_c(z'^+,\b')\,U_F(+\infty,z^+;\b)\Psi(z^+,\b) \big\rangle
\end{align} 
For later use, we generalize these target-averaged operators by allowing an arbitrary Dirac matrix $\Gamma$ between the quark fields.
These generalized operators then can be defined as 
\begin{align}
\label{eq:unshifted_gen_11}
\T_{11}[\Gamma](\k) &\equiv \int_{\b\b'}\,e^{i\k\cdot(\b'-\b)}
\int_{z^+,z'^+}
\big\langle \, 
\bar\Psi(z'^+,\b')U^\dagger_F(+\infty,z'^+;\b')\, \Gamma\,  U_F(+\infty,z^+;\b)\Psi(z^+,\b) \, 
\big\rangle,
\\
\label{eq:unshifted_gen_i12}
\T_{i,12}[\Gamma](\k) &\equiv \int_{\b\b'}e^{i\k\cdot(\b'-\b)}
\int_{z^+,z'^+,z''^+}
\big\langle \, 
\bar\Psi(z''^+,\b')U^\dagger_F(+\infty,z''^+;\b')\, \Gamma \, t^d \, U_A(+\infty,z'^+;\b)_{dc}
\notag\\
&\hspace{8cm}\times 
\,g{F_i^{\; -}}_c(z'^+,\b)\,U_F(+\infty,z^+;\b)\Psi(z^+,\b)\big\rangle,
\\
\label{eq:unshifted_gen_i21}
\T_{i,21}[\Gamma](\k) &\equiv \int_{\b\b'}e^{i\k\cdot(\b'-\b)}
\int_{z^+,z'^+,z''^+}
\big\langle
\bar\Psi(z''^+,\b')U^\dagger_F(+\infty,z''^+;\b')\, \Gamma \, t^d\, U_A(+\infty,z'^+;\b')_{dc}
\notag\\
& \hspace{8cm}\times 
g{F_i^{\; -}}_c(z'^+,\b')\,U_F(+\infty,z^+;\b)\Psi(z^+,\b) \big\rangle
\end{align} 

The original operators are recovered by setting $\Gamma=\gamma^-$. Following the procedure discussed in Ref.~\cite{Altinoluk:2023qfr}, we can express the generalized operators defined through CGC-like target averaging in terms of quantum matrix elements in the target state $|P_{\rm tar}\rangle$, with momentum $P_{\rm tar}^\mu$. For a target-averaged operator $\langle\mathcal{\hat O}\rangle$, this identification reads
\begin{equation}
\langle\mathcal{\hat O}\rangle
=
\lim_{P_{\rm tar}'\to P_{\rm tar}}
\frac{
\langle P_{\rm tar}'|\hat{\mathcal{O}}|P_{\rm tar}\rangle
}{
\langle P_{\rm tar}'|P_{\rm tar}\rangle
},
\label{def_average}
\end{equation}
where $\hat{\mathcal{O}}$ is the corresponding quantum operator. We use the target-state normalization
\begin{equation}
\langle P_{\rm tar}'|P_{\rm tar}\rangle
=
2P_{\rm tar}^-(2\pi)^3
\delta(P_{\rm tar}^{\prime-}-P_{\rm tar}^-)
\delta^{(2)}(\P_{\rm tar}'-\P_{\rm tar}).
\label{norm_states}
\end{equation}

To simplify these matrix elements, we use translation invariance to express their coordinate dependence through relative separations. Under a translation by $a^\mu$, a local operator satisfies
\begin{equation}
\hat{\mathcal{O}}(x)
=
e^{ia^\mu\hat P_\mu}
\hat{\mathcal{O}}(x-a)
e^{-ia^\mu\hat P_\mu},
\end{equation}
where $\hat P_\mu$ is the momentum operator. A simultaneous translation of all operator coordinates therefore gives
\begin{align}
&\langle P_{\rm tar}'|
\hat{\mathcal{O}}_1(x_1)\cdots\hat{\mathcal{O}}_n(x_n)
|P_{\rm tar}\rangle
\nonumber\\
&\qquad=
e^{ia^\mu(P_{\rm tar}'-P_{\rm tar})_\mu}
\langle P_{\rm tar}'|
\hat{\mathcal{O}}_1(x_1-a)\cdots\hat{\mathcal{O}}_n(x_n-a)
|P_{\rm tar}\rangle.
\label{transl_correlator}
\end{align}
For the generalized operators considered here, this translation acts on both the field insertions and the Wilson-line paths. 
To express the generalized operators in terms of relative coordinates, we first translate all field coordinates by the quark-field coordinate, keeping $P_{\rm tar}'\neq P_{\rm tar}$ at this stage. The quark field is then evaluated at the origin, while the dependence on its original coordinate appears through the translation phase. Integrating over this common longitudinal and transverse position generates the momentum-conserving delta functions that cancel those in the target-state normalization. Taking the forward limit after this cancellation leaves the overall factor $1/(2P_{\rm tar}^-)$. The generalized operators can therefore be written as
\begin{align}
\label{eq:T_gen_shifted_11}
\T_{11}[\Gamma](\k) &= \int_{\Delta\b}\,\frac{e^{i\k\cdot\Delta\b}}{2P^-_{\rm tar}}
\int_{\Delta z^+}
\langle P_{\rm tar}| \, 
\bar\Psi(\Delta z^+,\Delta \b)
U^\dagger_F(+\infty,\Delta z^+;\Delta \b)\, \Gamma \, U_F(+\infty,0;{\bf 0})\Psi(0,{\bf 0}) \, 
| P_{\rm tar}\rangle,
\\
\label{eq:T_gen_shifted_12}
\T_{i,12}[\Gamma](\k) &=  \int_{\Delta \b}\frac{e^{i\k\cdot\Delta \b}}{2P^-_{\rm tar}}
\int_{\Delta z^+,\Delta z'^+}
\langle P_{\rm tar}| \, 
\bar\Psi(\Delta z'^+,\Delta \b)U^\dagger_F(+\infty,\Delta z'^+;\Delta \b) \, \Gamma \, t^d \, U_A(+\infty,\Delta z^+;{\bf 0})_{dc}
\notag\\
&\hspace{8cm}\times 
\,g{F_i^{\; -}}_c(\Delta z^+,{\bf 0})\,U_F(+\infty,0;{\bf 0})\Psi(0,{\bf 0})\, |P_{\rm tar}\rangle,
\\
\label{eq:T_gen_shifted_21}
\T_{i,21}[\Gamma](\k) &\equiv \int_{\Delta \b}\frac{e^{i\k\cdot\Delta \b}}{2P^-_{\rm tar}}
\int_{\Delta z^+,\Delta z'^+}
\langle P_{\rm tar}|
\bar\Psi(\Delta z'^+,\Delta \b)U^\dagger_F(+\infty,\Delta z'^+;\Delta \b)\, \Gamma\,  t^d\, U_A(+\infty,\Delta z^+;\Delta \b)_{dc}
\notag\\
& \hspace{8cm}\times 
g{F_i^{\; -}}_c(\Delta z^+,\Delta \b)\,U_F(+\infty,0;{\bf 0})\Psi(0,{\bf 0})\, |P_{\rm tar}\rangle
\end{align} 
where we have introduced the following notations to define the relative coordinates as 
\begin{align}
\Delta\b\equiv \b'-\b\; , \qquad \Delta z^+\equiv z'^+-z^+\; , \qquad \Delta z'^+\equiv z''^+-z^+\; . 
\end{align}
Defining the longitudinal hard coefficients as
\begin{align}
\label{def:C_L}
\mathcal{C}_L(z,\k,\P)& \equiv \frac{4\, z^3(1-z)^2\, Q^2\, C_F}{[\P^2+\Qb^2]^2}\bigg[ 1-4z\frac{\P\cdot\k}{[\P^2+\Qb^2]}\bigg]
\\
\mathcal{C}^i_L(z,\P)&\equiv 16\, z^3(1-z)^2\, Q^2\bigg[\frac{C_A}{2}-C_F\bigg]\frac{\P^i}{[\P^2+\Qb^2]^3}
\label{def:C_L_i}
\end{align}
such that the NEik quark-gluon DIS dijet production via longitudinally polarized photon can be organized as  
\begin{equation}
\label{eq:qg_Xsec_NLP_3}
(2\pi)^5 2z(1-z)\frac{\dd\sigma^{\gamma^*_L\to q_1g_2}}{\dd^2\k\, \dd z\, \dd^2\P} 
 = 
\frac{e^2\, e^2_f\, g^2}{(2q^+)} \bigg[ \mathcal{C}_L(z,\k,\P) \, \T_{11}[\gamma^-](\k)+ \frac{1}{2}\, \mathcal{C}^i_L(z,\P)\Big( \T_{i,12}[\gamma^-](\k)+\T_{i,21}[\gamma^-](\k)\Big)\bigg] 
\end{equation}
Eq. \eqref{eq:qg_Xsec_NLP_3} provides the final expression of the quark-gluon dijet cross section for the production via longitudinal photon written in terms of the generalized target averaged color operators listed in Eqs. \eqref{eq:T_gen_shifted_11}, \eqref{eq:T_gen_shifted_12} and \eqref{eq:T_gen_shifted_21}. 
\\

We now turn to the quark–gluon dijet production cross section for a transversely polarized virtual photon, at NEik and at NLP accuracy in the correlation limit. Unlike the longitudinal case, the ``inside'' amplitude also contributes. The cross section therefore contains the squares of the ``before'' and ``inside'' amplitudes, together with their interference terms:
\begin{align}
\label{eq:Transverse_X_sec_1}
(2\pi)^5 2z(1-z)\frac{\dd\sigma^{\gamma^*_T\to q_1g_2}}{\dd^2\k\, \dd z\, \dd^2\P} 
 &= \frac{1}{2}\sum_{\lambda}\sum_{h_1,\lambda_2}\sum_{\rm col.}
 \Big\langle \big|
 \mathcal{M}^{\rm bef}_{\gamma_{T}^*\to q_1 g_2}+\mathcal{M}^{\rm in}_{\gamma_{T}^*\to q_1 g_2}  \big|^2\Big\rangle 
 \nonumber\\
 &= \frac{1}{2}\sum_{\lambda, h_1,\lambda_2}\sum_{\rm col.}
 \Big\langle \big|\mathcal{M}^{\rm in}_{\gamma_{T}^*\to q_1 g_2}  \big|^2+  \big|\mathcal{M}^{\rm bef}_{\gamma_{T}^*\to q_1 g_2}  \big|^2
 + \big(\mathcal{M}^{\rm in}_{\gamma_{T}^*\to q_1 g_2}\big)^\dagger \big(\mathcal{M}^{\rm bef}_{\gamma_{T}^*\to q_1 g_2}\big)
 \nonumber\\
 & \hspace{7.3cm}
 +
 \big(\mathcal{M}^{\rm bef}_{\gamma_{T}^*\to q_1 g_2}\big)^\dagger \big(\mathcal{M}^{\rm in}_{\gamma_{T}^*\to q_1 g_2}\big) \Big\rangle
\end{align}
The factor of $1/2$ and the summation over $\lambda$ corresponds  to averaging over the transverse polarizations of the incoming photon. We evaluate each contribution following the same procedure as in the longitudinal case. The required Dirac algebra is presented in Appendix~\ref{sec:app_1}. Here, we quote the results entering the cross section at NLP accuracy.

We begin with the contribution in which the incoming photon interacts with the target through a $t$-channel quark exchange inside the shockwave in both the amplitude and its complex conjugate. This contribution, corresponding to the square of the ``inside'' amplitude, reads
\begin{align}
\label{eq:Trans_in_square_1}
\frac{1}{2}\sum_{\lambda,h_1,\lambda_2}\sum_{\rm col.}\Big\langle \big| {\mathcal M}_{\gamma^*_T\to q_1g_2}^{\rm in}\big|^2\Big\rangle 
&=
\frac{e^2\, e_f^2\, g^2}{(2q^+)^2}\int_{\b'\b}e^{-i\k\cdot(\b-\b')} \, \frac{1}{2}\sum_{\lambda,h_1,\lambda_2}\sum_{\rm col.} \int_{z^+,z'^+}
 \Big\langle 
\bar{\Psi}(z'^+,\b')U^\dagger_F(+\infty, z'^+;\b') t^{a_2} 
\nonumber\\
& \hspace{2cm}\times 
\gamma^0 \frac{(1-z)^2\big(\Gamma_T^{\rm in}\big)^\dagger}{[\P^2+(1-z)^2m^2]}\gamma^0\frac{\gamma^-\gamma^+}{2} u(1)\bar{u}(1) \frac{\gamma^+\gamma^-}{2} \frac{(1-z)^2\Gamma_T^{\rm in}}{[\P^2+(1-z)^2m^2]}
\nonumber \\
& \hspace{7.5cm}\times 
t^{a_2} U_F(+\infty, z^+; \b) \Psi(z^+,\b)\Big\rangle
\end{align}
where we have used the amplitude given in Eq. \eqref{in_S_new_var}. The corresponding Dirac structure, evaluated explicitly in Appendix \ref{app:inside-inside_cont} and given in Eq. \eqref{app:DT_in_in}, and it reads  
\begin{align}
\label{eq:Trans_in_square_dirac_fin}
\frac{1}{2}\sum_{\lambda,h_1,\lambda_2}(1-z)^4\gamma^0\big(\Gamma_T^{\rm in}\big)^\dagger \gamma^0\frac{\gamma^-\gamma^+}{2}u(1)\bar{u}(1)\frac{\gamma^+\gamma^-}{2}\Gamma_T^{\rm in}= (2q^+) z\big[ (1+z^2)\P^2+(1-z)^4m^2\big]\gamma^-
\end{align}
Substituting Eq. \eqref{eq:Trans_in_square_dirac_fin} into Eq. \eqref{eq:Trans_in_square_1} and performing the summation over the color using Eq. \eqref{eq:color_sum_bef}, we obtain
\begin{align}
\label{eq:trans_in_square_fin}
\frac{1}{2}\sum_{\lambda,h_1,\lambda_2}\sum_{\rm col.}\Big\langle \big| {\mathcal M}_{\gamma^*_T\to q_1g_2}^{\rm in}\big|^2\Big\rangle 
&= \frac{e^2\, e_f^2\, g^2}{(2q^+)} \, C_F\, \frac{z\big[ (1+z^2)\P^2+(1-z)^4m^2\big]}{[\P^2+(1-z)^2m^2]^2}\T_{11}[\gamma^-](\k)
\end{align}
where the target-averaged operator $\T_{11}[\gamma^-](\k)$ is defined in Eq. \eqref{eq:T_gen_shifted_11} after changing from the CGC averaging to quantum expectation value in the target state. 

We next consider the square of the ``before'' amplitude, including both the LP and NLP contributions derived in previous subsection. This contribution reads 
\begin{align}
\label{eq:trans_bef_square_1}
&
\frac{1}{2}\sum_{\lambda, h_1,\lambda_2}\sum_{\rm col.}\Big\langle \big| \mathcal{M}_{\gamma^*_T\to q_1g_2}^{\rm bef}\big|^2\Big\rangle
= \frac{e^2\, e_f^2\, g^2}{(2q^+)^2}\int_{\b'\b}e^{-i\k\cdot(\b-\b')} \, \frac{1}{2}\sum_{\lambda,h_1,\lambda_2}\sum_{\rm col.} \int_{z^+,z'^+} 
\Big\langle \bar{\Psi}(z'^+, \b') U^\dagger_F(+\infty, z'^+;\b')t^{a_2}
\nonumber\\
& \times 
\bigg\{ \gamma^0\bigg[ \frac{\Gamma_T^{\rm bef}(\P)}{[\P^2+\Qb^2]}+z\k^{i'}\Gamma_T^{{\rm bef, NLP}, i'}\bigg]^\dagger \!\gamma^0
+\gamma^0\Big(\Gamma_T^{{\rm bef, NLP},i'}\Big)^\dagger\gamma^0
\int_{z''^+} t^{d'}U_A(+\infty, z''^+;\b')_{d'c'} \, g{F_{i'}^{\; -}}_{c'}(z''^+,\b')\bigg\}
\nonumber\\
& \hspace{0.5cm} 
\times 
\frac{\gamma^-\gamma^+}{2}u(1)\bar{u}(1)\frac{\gamma^+\gamma^-}{2}
\nonumber\\
& \times 
\bigg\{ \bigg[ \frac{\Gamma_T^{\rm bef}(\P)}{[\P^2+\Qb^2]}+ z\k^i\Gamma_T^{{\rm bef, NLP}, i}\bigg]
+ \Gamma_T^{{\rm bef, NLP}i}\int_{z'''^+} t^dU_A(+\infty, z'''^+;\b)_{dc}\, g{F_i^{\; -}}_c
(z'''^+,\b)\bigg\}
\nonumber \\
& \hspace{0.5cm}
\times 
 t^{a_2}U_F(+\infty, z^+;\b)\Psi(z^+, \b) \Big\rangle
\end{align}
After summing over the quark helicity $h_1$ and the colors, and neglecting terms quadratic in the NLP amplitude to obtain the cross section at NLP accuracy, the square of the ``before'' contribution can be organized as
\begin{align}
\label{eq:trans_bef_square_2}
&
\frac{1}{2}\sum_{\lambda, h_1,\lambda_2}\sum_{\rm col.}\Big\langle \big| \mathcal{M}_{\gamma^*_T\to q_1g_2}^{\rm bef}\big|^2\Big\rangle
= \frac{e^2\, e_f^2\, g^2}{(2q^+)^2}\int_{\b'\b}e^{-i\k\cdot(\b-\b')} \int_{z^+,z'^+} \, (2q^+)\,  z\; 
\Big\langle  \bar{\Psi}(z'^+,\b')U^\dagger_F(+\infty, z'^+;\b')\,
\nonumber\\
& \times 
 \frac{1}{2}\sum_{\lambda,\lambda_2}
\bigg\{ 
C_F \, \gamma^0 \frac{\big[\Gamma_T^{\rm bef}(\P)\big]^\dagger}{[\P^2+\Qb^2]^2}\, \gamma^0 \, \frac{\gamma^-}{2} \Gamma_T^{\rm bef}(\P)
\nonumber\\
& \hspace{1.7cm}
+ 
\gamma^0 \frac{\big[ \Gamma_T^{\rm bef}(\P)\big]^\dagger}{[\P^2+\Qb^2]}\, \gamma^0\frac{\gamma^-}{2}\Gamma_T^{{\rm bef, NLP}, i}
\Big[ C_F\, z\, \k^i+\Big(C_F-\frac{C_A}{2}\Big) 
\int_{z''^+} t^d \, U_A(+\infty, z''^+;\b)_{dc}\, g{F_{i}^{\; -}}_c(z''^+, \b)\Big]
\nonumber\\
& \hspace{1.7cm}
+ 
\gamma^0 \big[ \Gamma_T^{{\rm bef, NLP}, i}\big]^\dagger\gamma^0\frac{\gamma^-}{2}\frac{\Gamma_T^{\rm bef}(\P)}{[\P^2+\Qb^2]}
\Big[ C_F\, z\, \k^i+\Big(C_F-\frac{C_A}{2}\Big) 
\int_{z''^+} t^d \, U_A(+\infty, z''^+;\b')_{dc}\, g{F_{i}^{\; -}}_c(z''^+, \b')\Big] \bigg\} 
\nonumber\\
& \times 
U_F(+\infty, z^+;\b)\Psi(z^+,\b)\Big\rangle+ \mathcal{O}(\rm NNLP)
\end{align}
The Dirac algebra for this contribution is presented in Appendix~\ref{app:before-before_dirac}. Here, we quote the results for the structures entering the LP term and its interference with the NLP terms. After summing over the outgoing gluon polarization and averaging over the transverse polarizations of the incoming photon, we obtain (see Eqs. \eqref{D_bef_LP_dag_LP}, \eqref{D_bef_LP_dag_NLP} and \eqref{D_bef_NLP_dag_LP} respectively)
\begin{align}
\label{eq:Dirac_T_bef_LP_dag_LP_fin}
\frac{1}{2}\sum_{\lambda, \lambda_2} (2q^+)\, z\, \gamma^0 \frac{\big[\Gamma_T^{\rm bef}(\P)\big]^\dagger}{[\P^2+\Qb^2]^2}\, \gamma^0 \, \frac{\gamma^-}{2} \Gamma_T^{\rm bef}(\P) 
&=
\frac{(2q^+)\, z}{[\P^2+\Qb^2]^2}\Big\{ \big[ z^2+(1-z)^2
\big] \P^2+m^2\Big\}\, \gamma^-
\\
\label{eq:Dirac_T_bef_LP_dag_NLP_fin}
\frac{1}{2}\sum_{\lambda, \lambda_2} (2q^+)\, z\, 
\gamma^0 \frac{\big[ \Gamma_T^{\rm bef}(\P)\big]^\dagger}{[\P^2+\Qb^2]}\, \gamma^0\frac{\gamma^-}{2}\Gamma_T^{{\rm bef, NLP}, i} &=
\frac{(2q^+)\, z}{[\P^2+\Qb^2]^2}\Big\{
\Big[ [z^2+(1-z)^2][\Qb^2-\P^2]-2m^2\Big] \frac{\P^i}{[\P^2+\Qb^2]}\gamma^-
\nonumber\\
& \hspace{6cm}
+
(1-2z)\P^j\, \gamma^-\frac{[\gamma^i,\gamma^j]}{2}\Big\}
\\
\label{eq:Dirac_T_bef_NLP_dag_LP_fin}
\frac{1}{2}\sum_{\lambda, \lambda_2} (2q^+)\, z\, 
\gamma^0 \, \big[\Gamma_T^{{\rm bef, NLP}, i}\big]^\dagger
\gamma^0 \, \frac{\gamma^-}{2} \, \frac{\Gamma_T^{\rm bef}(\P)}{[\P^2+\Qb^2]}
&= \frac{(2q^+)\, z}{[\P^2+\Qb^2]^2}\Big\{
\Big[ [z^2+(1-z)^2][\Qb^2-\P^2]-2m^2\Big]\frac{\P^i}{[\P^2+\Qb^2]}\gamma^-
\nonumber\\
& \hspace{6cm}
-
(1-2z)\P^j\, \gamma^-\frac{[\gamma^i,\gamma^j]}{2}\Big\}
\end{align}
Substituting these Dirac structures  into the expression given in Eq. \eqref{eq:trans_bef_square_2} and collecting the LP and NLP terms, we obtain the square of the ``before'' amplitude at NLP accuracy as 
\begin{align}
\label{eq:Trans_bef_square_fin}
&
\frac{1}{2}\sum_{\lambda,h_1,\lambda_2}\sum_{\rm col.}\Big\langle \big| {\mathcal M}_{\gamma^*_T\to q_1g_2}^{\rm bef}\big|^2\Big\rangle 
=
\frac{e^2\, e_f^2\, g^2}{(2q^+)^2} \, 
\frac{(2q^+)\, z}{[\P^2+\Qb^2]^2}\, 
\nonumber \\
& \hspace{1cm}
\times 
\bigg\{ C_F\, 
\Big\lgroup
\Big[ \Big( z^2+(1-z)^2\Big)\P^2+m^2\Big] 
+2\, z\, \frac{\P\cdot\k}{[\P^2+\Qb^2]}\Big[ \Big( z^2+(1-z)^2 \Big)(\Qb^2-\P^2)-2m^2\Big]
\Big\rgroup \T_{11}[\gamma^-](\k)
\nonumber\\
& \hspace{1.5cm}
+ 
\Big(C_F-\frac{C_A}{2}\Big) 
\Big[ \Big( z^2+(1-z)^2\big)(\Qb^2-\P^2)-2m^2\Big]
\, \frac{2\, \P^i}{[\P^2+\Qb^2]}\,\frac{1}{2} \Big[\T_{i, 12}[\gamma^-](\k)+\T_{i,21}[\gamma^-](\k)\Big]
\nonumber\\
& \hspace{1.5cm}
+ 
\Big(C_F-\frac{C_A}{2}\Big)\,  (1-2z)\, 2\, \P^j\, \frac{1}{2}
\bigg\lgroup\T_{i,12}\bigg[\frac{\gamma^-[\gamma^i,\gamma^j]}{2}\bigg](\k)-\T_{i,21}\bigg[\frac{\gamma^-[\gamma^i,\gamma^j]}{2}\bigg](\k)\bigg\rgroup
\bigg\}
+ \mathcal{O}(\rm NNLP) 
%
\end{align}
The terms proportional to $\T_{11}[\gamma^-](\k)$ contain the LP contribution and its kinematic NLP correction, while the field-strength insertions, that are encoded in functions $\T_{i,12}[\Gamma](\k)$ and $\T_{i,21}[\Gamma](\k)$,  give the genuine twist-three contributions. The commutator terms in Eqs.~\eqref{eq:Dirac_T_bef_LP_dag_NLP_fin} and \eqref{eq:Dirac_T_bef_NLP_dag_LP_fin} enter with opposite signs. They therefore cancel in the kinematic correction and combine into the difference of field-strength insertions displayed in the last term of Eq.~\eqref{eq:Trans_bef_square_fin}.

We finally consider the interference between the ``inside'' and ``before'' amplitudes. The ``inside'' amplitude is independent of the transverse separation $\r$ and therefore requires no expansion. To compute the interference contribution through NLP, we combine it with both the LP and NLP terms of the ``before'' amplitude. We first evaluate $\big(\mathcal{M}^{\rm in}\big)^\dagger\mathcal{M}^{\rm bef}$ and then include its complex conjugate:
\begin{align}
\label{eq:inter_in_dag_bef_1}
&
\frac{1}{2}\sum_{\lambda,h_1,\lambda_2}\sum_{\rm col.}\Big\langle \big( \mathcal{M}^{\rm in}_{\gamma^*_T\to q_1g_2}\big)^\dagger  \big(\mathcal{M}^{\rm bef}_{\gamma^*_T\to q_1g_2}\big) 
\Big\rangle 
= 
\frac{e^2\, e_f^2\, g^2}{(2q^+)^2} \, 
\frac{1}{2}\sum_{\lambda,h_1,\lambda_2}\sum_{\rm col.}
\int_{\b\b'}e^{i\k\cdot(\b'-\b)} 
\nonumber\\
&\times 
\int_{z^+,z'^+} 
\Big\langle \bar{\Psi}(z'^+,\b')U^\dagger_F(+\infty, z'^+;\b')\, t^{a_2} 
\gamma^0\, \frac{(1-z)^2\,  (\Gamma_{T}^{\rm in})^\dagger}{[\P^2+(1-z)^2m^2]} \, \gamma^0 \, \frac{\gamma^-\gamma^+}{2}u(1)\bar{u}(1)\frac{\gamma^+\gamma^-}{2} 
\\
& \times 
\bigg\{ 
\frac{\Gamma_T^{\rm bef}(\P)}{[\P^2+\Qb^2]}
+
\Gamma_T^{{\rm bef, NLP},i}
\Big[ z\k^i+\int_{z''^+} t^d\, U_A(+\infty, z''^+;\b)_{dc}\, g{F_i^{\; -}}_c(z''^+,\b) \Big] 
\bigg\} \, t^{a_2}\, U_F(+\infty, z^+;\b)\, \Psi(z^+,\b) 
\Big\rangle+\mathcal{O}({\rm NNLP}) \nonumber
\end{align}
After summing over the outgoing quark helicity and the colors, Eq.~\eqref{eq:inter_in_dag_bef_1} becomes
\begin{align}
\label{eq:inter_in_dag_bef_2}
& \hspace{0cm}
\frac{1}{2}\sum_{\lambda,h_1,\lambda_2}\sum_{\rm col.}\Big\langle \big( \mathcal{M}^{\rm in}_{\gamma^*_T\to q_1g_2}\big)^\dagger  \big(\mathcal{M}^{\rm bef}_{\gamma^*_T\to q_1g_2}\big) 
\Big\rangle 
= 
\frac{e^2\, e_f^2\, g^2}{(2q^+)^2} \, 
\int_{\b\b'}e^{i\k\cdot(\b'-\b)} 
\int_{z^+,z'^+} 
\frac{(2q^+)\, z}{[\P^2+(1-z)^2m^2]}
\nonumber\\
&\hspace{1.2cm}
\times 
\frac{1}{2}\sum_{\lambda,\lambda_2}
 \Big\langle \bar{\Psi}(z'^+,\b')U^\dagger_F(+\infty, z'^+;\b')\, 
 \Big\{C_F\, \frac{(1-z)^2}{[\P^2+\Qb^2]}\, 
\gamma^0 (\Gamma_T^{\rm in})^\dagger \gamma^0\, \frac{\gamma^-}{2} \, 
\Gamma_T^{\rm bef}(\P)
\nonumber\\
& \hspace{2cm}
+ 
(1-z)^2\, \gamma^0\, (\Gamma_T^{\rm in})^\dagger\, \gamma^0\, \frac{\gamma^-}{2}\, \Gamma_T^{{\rm bef, NLP},i}
\Big[ C_F\, z\, \k^i+\Big(C_F-\frac{C_A}{2}\Big) \int_{z''^+}t^d \, U_A(+\infty,z''^+;\b)_{dc}\, g{F_i^{\; -}}_c(z''^+,\b)\Big]\Big\}
\nonumber\\
& \hspace{2.4cm}
\times
U_F(+\infty, z^+;\b)\Psi(z^+,\b)\Big\rangle
+\mathcal{O}({\rm NNLP})
 \end{align}
The Dirac structures entering the LP and NLP interference terms are evaluated in Appendix~\ref{sec:app_bef_in_interfrence}. After summing over the outgoing gluon polarization and averaging over the transverse polarizations of the incoming photon, we obtain
\begin{align}
\label{eq:dirac_inter_in_dag_LP_bef}
\frac{(1-z)^2}{[\P^2+\Qb^2]}\, \frac{1}{2}\sum_{\lambda,\lambda_2}
\gamma^0\, (\Gamma_T^{\rm in})^\dagger\, \gamma^0\, \frac{\gamma^-}{2}\, \Gamma_T^{\rm bef}(\P)&= \frac{\gamma^-}{[\P^2+\Qb^2]}
\Big[-z^2\, \P^2+ m\big[(1-z)^3-1\big] \P^j\gamma^j\Big] 
\\
\label{eq:dirac_inter_indeg_NLP_bef}
(1-z)^2\, \frac{1}{2}\sum_{\lambda,\lambda_2} 
\gamma^0\, (\Gamma_T^{\rm in})^\dagger\, \gamma^0\, \frac{\gamma^-}{2} \, \Gamma_T^{{\rm bef, NLP}i}
&=\frac{\gamma^-}{[\P^2+\Qb^2]}
\bigg\{ -z^2\frac{[\Qb^2-\P^2]}{[\P^2+\Qb^2]}\, \P^i
+ z^2\, \P^j\, \frac{[\gamma^i,\gamma^j]}{2}
\nonumber\\
& \hspace{2.4cm}
+ 2m\, \frac{[1-(1-z)^3]}{[\P^2+\Qb^2]}\P^i\P^j\gamma^j
+ m\, (1-z)^3\, \gamma^i \bigg\}
\end{align}
Using the generalized target-averaged operators defined in Eqs. \eqref{eq:T_gen_shifted_11}, \eqref{eq:T_gen_shifted_12} and \eqref{eq:T_gen_shifted_21} together with the simplified Dirac structures given in Eqs. \eqref{eq:dirac_inter_in_dag_LP_bef} and \eqref{eq:dirac_inter_indeg_NLP_bef} identities, the first interference contribution at NLP accuracy reads
\begin{align}
\label{eq:inter_in_dag_bef_fin}
& \hspace{-0.3cm}
\frac{1}{2}\sum_{\lambda,h_1,\lambda_2}\sum_{\rm col.}\Big\langle \big( \mathcal{M}^{\rm in}_{\gamma^*_T\to q_1g_2}\big)^\dagger  \big(\mathcal{M}^{\rm bef}_{\gamma^*_T\to q_1g_2}\big) 
\Big\rangle 
= 
\frac{e^2\, e_f^2\, g^2}{(2q^+)^2} \, \frac{(2q^+)\, z}{[\P^2+(1-z)^2m^2]}\, \frac{1}{[\P^2+\Qb^2]}
\nonumber\\
& \hspace{2cm}
\times 
\bigg\{ \T_{11}[\gamma^-](\k)\bigg\lgroup- C_F\, z^2\, \P^2 - C_F\, z^3\, (\P\cdot\k)\, \frac{[\Qb^2-\P^2]}{[\P^2+\Qb^2]}  \bigg\rgroup
+
\T_{11}\bigg[\frac{\gamma^-[\gamma^i,\gamma^j]}{2}\bigg](\k)\, C_F\, z^3\, \k^i\P^j
\nonumber\\
& \hspace{2.5cm}
+ 
\T_{11}[\gamma^-\gamma^j](\k) \, C_F\, m\bigg\lgroup 
\big[ (1-z)^3-1\big] \P^j+z\, (1-z)^3\k^j + \frac{2z\, \big[1-(1-z)^3\big]}{[\P^2+\Qb^2]}\, (\P\cdot\k) \, \P^j \bigg\rgroup
\nonumber\\
& \hspace{2.5cm}
+ \Big(C_F-\frac{C_A}{2}\Big) 
\bigg\lgroup 
-\T_{i,12}[\gamma^-](\k)\, z^2\, \P^i\, \frac{[\Qb^2-\P^2]}{[\P^2+\Qb^2]} 
+ \T_{i,12}\bigg[\frac{\gamma^-[\gamma^i,\gamma^j]}{2}\bigg](\k)\, z^2\, \P^j
\nonumber\\
& \hspace{5.2cm}
+ \T_{i,12}[\gamma^-\gamma^j](\k)\,  
\bigg[2m \, \frac{\big[1-(1-z)^3\big]}{[\P^2+\Qb^2]}\, \P^i\P^j+ m\, (1-z)^3\delta^{ij}\bigg]\bigg\rgroup \bigg\} +\mathcal{O}({\rm NNLP})
\end{align}

The remaining interference contribution, $(\mathcal{M}^{\rm bef})^\dagger\mathcal{M}^{\rm in}$, is the complex conjugate of the preceding result. In this case, the NLP field-strength insertion belongs to the conjugate amplitude. Using the amplitudes derived in the previous subsection, we obtain
\begin{align}
\label{eq:intereference_bef_dag_in_1}
\frac{1}{2}\sum_{\lambda,h_1,\lambda_2}\sum_{\rm col.} \Big\langle \big(\mathcal{M}^{\rm bef}_{\gamma^*_T\to q_1g_2}\big)^\dagger \big(\mathcal{M}^{\rm in}_{\gamma^*_T\to q_1g_2} \big) \Big\rangle
&= \frac{e^2\, e_f^2\, g^2}{(2q^+)^2} \, \frac{1}{2}\sum_{\lambda,h_1,\lambda_2}\sum_{\rm col.}
\int_{\b\b'}e^{i\k\cdot(\b'-\b)}
\nonumber\\
& \hspace{-4cm} 
\times
\int_{z^+,z'^+} \Big\langle \bar{\Psi}(z'^+,\b')\, U^\dagger_F(+\infty,z'^+;\b')\,  t^{a_2}
\nonumber\\
& \hspace{-4cm}
 \times 
\bigg\{ \gamma^0\, \frac{\big(\Gamma^{\rm bef}_T(\P)\big)^\dagger}{[\P^2+\Qb^2]}\, \gamma^0
+
\gamma^0\, \Big( \Gamma_T^{{\rm bef, NLP}, i'}\Big)^\dagger\gamma^0
\bigg[ z\, \k^{i'}+ \int_{z''^+}t^{d'}\, U_A(+\infty, z''^+;\b')_{d'c'}\, g{F_{i'}^{\; -}}_{c'}(z''^+,\b')\bigg]\bigg\}
\nonumber\\
& \hspace{-4cm} 
\times 
\frac{\gamma^-\gamma^+}{2}u(1)\bar{u}(1)\frac{\gamma^+\gamma^-}{2} 
\, \frac{(1-z)^2\, \Gamma_T^{\rm in}}{[\P^2+(1-z)^2m^2]} \, 
t^{a_2}\, U_F(+\infty, z^+; \b)\Psi(z^+,\b)\Big\rangle
\end{align}
After summing over the outgoing quark helicity and the colors, this expression becomes
\begin{align}
\label{eq:intereference_bef_dag_in_2}
&
\frac{1}{2}\sum_{\lambda,h_1,\lambda_2}\sum_{\rm col.} \Big\langle \big(\mathcal{M}^{\rm bef}_{\gamma^*_T\to q_1g_2}\big)^\dagger \big(\mathcal{M}^{\rm in}_{\gamma^*_T\to q_1g_2} \big) \Big\rangle
= \frac{e^2\, e_f^2\, g^2}{(2q^+)^2} \, \int_{\b\b'}e^{i\k\cdot(\b'-\b)}\, \int_{z^+,z'^+} \, \frac{(2q^+)\, z}{[\P^2+(1-z)^2m^2]} 
\nonumber\\
&\hspace{1.2cm} 
\times 
\frac{1}{2}\sum_{\lambda,\lambda_2} 
\Big\langle \bar{\Psi}(z'^+,\b')U^\dagger_F(+\infty, z'^+;\b')
\Big\{ C_F\, \frac{(1-z)^2}{[\P^2+\Qb^2]} \, \gamma^0\big( \Gamma_T^{\rm bef}(\P)\big)^\dagger\, \gamma^0\, \frac{\gamma^-}{2}\, \Gamma_T^{\rm in} 
\nonumber\\
& \hspace{2cm}
+ (1-z)^2\, \gamma^0\, \big(\Gamma_T^{{\rm bef, NLP},i}\big)^\dagger\, \gamma^0 \, \frac{\gamma^-}{2}\, \Gamma_T^{\rm in} 
\Big[ C_F\, z\, \k^i+\Big(C_F-\frac{C_A}{2}\Big) 
\int_{z''^+}t^d \, U_A(+\infty,z''^+;\b')_{dc}\, g{F_i^{\; -}}_c(z''^+,\b')\Big]\Big\}
\nonumber\\
& \hspace{2.4cm}
\times
U_F(+\infty, z^+;\b)\Psi(z^+,\b)\Big\rangle
+\mathcal{O}({\rm NNLP})
 \end{align}
Evaluating the Dirac structures and using the generalized target-averaged operators defined above, we obtain
\begin{align}
\label{eq:intereference_bef_dag_in_fin}
& \hspace{-1cm}
\frac{1}{2}\sum_{\lambda,h_1,\lambda_2}\sum_{\rm col.} \Big\langle \big(\mathcal{M}^{\rm bef}_{\gamma^*_T\to q_1g_2}\big)^\dagger \big(\mathcal{M}^{\rm in}_{\gamma^*_T\to q_1g_2} \big) \Big\rangle
= \frac{e^2\, e_f^2\, g^2}{(2q^+)^2} \, \frac{(2q^+)\, z}{[\P^2+(1-z)^2m^2]} \, \frac{1}{[\P^2+\Qb^2]}
\nonumber\\
& \hspace{1cm} 
\times 
\bigg\{ 
-C_F\, z^2\, \bigg[ \P^2+z\frac{[\Qb^2-\P^2]}{[\P^2+\Qb^2]}(\P\cdot \k)\bigg] \, \T_{11}[\gamma^-](\k) - C_F\, z^3\, \k^i \, \P^j\, \T_{11}\bigg[\frac{\gamma^-[\gamma^i,\gamma^j]}{2}\bigg](\k)
\nonumber\\
&  \hspace{1.7cm}
+ C_F\, m\bigg\lgroup -\big[ (1-z)^3-1\big]\, \P^j\, \bigg[ 1-2z\, \frac{\P\cdot\k}{[\P^2+\Qb^2]}\bigg] -z\, (1-z)^3\, \k^j\bigg\rgroup
\T_{11}[\gamma^-\gamma^j](\k)
\nonumber\\
& \hspace{1.7cm}
- \Big(C_F-\frac{C_A}{2}\Big) \, z^2\, \frac{[\Qb^2-\P^2]}{[\P^2+\Qb^2]} \, \P^i\, \T_{i,21}[\gamma^-](\k) 
- \Big(C_F-\frac{C_A}{2}\Big) \, z^2\, \P^j\, \T_{i,21}\bigg[ \frac{\gamma^-[\gamma^i,\gamma^j]}{2}\bigg](\k)
\nonumber\\
& \hspace{1.7cm}
+ \Big(C_F-\frac{C_A}{2}\Big) \, m\, \bigg\lgroup 2\Big[(1-z)^3-1\Big] \frac{\P^i\P^j}{[\P^2+\Qb^2]}-(1-z)^3\, \delta^{ij}\bigg\rgroup \, \T_{i,21}[\gamma^-\gamma^j](\k)\bigg\}
+\mathcal{O}({\rm NNLP})
\end{align}
It is convenient to combine the two interference contributions. In the sum, the two-body terms proportional to $\T_{11}[\gamma^-\gamma^j]$ and $\T_{11}[\gamma^-[\gamma^i,\gamma^j]/2]$ cancel. The corresponding three-body terms instead enter through differences of $\T_{i,12}$ and $\T_{i,21}$, while those with the Dirac structure $\gamma^-$ enter through their sum. The result reads 
\begin{align}
\label{eq:sum_interfrence}
&
\frac{1}{2}\sum_{\lambda,h_1,\lambda_2}\sum_{\rm col.} \Big\langle
\big( \mathcal{M}^{\rm in}_{\gamma^*_T\to q_1g_2}\big)^\dagger  \big(\mathcal{M}^{\rm bef}_{\gamma^*_T\to q_1g_2}\big) 
+
 \big(\mathcal{M}^{\rm bef}_{\gamma^*_T\to q_1g_2}\big)^\dagger \big(\mathcal{M}^{\rm in}_{\gamma^*_T\to q_1g_2} \big) 
 \Big\rangle 
 =
 \frac{e^2\, e_f^2\, g^2}{q^+}\, \frac{z}{[\P^2+(1-z)^2m^2]}\frac{1}{[\P^2+\Qb^2]}
 \nonumber\\
 & \times 
 \bigg\{ -z^2\, C_F\, \bigg[ \P^2+z\frac{[\Qb^2-\P^2]}{[\P^2+\Qb^2]}(\P\cdot\k)\bigg] 
 \, \T_{11}[\gamma^-](\k)
 -
 z^2\, \Big(C_F-\frac{C_A}{2}\Big)\,  \frac{[\Qb^2-\P^2]}{[\P^2+\Qb^2]}\, \P^i\, \frac{1}{2}
 \Big(\T_{i,12}[\gamma^-](\k)+\T_{i,21}[\gamma^-](\k)\Big)
 \nonumber\\
 & \hspace{0.7cm}
 + 
 z^2\, \Big(C_F-\frac{C_A}{2}\Big)\, \P^j\, \frac{1}{2}\bigg( \T_{i,12}\bigg[\frac{\gamma^-[\gamma^i,\gamma^j]}{2}\bigg](\k)-\T_{i,21}\bigg[\frac{\gamma^-[\gamma^i,\gamma^j]}{2}\bigg](\k)\bigg)
 \nonumber\\
 & \hspace{0.7cm}
 + 
\Big(C_F-\frac{C_A}{2}\Big)\, m\, \bigg[ -2\big[(1-z)^3-1\big]\frac{\P^i\P^j}{[\P^2+\Qb^2]}+(1-z)^3\delta^{ij}\bigg] \, \frac{1}{2} \,  
\Big( \T_{i,12}[\gamma^-\gamma^j](\k)-\T_{i,21}[\gamma^-\gamma^j](\k)\Big)\bigg\}
 + \mathcal{O}({\rm NNLP})
\end{align}
The transverse cross section at NLP accuracy follows by substituting the ``inside'' square in Eq.~\eqref{eq:trans_in_square_fin}, the ``before'' square in Eq.~\eqref{eq:Trans_bef_square_fin}, and the combined interference contribution in Eq.~\eqref{eq:sum_interfrence} into Eq.~\eqref{eq:Transverse_X_sec_1}. All contributions are now expressed in terms of the generalized target-averaged operators defined above. The coefficient of the two-body operator $\T_{11}[\gamma^-]$ contains both the LP result and its kinematic NLP correction, while the three-body operators describe the genuine twist-three contributions. Together with the longitudinal result in Eq.~\eqref{eq:qg_Xsec_NLP_3}, this completes the quark–gluon dijet cross-section calculation at NEik and  NLP accuracy in the correlation limit.

\subsection{NEik quark-gluon dijet production cross section in terms of TMD correlators}
\label{sec:TMD_correlators}

In this subsection, we express the quark–gluon dijet production cross section in terms of the TMD correlators introduced in Ref.~\cite{Rodini:2022wki}. We first establish the correspondence between their coordinate-space definitions and the generalized target-averaged operators obtained in the previous subsection, and then introduce the longitudinal momentum-fraction representation. We use our light-cone conventions, in which the target carries a large minus momentum $P_{\rm tar}^-$, and choose future-pointing gauge links, consistently with the operator structures appearing in our calculation.

We begin with the two-body correlator, which for the projection $\Gamma=\gamma^-$ describes the twist-two contribution. In our notation, its coordinate-space definition reads
\begin{align}
\label{eq:def_twist_2_TMD_corr_1}
\tilde{\Phi}_{11}^{[\Gamma]}(z_1^+,z_2^+;\Delta\b)\equiv 
\big\langle P_{\rm tar},s\big| T\Big\{ \bar{\Psi}(z_1^+, \Delta\b) U_F^\dagger(+\infty,z_1^+;\Delta\b)\, \frac{\Gamma}{2}\, U_F(+\infty, z_2^+, {\bf 0})\Psi(z_2^+,{\bf 0})\Big\}
\big| P_{\rm tar}, s\big\rangle 
\end{align}
Here, $T$ denotes time ordering, and $s$ labels the target polarization. Translation invariance allows us to shift both longitudinal coordinates by $-z_2^+$, placing the quark field at the origin:
\begin{align}
\label{eq:def_twist_2_TMD_corr_2}
\tilde{\Phi}_{11}^{[\Gamma]}(z_1^+,z_2^+;\Delta\b)=
\big\langle P_{\rm tar},s\big| T\Big\{ \bar{\Psi}(z_1^+-z_2^+, \Delta\b) U_F^\dagger(+\infty,z_1^+-z_2^+;\Delta\b)\, \frac{\Gamma}{2}\, U_F(+\infty,0+, {\bf 0})\Psi(0,{\bf 0})\Big\}
\big| P_{\rm tar}, s\big\rangle 
\end{align}
Comparing Eq.~\eqref{eq:def_twist_2_TMD_corr_2} with the generalized target-averaged operator in Eq.~\eqref{eq:T_gen_shifted_11}, and identifying the operator products after removing the explicit time ordering, we obtain
\begin{equation}
\T_{11}[\Gamma](\k)
=
\frac{1}{P_{\rm tar}^-}
\int_{\Delta\b}e^{i\k\cdot\Delta\b}
\int_{\Delta z^+}
\tilde{\Phi}_{11}^{[\Gamma]}(\Delta z^+,0;\Delta\b).
\label{eq:T_11_in_TMD_corr}
\end{equation}
The prefactor follows from the target-state normalization in Eq.~\eqref{eq:T_gen_shifted_11} and the factor $1/2$ included in the Dirac projection defining $\tilde{\Phi}_{11}^{[\Gamma]}$.

Following the Fourier convention of Ref.~\cite{Rodini:2022wki}, we introduce the longitudinal momentum-fraction representation through
\begin{equation}
\tilde{\Phi}_{11}^{[\Gamma]}(z_1^+,z_2^+;\Delta\b)
=
P_{\rm tar}^-
\int_{-1}^{1}\dd\rmx\,
e^{i\rmx P_{\rm tar}^-(z_1^+-z_2^+)}
\Phi_{11}^{[\Gamma]}(\rmx;\Delta\b).
\label{eq:twist_2_TMD_corr_mom_frac}
\end{equation}
Substituting this representation into Eq.~\eqref{eq:T_11_in_TMD_corr} and performing the longitudinal-coordinate integration gives
\begin{align}
\T_{11}[\Gamma](\k)
&=
\int_{\Delta\b}e^{i\k\cdot\Delta\b}
\int_{-1}^{1}\dd\rmx\,
2\pi\delta(\rmx P_{\rm tar}^-)\,
\Phi_{11}^{[\Gamma]}(\rmx;\Delta\b)
\nonumber\\
&=
\frac{2\pi}{P_{\rm tar}^-}
\int_{\Delta\b}e^{i\k\cdot\Delta\b}
\Phi_{11}^{[\Gamma]}(0;\Delta\b).
\label{eq:T_11_integrated}
\end{align}
The last line expresses the formal zero-longitudinal-momentum-fraction projection associated with the absence of longitudinal Fourier phases in our target-averaged operator.
%
%
%
%
%
%

We next consider the three-body correlators associated with the genuine twist-three contributions. Starting with the correlator in which the field-strength insertion lies on the same transverse side as the antiquark field, the definition of Ref.~\cite{Rodini:2022wki}, adapted to our conventions, reads
\begin{align}
\tilde{\Phi}_{i,21}^{[\Gamma]}
(z_1^+,z_2^+,z_3^+;\Delta\b)
&\equiv
\big\langle P_{\rm tar},s\big|
T\Big\{
\bar\Psi(z_1^+,\Delta\b)
U_F(z_1^+,z_2^+;\Delta\b)\,
gF_i^{\;-}(z_2^+,\Delta\b)
\nonumber\\
&\hspace{5cm}
\times
\frac{\Gamma}{2}\,
U_F(z_2^+,+\infty;\Delta\b)
U_F(+\infty,z_3^+;{\bf 0})
\Psi(z_3^+,{\bf 0})
\Big\}
\big|P_{\rm tar},s\big\rangle,
\label{def:twist_three_TMD_corr_org}
\end{align}
where $F_i^{\;-}=t^aF_{i,a}^{\;-}$, and the target-spin average is understood. Using the composition property of the fundamental Wilson lines and the relation between the fundamental and adjoint representations, we have
\begin{align}
&U_F(z_1^+,z_2^+;\Delta\b)\,
gF_i^{\;-}(z_2^+,\Delta\b)\,
U_F(z_2^+,+\infty;\Delta\b)
=
U_F^\dagger(+\infty,z_1^+;\Delta\b)\,
t^bU_A(+\infty,z_2^+;\Delta\b)_{ba}\,
gF_{i,a}^{\;-}(z_2^+,\Delta\b).
\end{align}
The correlator in Eq. \eqref{def:twist_three_TMD_corr_org} can therefore be written as
\begin{align}
\tilde{\Phi}_{i,21}^{[\Gamma]}
(z_1^+,z_2^+,z_3^+;\Delta\b)
&=
\big\langle P_{\rm tar},s\big|
T\Big\{
\bar\Psi(z_1^+,\Delta\b)
U_F^\dagger(+\infty,z_1^+;\Delta\b)\,
\frac{\Gamma}{2}
\nonumber\\
& \hspace{2.5cm}
\times
t^bU_A(+\infty,z_2^+;\Delta\b)_{ba}\,
gF_{i,a}^{\;-}(z_2^+,\Delta\b)\,
U_F(+\infty,z_3^+;{\bf 0})
\Psi(z_3^+,{\bf 0})
\Big\}
\big|P_{\rm tar},s\big\rangle.
\label{eq:twist_three_TMD_corr_1}
\end{align}
A simultaneous longitudinal translation by $-z_3^+$ places the quark field at the origin, giving
\begin{align}
\tilde{\Phi}_{i,21}^{[\Gamma]}
(z_1^+,z_2^+,z_3^+;\Delta\b)
&=
\big\langle P_{\rm tar},s\big|
T\Big\{
\bar\Psi(z_1^+-z_3^+,\Delta\b)
U_F^\dagger(+\infty,z_1^+-z_3^+;\Delta\b)\,
\frac{\Gamma}{2}
\nonumber\\
&\hspace{1.5cm}
\times
t^bU_A(+\infty,z_2^+-z_3^+;\Delta\b)_{ba}\,
gF_{i,a}^{\;-}(z_2^+-z_3^+,\Delta\b)
U_F(+\infty,0;{\bf 0})\Psi(0,{\bf 0})
\Big\}
\big|P_{\rm tar},s\big\rangle.
\label{eq:twist_three_TMD_corr_2}
\end{align}
Comparing this expression with Eq.~\eqref{eq:T_gen_shifted_21}, with the same treatment of time ordering as in the two-body case, we obtain
\begin{equation}
\T_{i,21}[\Gamma](\k)
=
\frac{1}{P_{\rm tar}^-}
\int_{\Delta\b}e^{i\k\cdot\Delta\b}
\int_{\Delta z^+,\Delta z'^+}
\tilde{\Phi}_{i,21}^{[\Gamma]}
(\Delta z'^+,\Delta z^+,0;\Delta\b).
\label{eq:T_21_in_TMD_corr}
\end{equation}
Here, $\Delta z'^+$ denotes the longitudinal position of the antiquark relative to the quark, while $\Delta z^+$ denotes that of the field-strength insertion.

Following the convention of Ref.~\cite{Rodini:2022wki}, the longitudinal momentum-fraction representation of the three-body correlator is defined by
\begin{equation}
\tilde{\Phi}_{i,21}^{[\Gamma]}
(z_1^+,z_2^+,z_3^+;\Delta\b)
=
(P_{\rm tar}^-)^2
\int[\dd\rmx]\,
e^{-iP_{\rm tar}^-(\rmx_1z_1^++\rmx_2z_2^++\rmx_3z_3^+)}
\Phi_{i,21}^{[\Gamma]}
(\rmx_1,\rmx_2,\rmx_3;\Delta\b),
\label{eq:twist_3_TMD_corr_mom_frac}
\end{equation}
where
\begin{equation}
\int[\dd\rmx]
\equiv
\int_{-1}^{1}\dd\rmx_1
\int_{-1}^{1}\dd\rmx_2
\int_{-1}^{1}\dd\rmx_3\,
\delta(\rmx_1+\rmx_2+\rmx_3).
\label{eq:twist_3_fraction_measure}
\end{equation}
The constraint $\rmx_1+\rmx_2+\rmx_3=0$ follows from longitudinal translation invariance.
The sign of the Fourier phase differs from that in the two-body definition in Eq.~\eqref{eq:twist_2_TMD_corr_mom_frac}. We follow Ref.~\cite{Rodini:2022wki}, where this choice is made to agree with the usual conventions for twist-two and twist-three collinear distributions. The phases coincide upon setting $\rmx_1=-\rmx$, $\rmx_2=0$ and $\rmx_3=\rmx$, consistently with $\rmx_1+\rmx_2+\rmx_3=0$. This correspondence concerns only the Fourier conventions and does not identify the two-body and three-body correlators.
Substituting Eq.~\eqref{eq:twist_3_TMD_corr_mom_frac} into Eq.~\eqref{eq:T_21_in_TMD_corr}, we obtain
\begin{align}
\T_{i,21}[\Gamma](\k)
&=
\frac{1}{P_{\rm tar}^-}
\int_{\Delta\b}e^{i\k\cdot\Delta\b}
\int_{\Delta z^+,\Delta z'^+}
(P_{\rm tar}^-)^2
\int[\dd\rmx]\,
e^{-iP_{\rm tar}^-(\rmx_1\Delta z'^++\rmx_2\Delta z^+)}
\Phi_{i,21}^{[\Gamma]}
(\rmx_1,\rmx_2,\rmx_3;\Delta\b).
\end{align}
The two longitudinal-coordinate integrations give
\begin{equation}
\int_{\Delta z^+,\Delta z'^+}
e^{-iP_{\rm tar}^-(\rmx_1\Delta z'^++\rmx_2\Delta z^+)}
=
\frac{(2\pi)^2}{(P_{\rm tar}^-)^2}
\delta(\rmx_1)\delta(\rmx_2).
\end{equation}
Together with the constraint in Eq.~\eqref{eq:twist_3_fraction_measure}, these delta functions formally select vanishing longitudinal momentum fractions:
\begin{align}
\T_{i,21}[\Gamma](\k)
&=
\frac{(2\pi)^2}{P_{\rm tar}^-}
\int_{\Delta\b}e^{i\k\cdot\Delta\b}
\int[\dd\rmx]\,
\delta(\rmx_1)\, \delta(\rmx_2)\,
\Phi_{i,21}^{[\Gamma]}
(\rmx_1,\rmx_2,\rmx_3;\Delta\b)
\nonumber\\
&=
\frac{(2\pi)^2}{P_{\rm tar}^-}
\int_{\Delta\b}e^{i\k\cdot\Delta\b} \, 
\Phi_{i,21}^{[\Gamma]}(0,0,0;\Delta\b).
\label{eq:T_21_integrated}
\end{align}

The same procedure applies to the correlator $\tilde{\Phi}_{i,12}^{[\Gamma]}$, in which the field-strength insertion lies on the same transverse side as the quark field. Using the same longitudinal Fourier convention, we find
\begin{equation}
\T_{i,12}[\Gamma](\k)
=
\frac{(2\pi)^2}{P_{\rm tar}^-}
\int_{\Delta\b}e^{i\k\cdot\Delta\b}
\Phi_{i,12}^{[\Gamma]}(0,0,0;\Delta\b).
\label{eq:T_12_integrated}
\end{equation}
As in the two-body case, the zero-fraction arguments in Eqs.~\eqref{eq:T_21_integrated} and \eqref{eq:T_12_integrated} denote the formal projection associated with the absence of longitudinal Fourier phases in our operators. 
Using Eqs.~\eqref{eq:T_11_integrated}, \eqref{eq:T_21_integrated} and \eqref{eq:T_12_integrated}, we can express the longitudinal cross section in Eq.~\eqref{eq:qg_Xsec_NLP_3} in terms of TMD correlators:
\begin{align}
&(2\pi)^5\,2z(1-z)
\frac{\dd\sigma^{\gamma_L^*\to q_1g_2}}
{\dd^2\k\,\dd z\,\dd^2\P}
=
\frac{\alpha_{\rm em}\alpha_s e_f^2(4\pi)^2}{2q^+}\,
\frac{2\pi}{P_{\rm tar}^-}
\int_{\Delta\b}e^{i\k\cdot\Delta\b}
\nonumber\\
&\qquad\times
\Bigg\{
\mathcal{C}_L(z,\k,\P)\,
\Phi_{11}^{[\gamma^-]}(0;\Delta\b)
+
2\pi\,\mathcal{C}_L^i(z,\P)\,
\frac{1}{2}\Big[ 
\Phi_{i,12}^{[\gamma^-]}(0,0,0;\Delta\b)
+
\Phi_{i,21}^{[\gamma^-]}(0,0,0;\Delta\b)\Big]
\Bigg\}
+\mathcal{O}(\mathrm{NNLP}),
\label{eq:qg_L_Xsec_TMD_corr}
\end{align}
where we have used $e^2=4\pi\alpha_{\rm em}$ and $g^2=4\pi\alpha_s$. The hard coefficients $\mathcal{C}_L$ and $\mathcal{C}_L^i$ are those defined in Eqs. \eqref{def:C_L} and \eqref{def:C_L_i} in the previous subsection, and the zero-fraction arguments carry the formal interpretation discussed above.

The first term contains the twist-two correlator, with both the LP contribution and its kinematic NLP correction included in $\mathcal{C}_L$. The second term describes the genuine twist-three contribution and involves the symmetric combination of the two three-body correlators. The additional factor $2\pi$ in this term follows from the extra longitudinal-coordinate integration in the three-body operator.

The transverse cross section can be expressed in terms of TMD correlators in a similar manner. Combining the contributions obtained in the previous subsection, we first organize the result according to the Dirac structures of the generalized target-averaged operators. The two-body contribution involves the projection $\gamma^-$, while the three-body contributions also contain the projections $\gamma^-[\gamma^i,\gamma^j]/2$ and $\gamma^-\gamma^j$. The transverse cross section then can be written as 
\begin{align}
(2\pi)^5\, 2z(1-z)\, \frac{\dd \sigma^{\gamma^*_T\to q_1g_2}}{\dd^2 \k\, \dd z\, \dd^2\P}
&= \frac{e^2\, e_f^2\, g^2}{(2q^+)}\Big\{ 
\C_T(z,\k,\P)\, \T_{11}[\gamma^-](\k)
+ 
\C^{{\rm unpol.}, i}_T(z,\P) \, \frac{1}{2} \Big[ \T_{i,12}[\gamma^-](\k)+\T_{i,21}[\gamma^-](\k)\Big] 
\nonumber\\
& \hspace{1.5cm}
+ 
\C^{{\rm h. \, pol.},j}_T(z,\P)\, \frac{1}{2} \, \Big[ \T_{i,21}\bigg[ \frac{\gamma^-[\gamma^i,\gamma^j]}{2}\bigg](\k)- \T_{i,12}\bigg[ \frac{\gamma^-[\gamma^i,\gamma^j]}{2}\bigg](\k)
\Big]
\nonumber\\
& \hspace{1.5cm}
+ m\, \C^{{\rm h.\, flip}, ij}_T(z,\P)\, \frac{1}{2}\, \Big[ \T_{i,21}[\gamma^-\gamma^j](\k)- \T_{i,12}[\gamma^-\gamma^j](\k)\Big] 
    \Big\}
\end{align}
after introducing the corresponding hard coefficients as follows: 
\begin{align}
\C_T(z,\k,\P)
&= C_F\,z\,
\Bigg\{
\frac{(1+z^2)\P^2+(1-z)^4m^2}{[\P^2+(1-z)^2m^2]^2}
+\frac{\big(z^2+(1-z)^2\big)\P^2+m^2}{[\P^2+\Qb^2]^2}
-\frac{2z^2\P^2}{[\P^2+(1-z)^2m^2][\P^2+\Qb^2]}
\nonumber\\
&\hspace{1.5cm}
+\frac{2z\,(\P\cdot\k)}{[\P^2+\Qb^2]^3}
\Big[\big(z^2+(1-z)^2\big)(\Qb^2-\P^2)-2m^2\Big]
-\frac{2z^3\,(\P\cdot\k)(\Qb^2-\P^2)}{[\P^2+(1-z)^2m^2][\P^2+\Qb^2]^2}
\Bigg\},
\\[2mm]
\C_T^{{\rm unpol.},i}(z,\P)
&=2\left(C_F-\frac{C_A}{2}\right)
\frac{z\,\P^i}{[\P^2+\Qb^2]^2}
\left\{
\frac{\big(z^2+(1-z)^2\big)(\Qb^2-\P^2)-2m^2}{[\P^2+\Qb^2]}
-\frac{z^2(\Qb^2-\P^2)}{[\P^2+(1-z)^2m^2]}
\right\},
\\[2mm]
\C_T^{{\rm h.\,pol.},j}(z,\P)
&=-2\left(C_F-\frac{C_A}{2}\right)
\frac{z\,\P^j}{[\P^2+\Qb^2]^2}
\left(1-2z+\frac{z^2{[\P^2+\Qb^2]}}{[\P^2+(1-z)^2m^2]}\right),
\\[2mm]
\C_T^{{\rm h.\,flip},ij}(z,\P)
&=-2\left(C_F-\frac{C_A}{2}\right)
\frac{z}{[\P^2+(1-z)^2m^2][\P^2+\Qb^2]}
\left\{
(1-z)^3\delta^{ij}
+\frac{2\big[1-(1-z)^3\big]}{[\P^2+\Qb^2]}\,\P^i\P^j
\right\}.
\end{align}
The labels ``unpol.'', ``h. pol.'' and ``h. flip'' distinguish the unpolarized, helicity-dependent and helicity-flip quark structures, respectively; the target remains unpolarized. Using the Dirac identity
\begin{equation}
\frac{\gamma^-[\gamma^i,\gamma^j]}{2}
=
i\epsilon^{ij}\gamma^-\gamma_5,
\end{equation}
together with the matching relations established above, the transverse cross section becomes
\begin{align}
(2\pi)^5\, 2z(1-z)\, \frac{\dd \sigma^{\gamma^*_T\to q_1g_2}}{\dd^2 \k\, \dd z\, \dd^2\P}
&=
\frac{\alpha_{\rm em}\, \alpha_s\, e_f^2\, (4\pi)^2}{2q^+}\, \frac{(2\pi)^2}{P^-_{tar}}
\int_{\Delta\b}\, e^{i\k\cdot\Delta\b}
\bigg\{ 
\frac{1}{2\pi}\, \C_T(z,\k,\P)\, \Phi_{11}^{[\gamma^-]}(0;\Delta\b)
\nonumber\\
& \hspace{2.5cm}
+ \C_T^{{\rm unpol.}, i}(z,\P)\, \frac{1}{2}\, 
\Big[ \Phi^{[\gamma^-]}_{i,21}(0,0,0;\Delta\b) + \Phi^{[\gamma^-]}_{i,12}(0,0,0;\Delta\b) \Big] 
\nonumber\\
& \hspace{2.5cm}
+\epsilon^{ij}\,  \C_T^{{\rm h.\, pol.}, j}(z,\P)\, \frac{i}{2}\, 
\Big[ \Phi^{[\gamma^-\gamma_5]}_{i,21}(0,0,0;\Delta\b) - \Phi^{[\gamma^-\gamma_5]}_{i,12}(0,0,0;\Delta\b) \Big] 
\nonumber\\
& \hspace{2.5cm}
+ m\, \C_T^{{\rm h. \, flip}, ij}(z,\P)\, \frac{1}{2}\, 
\Big[ \Phi^{[\gamma^-\gamma^j]}_{i,21}(0,0,0;\Delta\b) - \Phi^{[\gamma^-\gamma^j]}_{i,12}(0,0,0;\Delta\b) \Big] \bigg\}
\end{align}
The three-body correlators enter the transverse cross section through sums and differences of the TMD correlators $\Phi^{[\Gamma]}_{i, 21}$ and $\Phi^{[\Gamma]}_{i, 12}$. It is therefore convenient to introduce the combinations used in Ref.~\cite{Rodini:2022wki},
\begin{align}
\Phi_{\mu,\oplus}^{[\Gamma]}(x_1,x_2,x_3;\Delta\b)
&\equiv \frac{1}{2}
\Big[
\Phi_{\mu,21}^{[\Gamma]}(x_1,x_2,x_3;\Delta\b)
+
\Phi_{\mu,12}^{[\Gamma]}(-x_3,-x_2,-x_1;\Delta\b)
\Big],
\\
\Phi_{\mu,\ominus}^{[\Gamma]}(x_1,x_2,x_3;\Delta\b)
&\equiv \frac{i}{2}
\Big[
\Phi_{\mu,21}^{[\Gamma]}(x_1,x_2,x_3;\Delta\b)
-
\Phi_{\mu,12}^{[\Gamma]}(-x_3,-x_2,-x_1;\Delta\b)
\Big].
\end{align}
At vanishing longitudinal momentum fractions, the reversal of the arguments in the \(12\) correlators has no effect. The combinations appearing in our cross section can thus be expressed directly in terms of \(\Phi_{\oplus}^{[\Gamma]}\) and \(\Phi_{\ominus}^{[\Gamma]}\). In particular, the unpolarized quark structure involves \(\Phi_{\oplus}^{[\gamma^-]}\), while the helicity-dependent and helicity-flip structures involve \(\Phi_{\ominus}^{[\gamma^-\gamma_5]}\) and \(\Phi_{\ominus}^{[\gamma^-\gamma^j]}\), respectively. With these definitions, the transverse cross section takes the form
\begin{align}
(2\pi)^5\, 2z(1-z)\, \frac{\dd \sigma^{\gamma^*_L\to q_1g_2}}{\dd^2 \k\, \dd z\, \dd^2\P}
&=
\frac{ 4\, \alpha_{\rm em}\, \alpha_s\, e_f^2}{2q^+}\, \frac{(2\pi)^4}{P^-_{tar}}
\int_{\Delta\b}\, e^{i\k\cdot\Delta\b}
\nonumber\\
& 
\times
\bigg\{ 
\frac{1}{2\pi}\, \C_T(z,\k,\P)\, \Phi_{11}^{[\gamma^-]}(0;\Delta\b)
+
\C^{{\rm unpol.},i}_T(z,\P)\, \Phi^{[\gamma^-]}_{i, \oplus}(0,0,0;\Delta\b)
\nonumber\\
& \hspace{0.5cm}
+ 
\epsilon^{ij}\, \C^{{\rm h.\, pol.j}}_T(z,\P)\, \Phi^{[\gamma^-\gamma_5]}_{i,\ominus}(0,0,0;\Delta\b)
-
    i\, m\, \C_T^{{\rm h.\, flip}, ij}(z,\P)\, \Phi^{[\gamma^-\gamma^j]}_{i,\ominus}(0,0,0;\Delta\b) \bigg\}
\end{align} 
Up to this point, our results are valid for arbitrary spin of the target. For simplicity, we consider an unpolarized target for the rest of this section. In that case, we express the correlators in terms of scalar TMD distributions.
Following the parametrizations of Ref.~\cite{Rodini:2022wki}, adapted to our light-cone and transverse-index conventions, the two-body correlator reads
\begin{equation}
\Phi_{11}^{[\gamma^-]}(x;\Delta\b)
=
f_1(x;\Delta\b).
\end{equation}
For the three-body correlators entering the unpolarized and helicity-dependent contributions, we have
\begin{align}
\Phi_{i,\oplus}^{[\gamma^-]}(x_1,x_2,x_3;\Delta\b)
&=
-i\,\Delta\b^i M^2\,
f_{\oplus}^{\perp}(x_1,x_2,x_3;\Delta\b),
\\
\Phi_{i,\ominus}^{[\gamma^-\gamma_5]}(x_1,x_2,x_3;\Delta\b)
&=
-i\,\epsilon^{il}\Delta\b^l M^2\,
g_{\ominus}^{\perp}(x_1,x_2,x_3;\Delta\b),
\end{align}
where \(M\) denotes the target mass. Using \(\epsilon^{ij}\epsilon^{il}=\delta^{jl}\), the contraction appearing in the helicity-dependent contribution becomes
\begin{equation}
\epsilon^{ij}
\Phi_{i,\ominus}^{[\gamma^-\gamma_5]}(x_1,x_2,x_3;\Delta\b)
=
-i\,\Delta\b^j M^2\,
g_{\ominus}^{\perp}(x_1,x_2,x_3;\Delta\b).
\end{equation}
To express the cross sections in transverse-momentum space, we next specify the Fourier-transform conventions and the relation between the scalar distributions in coordinate and momentum space.

Adopting the convention in Ref. \cite{Boussarie:2023izj}, we define the transverse Fourier transform:
\begin{align}
\widetilde f(x;\b)
&=\int \dd^2\k\,e^{-i\k\cdot\b}\,f(x;\k),
&
f(x;\k)
&=\int\frac{\dd^2\b}{(2\pi)^2}\,
e^{i\k\cdot\b}\,\widetilde f(x;\b).
\end{align}
For a scalar distribution depending only on the magnitude of the transverse momentum, the angular integration gives
\begin{equation}
\widetilde f(x;b)
=
2\pi\int_0^\infty \dd k\,k\,J_0(kb)\,f(x;k),
\end{equation}
where \(b=|\b|\) and \(k=|\k|\). We also define its first Bessel moment,
\begin{equation}
\widetilde f^{(1)}(x;b)
\equiv
-\frac{2}{M^2}\frac{\partial}{\partial b^2}
\widetilde f(x;b)
=
\frac{2\pi}{bM^2}
\int_0^\infty \dd k\,k^2\,J_1(kb)\,f(x;k).
\end{equation}
These definitions imply
\begin{equation}
\int\frac{\dd^2\b}{(2\pi)^2}\,
e^{i\k\cdot\b}
\big[-i\,b^iM^2\,\widetilde f^{(1)}(x;b)\big]
=
k^i f(x;k).
\label{eq:FT_first_moment}
\end{equation}
The same relations apply to distributions depending on three longitudinal momentum fractions, which are unchanged by the transverse Fourier transform.

We define the momentum-space distributions \(f_{\oplus}^{\perp}\) and \(g_{\ominus}^{\perp}\) such that the coordinate-space scalar functions accompanying \(\Delta\b^i\) in the parametrizations above are their first Bessel moments. Thus, making the coordinate-space notation explicit, these parametrizations read
\begin{align}
\Phi_{11}^{[\gamma^-]}(x;\Delta\b)
&=\widetilde f_1(x;\Delta b),
\\
\Phi_{i,\oplus}^{[\gamma^-]}(x_1,x_2,x_3;\Delta\b)
&=-i\,\Delta \b^iM^2\,
\widetilde f_{\oplus}^{\perp(1)}
(x_1,x_2,x_3;\Delta b),
\\
\epsilon^{ij}
\Phi_{i,\ominus}^{[\gamma^-\gamma_5]}(x_1,x_2,x_3;\Delta\b)
&=-i\,\Delta \b^jM^2\,
\widetilde g_{\ominus}^{\perp(1)}
(x_1,x_2,x_3;\Delta b),
\end{align}
Their transverse Fourier transforms consequently give
\begin{align}
\int\dd^2\Delta\b\,e^{i\k\cdot\Delta\b}\,
\Phi_{11}^{[\gamma^-]}(x;\Delta\b)
&=(2\pi)^2 f_1(x;k),
\\
\int\dd^2\Delta\b\,e^{i\k\cdot\Delta\b}\,
\Phi_{i,\oplus}^{[\gamma^-]}(x_1,x_2,x_3;\Delta\b)
&=(2\pi)^2 \k^i
f_{\oplus}^{\perp}(x_1,x_2,x_3;k),
\\
\int\dd^2\Delta\b\,e^{i\k\cdot\Delta\b}\,
\epsilon^{ij}
\Phi_{i,\ominus}^{[\gamma^-\gamma_5]}(x_1,x_2,x_3;\Delta\b)
&=(2\pi)^2 \k^j
g_{\ominus}^{\perp}(x_1,x_2,x_3;k).
\end{align}
Here, $\Delta b=|\Delta\b|$ and $k=|\k|$. For the unpolarized target considered here, the scalar distributions depend only on these magnitudes, while the explicit factors $\Delta\b^i$ and $\k^i$ carry the transverse-vector dependence.

Using these relations, we can now express the longitudinal and transverse cross sections in terms of momentum-space TMD distributions.
Substituting these Fourier transforms into the expressions obtained above, the longitudinal cross section becomes
\begin{align}
(2\pi)^5\,2z(1-z)
\frac{\dd\sigma^{\gamma_L^*\to q_1g_2}}
{\dd^2\k\,\dd z\,\dd^2\P}
&=
\frac{4\,\alpha_{\rm em}\,\alpha_s\,e_f^2}{2q^+}
\frac{(2\pi)^5}{P^-_{\rm tar}}
\Big\{
\C_L(z,\k,\P)\,f_1(0;k)
\nonumber\\
&\hspace{5cm}
+2\pi\,\C_L^i(z,\P)\,\k^i\,
f_{\oplus}^{\perp}(0,0,0;k)
\Big\}
+\mathcal{O}({\rm NNLP}).
\label{eq:qg_long_TMD_fin}
\end{align}
Similarly, for a transversely polarized photon, we obtain
\begin{align}
& \hspace{-0.5cm}
(2\pi)^5\,2z(1-z)
\frac{\dd\sigma^{\gamma_T^*\to q_1g_2}}
{\dd^2\k\,\dd z\,\dd^2\P}
=
\frac{4\,\alpha_{\rm em}\,\alpha_s\,e_f^2}{2q^+}
\frac{(2\pi)^5}{P^-_{\rm tar}}
\Bigg\{
\C_T(z,\k,\P)\,f_1(0;k)
+2\pi\,\C_T^{{\rm unpol.},i}(z,\P)\,\k^i\,
f_{\oplus}^{\perp}(0,0,0;k)
\nonumber\\
&\hspace{0.5cm}
+2\pi\,\C_T^{{\rm h.\,pol.},j}(z,\P)\,\k^j\,
g_{\ominus}^{\perp}(0,0,0;k)
-2\pi i\,m\,\C_T^{{\rm h.\,flip},ij}(z,\P)
\int\frac{\dd^2\Delta\b}{(2\pi)^2}\,
e^{i\k\cdot\Delta\b}\,
\Phi_{i,\ominus}^{[\gamma^-\gamma^j]}
(0,0,0;\Delta\b)
\Bigg\}
+\mathcal{O}({\rm NNLP}).
\label{eq:qg_trans_TMD_fin}
\end{align}
Here, the vanishing longitudinal momentum fractions follow from the unrestricted longitudinal-coordinate integrations in the target operators. As in the matching relations above, this notation denotes the formal zero-fraction projection.

Equations~\eqref{eq:qg_long_TMD_fin} and \eqref{eq:qg_trans_TMD_fin} exhibit the two sources of NLP contributions. The kinematic corrections enter through the hard coefficients $\C_L$ and $\C_T$ multiplying the twist-two distribution $f_1$, while the genuine twist-three contributions involve three-body correlators. Both photon polarizations probe $f_{\oplus}^{\perp}$, whereas the transverse cross section also probes $g_{\ominus}^{\perp}$ and contains a mass-dependent helicity-flip contribution, whose tensor correlator has been left explicit. This contribution is included because we allow the quark mass to be of the same order as the hard transverse momentum $|\P|$. If instead one considers the regime $m\ll|\P|$, the helicity-flip contribution is power suppressed by an extra factor $m/|\P|$ and should thus be considered of order next-to-next-to-leading power (NNLP) instead of NLP, which is beyond the accuracy considered in the present study. In the regime $m\ll|\P|$, the hard coefficients can also be expanded around $m=0$.

\section{Quark-antiquark dijet production}
\label{sec:qbarq_production}

\begin{figure}[t]
\centering
\includegraphics[width=0.48\textwidth]{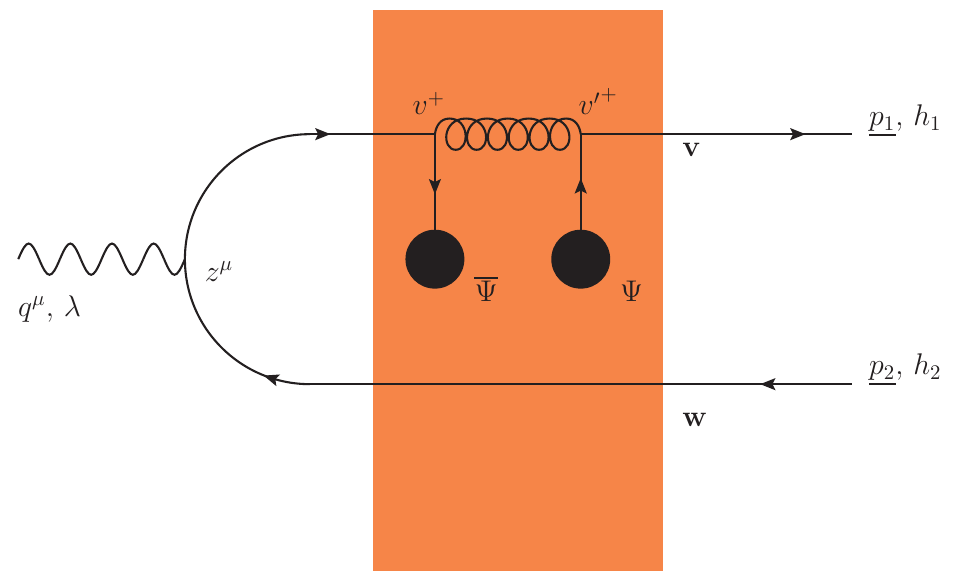}\hfill
\includegraphics[width=0.48\textwidth]{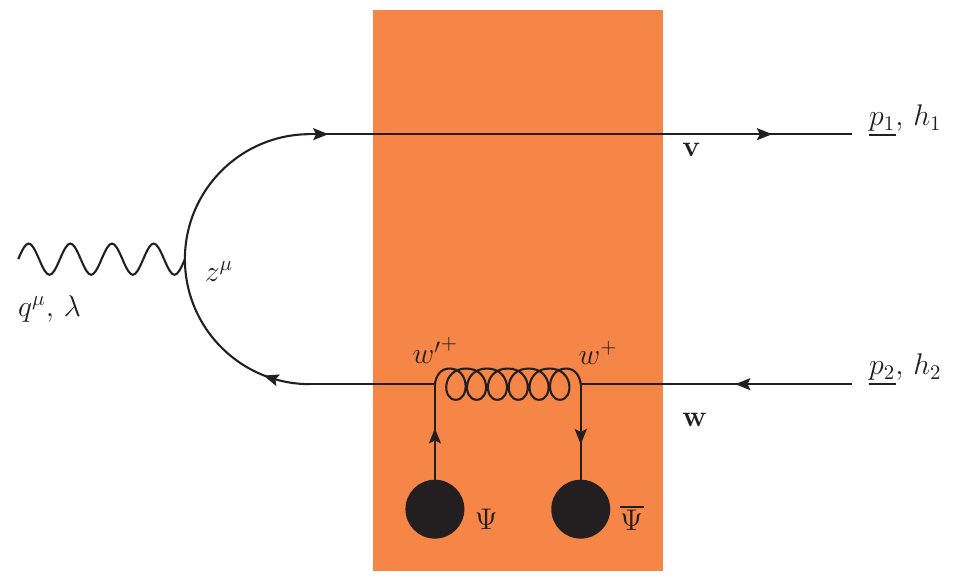}
\caption{Quark-background contributions to quark-antiquark dijet production, with the double quark-background insertions on the quark line (left) and the antiquark line (right).}
\label{fig:qqbar_quark_background}
\end{figure}

In this section, we study quark-antiquark dijet production in DIS, focusing on the quark-background contributions at NEik accuracy. In this channel, the quark background enters through two field insertions on either the quark or the antiquark line (see Fig. \ref{fig:qqbar_quark_background}). We first derive the corresponding scattering amplitudes from the $\S$-matrix in general kinematics, together with the eikonal amplitude needed to compute their interference. We then perform the correlation-limit expansion through NLP and evaluate the quark-background contribution to the cross section for longitudinally and transversely polarized photons. 
We then identify the three-body operator structures arising from this interference and express the results in terms of TMD correlators, following the approach used for quark–gluon production.
%

\subsection{Scattering amplitudes in general kinematics}
\label{subsec:qqbar_amp_gen_kin}

We start from the $\S$-matrix element for the production of a quark with momentum $\check p_1$\footnote{The four-vector $\check k^\mu$ denotes the on-shell counterpart of $k^\mu$, with the same plus and transverse components, $\check k^+=k^+$ and $\check\k=\k$, while its minus component is fixed by the on-shell condition, $\check k^-=(\k^2+m^2)/(2k^+)$.} and helicity $h_1$, and an antiquark with momentum $\check p_2$ and helicity $h_2$. At leading order in the electromagnetic coupling, with the interactions with the target encoded in the background-field propagators, the LSZ reduction formula gives \cite{Altinoluk:2024dba}
\begin{align}
\label{eq:S_matrix-qqbar_prod}
\S_{\gamma^*\to q_1{\bar q}_2}&=\lim_{x^+,y^+\to+\infty}
\int \dd^2\x \int\dd x^- \int\dd^2\y \int \dd y^- \, 
e^{ix\cdot \check{p}_1} \, e^{iy\cdot \check{p}_2}
\int \dd^4 z \, e^{-iz\cdot q} \, \epsilon^{\lambda}_\mu(q)
\nonumber\\
& \hspace{4cm}
\times 
(-i\, e\, e_f)\, {\bar u}(\check{p}_1,h_1)\, \gamma^+\, 
S_F(x,z)\, \gamma^\mu \big[ -S_F(z,y)\big]\, \gamma^+\, v(\check{p}_2,h_2)
\end{align}
To isolate the quark-background contributions at NEik accuracy, we use the eikonal quark propagator in the gluon background together with the correction containing two quark-background insertions. The eikonal propagator reads
\begin{align}
\label{eq:eik_q_prop_1}
S_F^{\rm Eik}(x,y)& =
\delta^{(3)}(\underline{x}-\underline{y})\, \text{sgn}(x^--y^-)\, \frac{\gamma^+}{4}
+ 
\int \frac{\dd^3\underline{p}}{(2\pi)^3}
\int \frac{\dd^3\underline{k}}{(2\pi)^3}
(2\pi) \delta(p^+-k^+)
\Big[ \theta(k^+)\theta(x^+-y^+)-\theta(-k^+)\theta(y^+-x^+)\Big]
\nonumber\\
& 
\times 
\frac{[p^+\gamma^-+i\gamma^l {\overrightarrow D}_{x^l}+m]}{2p^+}
\, \gamma^+\, e^{-ix\cdot\check{p}}
\int \dd^2\z \, e^{-i\z\cdot(\p-\k)}\, U_F(x^+,y^+;\z)\, 
e^{iy\cdot\check{k}}\, 
\frac{[k^+\gamma^--i\gamma^j{\overleftarrow D}_{y^j}+m]}{2k^+}
\end{align}
The quark-background correction describes the conversion of the propagating quark into a gluon and its subsequent conversion back into a quark. The adjoint and fundamental Wilson lines describe the multiple interactions of the intermediate gluon and the quark or antiquark with the target gluon background, respectively. This contribution is of NEik order and reads
 \begin{align}
 \label{eq:Neik_PBP_q_prop_1}
 S_F^{\rm NEik, \Psi\bar\Psi}(x,y) &=
 -g^2\, \delta^{ij}\, \int\frac{\dd^3 \underline{p}}{(2\pi)^3}\, \int\frac{\dd^3 \underline{k}}{(2\pi)^3}\, 
 \frac{2\pi\delta(p^+-k^+)}{2k^+}\, \int \dd z^+\int \dd z'^+\int\dd^2\z 
 \nonumber\\
 & 
 \times 
 \Big[ \theta(k^+)\theta(x^+-z'^+)\theta(z'^+-z^+)\theta(z^+-y^+) 
 - \theta(-k^+)\theta(y^+-z^+)\theta(z^+-z'^+)\theta(z'^+-x^+) \Big]
 \nonumber\\
&  \times 
 T\bigg\{ \frac{[p^+\gamma^-+i\gamma^l{\overrightarrow D}_{x^l}+m]}{2p^+}
 e^{-ix\cdot\check{p}}\, e^{-i\z\cdot(\p-\k)}\, 
 U_F(x^+,z'^+;\z)\, \gamma^+ t^a \gamma^i \frac{\gamma^-}{2}\, \Psi(z'^+,\z) 
 \, 
 \nonumber\\
 & \hspace{0.9cm}
 \times 
 U_A(z'^+,z^+;\z)_{ab}
 \bar{\Psi}(z^+,\z)\, \frac{\gamma^-}{2}t^b\gamma^j\gamma^+\, U_F(z^+,y^+;\z) 
 \, e^{iy\cdot\check{k}} \, \frac{[k^+\gamma^--i\gamma^m{\overleftarrow D}_{y^m}+m]}{2k^+}\bigg\}
 \end{align}
Substituting these propagators into Eq.~\eqref{eq:S_matrix-qqbar_prod} and expanding through NEik order, we obtain the eikonal contribution and two quark-background corrections, according to whether the pair of background-field insertions lies on the quark or the antiquark line:
\begin{align}
\label{eq:S_matrix_decomp}
\S_{\gamma^*\to q\bar q}=\S_{\gamma^*\to q\bar q}\big|_{\rm Eik}+ 
\S_{\gamma^*\to q\bar q}\big|_{\rm NEik}^{\rm \Psi\bar\Psi\, dec. \, on \, q}
+
\S_{\gamma^*\to q\bar q}\big|_{\rm NEik}^{\rm \Psi\bar\Psi\, dec. \, on \, \bar q}
\end{align}
The contribution containing a double quark exchange on both lines is beyond NEik accuracy and is neglected. Throughout this section, we consider only the NEik corrections induced by the quark background; interactions with the gluon background are included through the eikonal Wilson lines.
\begin{align}
\label{eq:S_matrix_Eik_explicit} 
\S_{\gamma^*\to q_1\bar q_2}\big|_{\rm Eik}& =\lim_{x^+,y^+\to+\infty}
\int \dd^2\x \int\dd x^- \int\dd^2\y \int \dd y^- \, 
e^{ix\cdot \check{p}_1} \, e^{iy\cdot \check{p}_2}
\int \dd^4 z \, e^{-iz\cdot q} \, \epsilon^{\lambda}_\mu(q)
\nonumber\\
& \hspace{4cm}
\times 
(-i\, e\, e_f)\, {\bar u}(\check{p}_1,h_1)\, \gamma^+\, 
S^{\rm Eik}_F(x,z)\, \gamma^\mu \big[ -S^{\rm Eik}_F(z,y)\big]\, \gamma^+\, v(\check{p}_2,h_2)
\end{align}

\begin{align}
\label{eq:S_matrix_NEik_on_q_explicit} 
\S_{\gamma^*\to q_1\bar q_2}\big|_{\rm NEik}^{\rm \Psi\bar\Psi \, dec . \, on\, q}& =\lim_{x^+,y^+\to+\infty}
\int \dd^2\x \int\dd x^- \int\dd^2\y \int \dd y^- \, 
e^{ix\cdot \check{p}_1} \, e^{iy\cdot \check{p}_2}
\int \dd^4 z \, e^{-iz\cdot q} \, \epsilon^{\lambda}_\mu(q)
\nonumber\\
& \hspace{3cm}
\times 
(-i\, e\, e_f)\, {\bar u}(\check{p}_1,h_1)\, \gamma^+\, 
S^{\rm NEik, \Psi\bar\Psi}_F(x,z)\, \gamma^\mu \big[ -S^{\rm Eik}_F(z,y)\big]\, \gamma^+\, v(\check{p}_2,h_2)
\end{align}

\begin{align}
\label{eq:S_matrix_NEik_on_barq_explicit} 
\S_{\gamma^*\to q_1{\bar q}_2}\big|_{\rm NEik}^{\rm \Psi\bar\Psi \, dec . \, on\, \bar q}& =\lim_{x^+,y^+\to+\infty}
\int \dd^2\x \int\dd x^- \int\dd^2\y \int \dd y^- \, 
e^{ix\cdot \check{p}_1} \, e^{iy\cdot \check{p}_2}
\int \dd^4 z \, e^{-iz\cdot q} \, \epsilon^{\lambda}_\mu(q)
\nonumber\\
& \hspace{3cm}
\times 
(-i\, e\, e_f)\, {\bar u}(\check{p}_1,h_1)\, \gamma^+\, 
S^{\rm Eik}_F(x,z)\, \gamma^\mu \big[ -S^{\rm NEik, \Psi\bar\Psi}_F(z,y)\big]\, \gamma^+\, v(\check{p}_2,h_2)
\end{align}
We next perform the integrations over the coordinates of the outgoing quark and antiquark in the LSZ formula. The instantaneous term is proportional to $\gamma^+$. In the LSZ reduction, this matrix multiplies the $\gamma^+$ from the external-state projector, so the contribution vanishes because $(\gamma^+)^2=0$. For the outgoing quark, the reduced propagator reads
\begin{align}
\label{eq:quark_prop_Eik+NEik}
S_F^{\rm Eik}(x,y)+S_F^{\rm NEik, \Psi\bar\Psi}(x,y)
& =
\delta^{(3)}(\underline{x}-\underline{y})\, \text{sgn}(x^--y^-)\, \frac{\gamma^+}{4}
+
\int\frac{\dd^3{\underline p}}{(2\pi)^3}\int\frac{\dd^3{\underline k}}{(2\pi)^3}
\, 2\pi\delta(p^+-k^+)\int\dd^2\z\, e^{-i\z\cdot(\p-\k)}
\nonumber\\
& \hspace{-2.5cm}
\times 
T\bigg\{ \frac{[p^+\gamma^-+i\gamma^l{\overrightarrow D}_{x^l}+m]}{2p^+} \, e^{-ix\cdot\check{p}}
\bigg\lgroup \big[ \theta(k^+)\theta(x^+-y^+)-\theta(-k^+)\theta(y^+-x^+)\big] \gamma^+\, U_F(x^+,y^+;\z)
\nonumber\\
& \hspace{-2.5cm}
+ \int\dd z^+\int\dd z'^+
\big[ \theta(k^+)\theta(x^+-z'^+)\theta(z'^+-z^+)\theta(z^+-y^+) - \theta(-k^+)\theta(y^+-z^+)\theta(z^+-z'^+)\theta(z'^+-x^+) \big]
\nonumber\\
& \hspace{-2.5cm}
\times 
\frac{(-g^2)\delta^{ij}}{2k^+} U_F(x^+,z'^+;\z)\, t^a\frac{\gamma^+\gamma^i\gamma^-}{2}\, \Psi(z'^+,\z)\, {U_A(z'^+,z^+;\z)}_{ab}\, 
\bar{\Psi}(z^+,\z)\frac{\gamma^-\gamma^j\gamma^+}{2}\, t^b\, U_F(z^+,y^+;\z)\bigg\rgroup
\nonumber\\
& \hspace{-2.5cm}
\times 
e^{iy\cdot\check{k}}\, \frac{[k^+\gamma^--i\gamma^m{\overleftarrow D}_{y^m}+m]}{2k^+}\bigg\}
\end{align} 
For the external-leg reduction, it is convenient to combine the eikonal propagator and its quark-background correction. Using Eqs.~\eqref{eq:eik_q_prop_1} and \eqref{eq:Neik_PBP_q_prop_1}, their sum can be written as
\begin{align}
\label{eq:reduced_prop_1}
& \hspace{-0.7cm}
\lim_{x^+\to+\infty}\int\dd^2\x\int\dd x^-\, e^{i\check{p}_1\cdot x}{\bar u}(\check{p}_1,h_1)\gamma^+
\, \Big[  S_F^{\rm Eik}(x,z)+S_F^{\rm NEik, \Psi\bar\Psi}(x,z)\Big]
\nonumber\\
& =
\theta(p_1^+)\int \frac{\dd^2\k_1}{(2\pi)^2}\int\dd^2\v \, e^{-i\v\cdot(\p_1-\k_1)} {\bar u}(\check{p}_1,h_1)\, T\bigg\{ 
\gamma^+U_F(+\infty, z^+;\v)+\int \dd v^+\int \dd v'^+
\theta(v'^+-v^+)\, \theta(v^+-z^+)
\nonumber\\
& \hspace{1.5cm}
\times 
\frac{(-g^2)\, \delta^{ij}}{2p_1^+}\, U_F(+\infty, v'^+;\v) \, t^a\, \frac{\gamma^+\gamma^i\gamma^-}{2}\, \Psi(v'^+, \v)\, {U_A(v'^+,v^+;\v)}_{ab}\, \bar\Psi(v^+,\v)\frac{\gamma^-\gamma^j\gamma^+}{2}\, t^b\, U_F(v^+,z^+;\v)\bigg\}
\nonumber\\
& \hspace{1.5cm}
\times 
e^{iz^-p_1^+}\, e^{-i\z\cdot\k_1}\, e^{iz^+(\k_1^2+m^2)/(2p_1^+)}\, \frac{[p_1^+\gamma^--i\gamma^m{\overleftarrow D}_{z^m}+m]}{2p_1^+}
\end{align}
Similarly, reducing the outgoing antiquark leg gives
\begin{align}
\label{eq:reduced_prop_2}
& \hspace{-0.7cm}
\lim_{y^+\to+\infty}\int\dd^2\y\int \dd y^-\, e^{iy\cdot\check{p}_2}\, (-1)\, \Big[  S_F^{\rm Eik}(z,y)+S_F^{\rm NEik, \Psi\bar\Psi}(z,y)\Big]
\, \gamma^+\, v(\check{p}_2,h_2)
\nonumber\\
& =
\theta(p_2^+) \int\frac{\dd^2\k_2}{(2\pi)^2}\int \dd^2\w \, e^{-i\w\cdot(\p_2-\k_2)}
\, T\bigg\{ 
\frac{[p_2^+\gamma^--i\gamma^l{\overrightarrow D}_{z^l}-m]}{2p_2^+}
\, e^{iz^-p_2^+}\, e^{-i\z\cdot\k_2}\, e^{iz^+(\k_2^2+m^2)/(2p_2^+)}
\nonumber\\
& \hspace{1.5cm}
\times 
\bigg\lgroup \gamma^+\, U_F(z^+,+\infty;\w)+\int \dd w^+\int \dd w'^+ 
\theta(w^+-w'^+)\, \theta(w'^+-z^+)\, \frac{g^2\, \delta^{ij}}{2p_2^+}\, 
U_F(z^+,w'^+;\w)\, t^a
\nonumber\\
& \hspace{2.2cm}
\times 
\frac{\gamma^+\gamma^i\gamma^-}{2}\, \Psi(w'^+,\w)\, {U_A(w'^+,w^+;\w)}_{ab}\, \bar{\Psi}(w^+,\w)
\, \frac{\gamma^-\gamma^j\gamma^+}{2}\, t^b\, U_F(w^+,+\infty,;\w)\bigg\rgroup \bigg\}
v(\check{p}_2,h_2)
\end{align}
The $\S$-matrix element is obtained by contracting these two reduced propagators with the photon vertex and integrating over the splitting coordinate $z$. Multiplying the two propagators, we include the product of their eikonal contributions and the terms with a quark-background correction on either the quark or the antiquark line. The term with corrections on both lines is beyond NEik accuracy and is neglected: 
\begin{align}
\label{eq:S-matrix_in_red_op}
\S_{\gamma^*\to q_1\bar q_2}&=\epsilon^\lambda_{\mu}(q)\, (-i\,e\, e_f) \int \dd^4z \, e^{-iz\cdot q}
\nonumber\\
&
\times 
\bigg\lgroup \lim_{x^+\to+\infty}\int\dd^2\x\int\dd x^-\, e^{i\check{p}_1\cdot x}{\bar u}(\check{p}_1,h_1)\gamma^+
\, \Big[  S_F^{\rm Eik}(x,z)+S_F^{\rm NEik, \Psi\bar\Psi}(x,z)\Big]
\bigg\rgroup 
\, \gamma^{\mu}
\nonumber \\
& 
\times 
\bigg\lgroup
\lim_{y^+\to+\infty}\int\dd^2\y\int \dd y^-\, e^{iy\cdot\check{p}_2}\, (-1)\, \Big[  S_F^{\rm Eik}(z,y)+S_F^{\rm NEik, \Psi\bar\Psi}(z,y)\Big]
\, \gamma^+\, v(\check{p}_2,h_2)
\bigg\rgroup
\end{align} 
To determine the relevant range of the photon-splitting coordinate, we consider target fields localized within a longitudinal interval of width $L^+$ around $x^+=0$. The contribution containing a double quark-background insertions is already of NEik order. If the photon also splits inside the target, the integration over its splitting time introduces an additional suppression by the target width, so that this contribution starts at NNEik order. A photon splitting after the target cannot contribute to the quark-background terms, since the required background-field insertions must occur after the splitting. Consequently, at the accuracy considered here, the photon splits before the target in both quark-background contributions.

For these terms, we may replace the upper boundary of the before-target splitting integral by $z^+=0$, since the resulting correction is beyond NEik accuracy. The Wilson-line endpoints associated with the splitting vertex can then be extended to $-\infty$, and the covariant derivatives at that vertex reduce to ordinary derivatives. Keeping the pure-gluon contribution at eikonal order, the before-target contribution to the $\S$-matrix therefore becomes
\begin{align}
\label{eq:S-matrix_in_red_op_explicit}
\S_{\gamma^*\to q_1\bar q_2}
&
= \epsilon^\lambda_{\mu}(q)\, (-i\,e\, e_f) 
\theta(p_1^+)\, \theta(p_2^+) 
\int\frac{\dd^2\k_1}{(2\pi)^2} \int\frac{\dd^2\k_2}{(2\pi)^2}
\int\dd^2\v \,  e^{-i\v\cdot(\p_1-\k_1)}
\int\dd^2\w \,  e^{-i\w\cdot(\p_2-\k_2)}
\int \dd^4z\, e^{-iz\cdot q}\, \theta(-z^+)
\nonumber\\
& 
\times 
{\bar u}(\check{p}_1,h_1)\, T\bigg\{ \gamma^+U_F(+\infty,-\infty;\v)+\int\dd v^+\int \dd v'^+ \, \theta(v'^+-v^+) \frac{(-g^2)\, \delta^{i'j'}}{2p_1^+} U_F(+\infty, v'^+;\v)\, t^{a'}\frac{\gamma^+\gamma^{i'}\gamma^-}{2}
\nonumber\\
& 
\times 
\Psi(v'^+,\v)\, {U_A(v'^+,v^+;\v)}_{a'b'}\, \bar\Psi(v^+,\v)\frac{\gamma^-\gamma^{j'}\gamma^+}{2}\, t^{b'}\, U_F(v^+, -\infty;\v)\bigg\}\,  e^{iz^-p_1^+}\, e^{-i\z\cdot\k_1}\, e^{iz^+(\k_1^2+m^2)/(2p_1^+)}
\nonumber\\
& 
\times 
\frac{[p_1^+\gamma^--\k_1^m\gamma^m+m]}{2p_1^+}\, \gamma^{\mu}\, \frac{[p_2^+\gamma^--\k_2^l\gamma^l-m]}{2p_2^+}
\, e^{iz^-p_2^+}\, e^{-i\z\cdot\k_2}\, e^{iz^+(\k_2^2+m^2)/(2p_2^+)}
T\bigg\{ \gamma^+U_F(-\infty,+\infty;\w)
\nonumber\\
& 
+ \int \dd w^+\int \dd w'^+\theta(w^+-w'^+)
\, \frac{g^2\, \delta^{ij}}{2p_2^+}\, U_F(-\infty,w'^+;\w)\, t^a\, \frac{\gamma^+\gamma^i\gamma^-}{2}\, 
\Psi(w'^+,\w)\, {U_A(w'^+,w^+;\w)}_{ab}\, \bar{\Psi}(w^+,\w)
\nonumber\\
& 
\times 
\frac{\gamma^-\gamma^j\gamma^+}{2}\, t^b\, U_F(w^+,+\infty;\w)\bigg\} v(\check{p}_2,h_2)
\end{align}
The dependence on the photon-splitting coordinate is now contained entirely in the phase factors. Integrating over \(z^-\) and \(\z\) enforces longitudinal and transverse momentum conservation at the photon vertex, while the integration over \(z^+<0\) gives the light-cone energy denominator:
\begin{align}
&
\int dz^-\, e^{-iz^-(q^+-p_1^+-p_2^+)}
\int \dd^2\z \, e^{i\z\cdot(\q-\k_1-\k_2)}
\int dz^+ \, \theta(-z^+)\, e^{-iz^+(q^--\check{k}_1^--\check{k}_2^-)}
\nonumber\\
& \hspace{6cm}
= (2\pi)\, \delta(q^+-p_1^+-p_2^+)\, (2\pi)^2\, \delta^{(2)}(\q-\k_1-\k_2)
\, \frac{(-i)}{[\check{k}_1^-+\check{k}_2^--q^--i0]}
\end{align}
Using \(p_1^++p_2^+=q^+\) and \(\k_1+\k_2=\q\), we introduce the relative transverse momentum
\begin{equation}
\K\equiv\k_1-\frac{p_1^+}{q^+}\q.
\end{equation}
The light-cone energy denominator can then be expressed as
\begin{align}
\frac{(-i)}{[\check{k}_1^-+\check{k}_2^--q^--i0]}
&=\frac{(-i)}{\left\{ \frac{q^+}{2p_1^+p_2^+}\left[ \Big(\k_1-\frac{p_1^+}{q^+}\q\Big)^2+m^2\right]-\frac{p_1^+\q^2}{2q^+p_2^+}+\frac{\q^2}{2p_2^+}-q^--i0\right\}}
\nonumber\\
& =
\frac{(-i)}{\Big[ \frac{q^+}{2p_1^+p_2^+}\big( \K^2+m^2\big)-\frac{q^2}{2q^+}-i0\Big]}
\end{align}
Substituting these results and using the transverse momentum delta function to perform the \(\k_2\) integration, the before-target contribution becomes
\begin{align}
\label{eq:S-matrix_in_red_op_explicit_2}
\S_{\gamma^*\to q_1\bar q_2} &=
\epsilon^{\lambda}_{\mu}(q)\, (-i\, e\, e_f) \, 2\pi\delta(p_1^++p_1^+-q^+)\, \theta(p_1^+) \, \theta(p_2^+)
\int \dd^2\v\,  e^{-i\v\cdot\big(\p_1-\frac{p_1^+}{q^+}\q\big)}
\int \dd^2\w\, e^{-i\w\cdot\big(\p_2-\frac{p_2^+}{q^+}\q\big)}
\nonumber\\
& \times 
\int \frac{\dd^2\K}{(2\pi)^2}\frac{(-i)\, e^{i\K\cdot(\v-\w)}}{\Big[ \frac{q^+}{2p_1^+p_2^+} (\K^2+m^2)-\frac{q^2}{2q^+}-i0\Big]}\, 
\bar{u}(\check{p}_1,h_1)\,
\nonumber\\
& 
\times  
T\, \bigg\{ U_F(\v)
-
\frac{g^2\delta^{i'j'}}{2p_1^+}\int \dd v^+\int \dd v'^+\, \theta(v'^+-v^+) U_F(+\infty, v'^+;\v)\, t^{a'}\, \frac{\gamma^+\gamma^{i'}\gamma^-}{2}\, \Psi(v'^+,\v)
\, {U_A(v'^+,v^+;\v)}_{a'b'}\, 
\nonumber\\
& 
\times 
\bar\Psi(v^+,\v)\, \frac{\gamma^-\gamma^{j'}}{2}\, t^{b'}\, U_F(v^+,-\infty;\v)\bigg\}\, \gamma^+\bigg[ p_1^+\gamma^--\gamma^{m}\Big(\K^m+\frac{p_1^+}{q^+}\q^m\Big)+m\bigg]\, 
\frac{\gamma^\mu}{(2p_1^+)(2p_2^+)}\, 
\nonumber\\
& 
\times 
\bigg[ p_2^+\gamma^--\gamma^{l}\Big(-\K^l+\frac{p_2^+}{q^+}\q^l\Big)-m\bigg]\, \gamma^+
\bigg\{ U_F^\dagger(\w)+\frac{g^2\delta^{ij}}{2p_2^+}\int \dd w^+\int \dd w'^+\, \theta(w^+-w'^+)
U_F(-\infty, w'^+;\w)\, 
\nonumber\\
& 
\times 
t^a\, \frac{\gamma^i\gamma^-}{2}\, \Psi(w'^+,\w)\, {U_A(w'^+,w^+;\w)}_{ab}\, \bar{\Psi}(w^+,\w)\frac{\gamma^-\gamma^j\gamma^+}{2}\, t^b\, U_F(w^+,+\infty;\w)\bigg\}v(\check{p}_2,h_2)
\end{align}
The remaining Dirac structure associated with the photon splitting can be evaluated separately for longitudinal and transverse polarizations. The details are presented in Appendix~\ref{subsec:appphoton-splitting_hard_factor}; here, we quote the results. For a longitudinally polarized photon, we use the effective polarization vector obtained after applying current conservation,
\begin{equation}
\label{eq:def_epsilon_L}
\epsilon^L_\mu(q)=\frac{Q}{q^+}g_\mu{}^+.
\end{equation}
The corresponding photon vertex reads
\begin{align} 
\label{gamma_vertex_L}
\epsilon^L_\mu(q)\gamma^+\Big[ p_1^+\gamma^--\gamma^m\Big(\K^m+\frac{p_1^+}{q^+}\q^m\Big)+m\Big]\, \gamma^\mu\, 
\Big[ p_2^+\gamma^--\gamma^l\Big(-\K^l+\frac{p_2^+}{q^+}\q^l\Big)-m\Big]\, \gamma^+
= \frac{Q}{q^+}\, (2p_1^+)\, (2p_2^+) \, \gamma^+
\end{align}
For a transversely polarized photon, we use
\begin{equation}
\label{eq:def_epsilon_T}
\epsilon^\lambda_\mu(q)
=
\varepsilon^\lambda_l
\left[g_\mu{}^l-\frac{\q^l}{q^+}g_\mu{}^+\right],
\end{equation}
with transverse indices raised and lowered according to our metric convention. Contracting this polarization vector with the photon vertex gives
\begin{align}
\label{gamma_vertex_T}
&
\epsilon^\lambda_\mu(q)\gamma^+\Big[ p_1^+\gamma^--\gamma^m\Big(\K^m+\frac{p_1^+}{q^+}\q^m\Big)+m\Big]\, \gamma^\mu\, 
\Big[ p_2^+\gamma^--\gamma^l\Big(-\K^l+\frac{p_2^+}{q^+}\q^l\Big)-m\Big]\, \gamma^+
\nonumber \\
& \hspace{5cm}
= -\varepsilon^l_{\lambda}\Big\{ 2(p_2^+-p_1^+)\K^l\gamma^++(2q^+)\K^m \gamma^+\frac{[\gamma^l,\gamma^m]}{2}+(2q^+)\,m\, \gamma^+\gamma^l\Big\}
\end{align} 
Both results can be expressed in terms of a common photon-splitting hard factor, defined by
\begin{align}
\Gamma^\lambda(\K)
=
\begin{cases}
\displaystyle
2Q\,\frac{p_1^+p_2^+}{(q^+)^2}\,\gamma^+,
& \lambda=L,
\\[3mm]
\displaystyle
-\varepsilon^l_\lambda
\left[
\frac{p_2^+-p_1^+}{q^+}\K^l\gamma^+
+\K^m\gamma^+\frac{[\gamma^l,\gamma^m]}{2}
+m\gamma^+\gamma^l
\right],
& \lambda=T.
\end{cases}
\label{eq:qqbar_photon_hard_factor}
\end{align}
With this definition, the contracted photon vertex is $2q^+\Gamma^\lambda(\K)$. This factor of $2q^+$ cancels the prefactor arising from the combination of the endpoint denominators and the light-cone energy denominator.

To obtain the complete result at the accuracy considered here, we also include photon splitting after the target. This contribution contains no interaction with the background fields. For a spacelike photon, the sum of the before- and after-target vacuum contributions vanishes by energy conservation. The after-target term therefore subtracts the no-interaction part of the eikonal before-target contribution, giving
\begin{equation}
U_F(\v)\,\Gamma^\lambda(\K)\,U_F^\dagger(\w)
\;\longrightarrow\;
U_F(\v)\,\Gamma^\lambda(\K)\,U_F^\dagger(\w)
-\Gamma^\lambda(\K).
\end{equation}
The quark-background contributions already vanish when the background fields are switched off and require no analogous subtraction. Combining the eikonal term with the corrections on the quark and antiquark lines, we obtain
\begin{align}
\S_{\gamma^*\to q_1\bar q_2}
& = 
(-i\, e\, e_f)\, 2\pi\delta(p_1^++p_2^+-q^+)\, \theta(p_1^+)\theta(p_2^+)
\int\dd^2 \v\, e^{-i\v\cdot\big(\p_1-\frac{p_1^+}{q^+}\q\big)}
\int\dd^2\w\, e^{-i\w\cdot\big(\p_2-\frac{p_2^+}{q^+}\q\big)}
\int \frac{\dd^2\K}{(2\pi)^2}\frac{e^{i\K\cdot(\v-\w)}}{[\K^2+\Qb^2]}
\nonumber\\
& 
\times 
(-i)\, {\bar u}(\check{p}_1,h_1)\, T\bigg\{ \Big[ U_F(\v)\Gamma^\lambda(\K)U^\dagger_F(\w)-\Gamma^\lambda(\K)\Big]
+
U_F(\v)\, \Gamma^{\lambda}(\K)\, \frac{g^2\, \delta^{ij}}{2p_2^+}\int\dd w^+\int \dd w'^+
\theta(w^+-w'^+)
\nonumber\\
& 
\times 
U_F(-\infty, w'^+;\w)\, t^a\, \frac{\gamma^i\gamma^-}{2}\, \Psi(w'^+,\w)\, {U_A(w'^+,w^+;\w)}_{ab}\, \bar\Psi(w^+,\w)\, \frac{\gamma^-\gamma^j\gamma^+}{2}\, t^b\, U_F(w^+, +\infty; \w)
\nonumber\\
&
-
\frac{g^2\delta^{ij}}{2p_1^+}\int \dd v^+\int \dd v'^+\theta(v'^+-v^+)\, U_F(+\infty, v'^+;\v)\, t^a
\frac{\gamma^+\gamma^i\gamma^-}{2}\, \Psi(v'^+, \v)\, {U_A(v'^+,v^+;\v)}_{ab}\, 
\nonumber\\
& 
\times \bar\Psi(v^+,\v)\, \frac{\gamma^-\gamma^j}{2}\, t^b\, U_F(v^+, -\infty;\v)\, \Gamma^\lambda(\K)\, U^\dagger_F(\w)\bigg\}v(\check{p}_2,h_2)
\end{align}
We define the scattering amplitudes by factoring out the longitudinal momentum-conserving delta function together with the normalization factor $2q^+$. For physical outgoing states, $p_1^+>0$ and $p_2^+>0$, so the corresponding step functions are unity. Separating the eikonal contribution from the quark-background corrections on the antiquark and quark lines, we write
\begin{align}
\S_{\gamma^*\to q_1\bar q_2}=(2q^+)\, 2\pi\delta(p_1^++p_2^+-q^+) \, i\Big[ {\mathcal M}_{\gamma_{\lambda}^*\to q_1\bar q_2}^{\rm Eik}
+
{\mathcal M}_{\gamma_\lambda^*\to q_1\bar q_2}^{\rm NEik, \Psi\bar\Psi \, on \, \bar q}
+
{\mathcal M}_{\gamma_\lambda^*\to q_1\bar q_2}^{\rm NEik, \Psi\bar\Psi \, on \, q}
\Big]
\end{align}
with
\begin{align}
i\,{\mathcal M}_{\gamma_\lambda^*\to q_1\bar q_2}^{\rm Eik}
&=
-\frac{e\,e_f}{2q^+}
\int\dd^2\v\,
e^{-i\v\cdot\left(\p_1-\frac{p_1^+}{q^+}\q\right)}
\int\dd^2\w\,
e^{-i\w\cdot\left(\p_2-\frac{p_2^+}{q^+}\q\right)}
\int\frac{\dd^2\K}{(2\pi)^2}\,
\frac{e^{i\K\cdot(\v-\w)}}{\K^2+\Qb^2}
\nonumber\\
& \hspace{5cm}
\times
\bar u(\check p_1,h_1)\,
T\Big\{
U_F(\v)\,\Gamma^\lambda(\K)\,U_F^\dagger(\w)
-\Gamma^\lambda(\K)
\Big\}
v(\check p_2,h_2),
\label{eq:qqbar_M_Eik}
\\[2mm]
i\,{\mathcal M}_{\gamma_\lambda^*\to q_1\bar q_2}
^{\rm NEik,\Psi\bar\Psi\,on\,\bar q}
&=
-\frac{e\,e_f\,g^2}{(2q^+)(2p_2^+)}
\int\dd^2\v\,
e^{-i\v\cdot\left(\p_1-\frac{p_1^+}{q^+}\q\right)}
\int\dd^2\w\,
e^{-i\w\cdot\left(\p_2-\frac{p_2^+}{q^+}\q\right)}
\int\frac{\dd^2\K}{(2\pi)^2}\,
\frac{e^{i\K\cdot(\v-\w)}}{\K^2+\Qb^2}
\nonumber\\
&\quad
\times
\delta^{ij}
\int\dd w^+\int\dd w'^+\,
\theta(w^+-w'^+)\,
\bar u(\check p_1,h_1)
\nonumber\\
&\quad
\times
T\Bigg\{
U_F(\v)\,\Gamma^\lambda(\K)\,
U_F(-\infty,w'^+;\w)\,t^a\,
\frac{\gamma^i\gamma^-}{2}\,
\Psi(w'^+,\w)\,
U_A(w'^+,w^+;\w)_{ab}
\nonumber\\
&\hspace{6cm}
\times
\bar\Psi(w^+,\w)\,
\frac{\gamma^-\gamma^j\gamma^+}{2}\,
t^b\,U_F(w^+,+\infty;\w)
\Bigg\}
v(\check p_2,h_2),
\label{eq:qqbar_M_NEik_barq}
\\[2mm]
i\,{\mathcal M}_{\gamma_\lambda^*\to q_1\bar q_2}
^{\rm NEik,\Psi\bar\Psi\,on\,q}
&=
\frac{e\,e_f\,g^2}{(2q^+)(2p_1^+)}
\int\dd^2\v\,
e^{-i\v\cdot\left(\p_1-\frac{p_1^+}{q^+}\q\right)}
\int\dd^2\w\,
e^{-i\w\cdot\left(\p_2-\frac{p_2^+}{q^+}\q\right)}
\int\frac{\dd^2\K}{(2\pi)^2}\,
\frac{e^{i\K\cdot(\v-\w)}}{\K^2+\Qb^2}
\nonumber\\
&\quad\times
\delta^{ij}
\int\dd v^+\int\dd v'^+\,
\theta(v'^+-v^+)\,
\bar u(\check p_1,h_1)
\nonumber\\
&\quad\times
T\Bigg\{
U_F(+\infty,v'^+;\v)\,t^a\,
\frac{\gamma^+\gamma^i\gamma^-}{2}\,
\Psi(v'^+,\v)\,
U_A(v'^+,v^+;\v)_{ab}\,
\bar\Psi(v^+,\v)
\nonumber\\
&\hspace{6.5cm}\times
\frac{\gamma^-\gamma^j}{2}\,
t^b\,U_F(v^+,-\infty;\v)\,
\Gamma^\lambda(\K)\,U_F^\dagger(\w)
\Bigg\}
v(\check p_2,h_2).
\label{eq:qqbar_M_NEik_q}
\end{align}
These amplitudes are valid in general transverse kinematics at the stated accuracy in the high-energy expansion. The quark-background contribution to the cross section at NEik order is obtained from the interference of each quark-background amplitude with the eikonal amplitude. Terms quadratic in the quark-background amplitudes are beyond this accuracy. In the next subsection, we express the amplitudes in the momentum and coordinate variables introduced in Section~II and perform the back-to-back correlation-limit expansion through NLP.

\subsection{Scattering amplitudes in the back-to-back correlation limit}
\label{sec:qbarq_prod_b2b_amplitudes}

We now express the amplitudes derived in the previous subsection in terms of the momentum and coordinate variables introduced in Sec.~\ref{sec:setup} and perform the correlation-limit expansion. The eikonal color structure vanishes at zero transverse separation, so its leading contribution is linear in $\r$. The quark-background operator structures have a nonvanishing term at $\r=0$. Their interference gives the three-body operator structure relevant at twist-three accuracy. We therefore expand the eikonal color structure to linear order in $\r$ and evaluate the quark-background operator structures at $\r=0$.

For the outgoing quark and antiquark with momenta $\underline p_1$ and $\underline p_2$, respectively, the momentum variables introduced in Sec.\ref{sec:setup} read
\begin{equation}
\k=\p_1+\p_2,
\qquad
\P=(1-z)\p_1-z\p_2,
\qquad
z=\frac{p_1^+}{q^+},
\end{equation}
where \(q^+=p_1^++p_2^+\). The corresponding transverse coordinates are
\begin{equation}
\v=\b+(1-z)\r,
\qquad
\w=\b-z\r.
\end{equation}
In these variables, the eikonal amplitude reads
\begin{align}
\label{eq:eik_amp_gen_kin_qbarq_prod}
i{\mathcal M}_{\gamma^*_\lambda\to q_1\bar q_2}^{\rm Eik} & = -\frac{e\, e_f}{2q^+}\int_{\b}\, e^{-i\b\cdot(\k-\q)}\int\frac{\dd^2\K}{(2\pi)^2}\frac{1}{[\K^2+\Qb^2]}\int_{\r}e^{-i\r\cdot(\P-\K)}
\, {\bar u}(\check{p}_1,h_1)\, \Gamma^{\lambda}(\K)\, v(\check{p}_2,h_2)
\nonumber\\
& \hspace{8cm}
\times 
T\Big[ U_F\big(\b+(1-z)\r\big)\, U^\dagger_F\big(\b-z\r\big)-\bm{1}_F \Big]
\end{align}
We keep the photon transverse momentum $\q$ explicit at this stage, so that the transverse momentum transferred by the target is $\k-\q$. Later, we consider the $\q=0$ limit, in which the back-to-back correlation limit takes the form $|\k|\ll|\P|$, as defined in Sec.\ref{sec:setup}.

Expanding the Wilson lines around \(\b\), the constant term cancels the no-interaction subtraction. Using \(U_F(\b)U_F^\dagger(\b)=\bm{1}_F\) and its transverse derivative, the terms linear in \(\r\) combine to give
\begin{align}
T\Big[
U_F\big(\b+(1-z)\r\big)
U_F^\dagger\big(\b-z\r\big)-\bm{1}_F
\Big]
&=
\r^j\,T\Big[
\big(\partial_{\b^j}U_F(\b)\big)U_F^\dagger(\b)
\Big]
+\mathcal O(\r^2).
\end{align}
The transverse derivative can be expressed through a field-strength insertion. Using the same Wilson-line identity and boundary convention as in the quark-gluon calculation, we obtain
\begin{align}
\r^j\,T\Big[
\big(\partial_{\b^j}U_F(\b)\big)U_F^\dagger(\b)
\Big]
&=
\r^j\,T\Bigg\{
\int_{v^+}
U_F(+\infty,v^+;\b)
\big[-igF_j^{\; -}(v^+,\b)\big]
U_F^\dagger(+\infty,v^+;\b)
\Bigg\}.
\end{align}
The relation between fundamental and adjoint Wilson lines then gives
\begin{align}
T\Big[
U_F\big(\b+(1-z)\r\big)
U_F^\dagger\big(\b-z\r\big)-\bm{1}_F
\Big]
&=
-i\r^j\,T\Bigg\{
\int_{v^+}
t^b\,U_A(+\infty,v^+;\b)_{ba}\,
g\,{F_j^{\; -}}_a(v^+,\b)
\Bigg\}
+\mathcal O(\r^2).
\end{align}
Substituting this expansion into Eq.~\eqref{eq:eik_amp_gen_kin_qbarq_prod}, the leading contribution to the eikonal amplitude becomes
\begin{align}
i{\mathcal M}_{\gamma^*_\lambda\to q_1\bar q_2}^{\rm Eik}
&=
\frac{i\,e\,e_f}{2q^+}
\int_{\b}e^{-i\b\cdot(\k-\q)}
T\Bigg\{
\int_{v^+}
t^b\,U_A(+\infty,v^+;\b)_{ba}\,
g\,{F_j^{\; -}}_a(v^+,\b)
\Bigg\}
\nonumber\\
&\quad\times
\int\frac{\dd^2\K}{(2\pi)^2}
\frac{\bar u(\check p_1,h_1)\,
\Gamma^\lambda(\K)\,v(\check p_2,h_2)}
{\K^2+\Qb^2}
\int_{\r}\r^j e^{-i\r\cdot(\P-\K)}
+\mathcal O({\rm NLP}).
\end{align}
The transverse-separation integral produces a derivative of the momentum delta function,
\begin{equation}
\int_{\r}\r^j e^{-i\r\cdot(\P-\K)}
=
i\partial_{\P^j}
\Big[(2\pi)^2\delta^{(2)}(\P-\K)\Big].
\end{equation}
Performing the \(\K\) integration, we therefore obtain
\begin{align}
i{\mathcal M}_{\gamma^*_\lambda\to q_1\bar q_2}^{\rm Eik}
&=
\frac{i\,e\,e_f}{2q^+}
\int_{\b}e^{-i\b\cdot(\k-\q)}
T\Bigg\{
\int_{v^+}
t^b\,U_A(+\infty,v^+;\b)_{ba}\,
g\,{F_j^{\; -}}_a(v^+,\b)
\Bigg\}
\nonumber\\
&\quad\times
i\,\partial_{\P^j}
\Bigg[
\frac{\bar u(\check p_1,h_1)\,
\Gamma^\lambda(\P)\,v(\check p_2,h_2)}
{\P^2+\Qb^2}
\Bigg]
+\mathcal O({\rm NLP}).
\end{align}
Here, the derivative originates from the transverse Fourier integral and acts on the hard kernel, with the external spinors held fixed. The remainder denotes contributions beyond the leading term in the correlation-limit expansion of the eikonal amplitude.

We next consider the NEik contribution with the double quark-background insertions on the quark line. Expressed in the back-to-back variables, the amplitude reads
\begin{align}
i{\mathcal M}_{\gamma^*_\lambda\to q_1\bar q_2}^{\rm NEik, \Psi\bar\Psi \, on \, q } &=
\frac{e\, e_f}{2q^+}\int\frac{\dd^2\K}{(2\pi)^2}\, \frac{1}{[\K^2+\Qb^2]}\int_{\b}e^{-i\b\cdot(\k-\q)}\, \int_{\r}\, e^{-i\r\cdot(\P-\K)} \, \frac{g^2\, \delta^{ij}}{2p_1^+}\int_{v^+,v'^+}\theta(v'^+-v^+)
\nonumber\\
& 
\times 
\bar{u}(\check{p}_1,h_1)\, T\, \Big\{ U_F\big(+\infty,v'^+;\b+(1-z)\r\big)\, t^a\, \frac{\gamma^+\gamma^i\gamma^-}{2}\, \Psi\big(v'^+,\b+(1-z)\r\big)\, {U_A\big(v'^+,v^+;\b+(1-z)\r\big)}_{ab}\, 
\nonumber\\
& 
\times 
\bar\Psi\big(v^+, \b+(1-z)\r\big)\, \frac{\gamma^-\gamma^j}{2}\, t^b\, U_F\big(v^+,-\infty;\b+(1-z)\r\big)\, \Gamma^\lambda(\K)\, U_F^\dagger\big(\b-z\r\big)\Big\}\, v(\check{p}_2,h_2)
\end{align}
For the interference with the leading eikonal amplitude at twist-three accuracy, only the zeroth-order term in the transverse-separation expansion is needed. Setting $\r=0$ in the operator structure, the integration over $\r$ gives $(2\pi)^2\delta^{(2)}(\P-\K)$, and the amplitude becomes
\begin{align}
i{\mathcal M}_{\gamma^*_\lambda\to q_1\bar q_2}^{\rm NEik, \Psi\bar\Psi \, on \, q } &=
\frac{e\, e_f}{2q^+}\, \frac{g^2\, \delta^{ij}}{2p_1^+}\, \int\frac{\dd^2\K}{(2\pi)^2}\, \frac{1}{[\K^2+\Qb^2]}
\int_{\b}e^{-i\b\cdot(\k-\q)}\, (2\pi)^2\delta^{(2)}(\P-\K)\int_{v^+,v'^+}\theta(v'^+-v^+)
\nonumber\\
& 
\times 
 {\bar u}(\check{p}_1,h_1)\, T\,\Big\{U_F(+\infty, v'^+;\b)\, t^a\, \frac{\gamma^+\gamma^i\gamma^-}{2}\, \Psi(v'^+,\b)\, {U_A(v'^+,v^+;\b)}_{ab}\, \bar\Psi(v^+,\b)\, \frac{\gamma^-\gamma^j}{2}\, t^b
\nonumber\\
& 
\times 
U_F(v^+,-\infty; \b)\, \Gamma^\lambda(\K)\, U^\dagger_F(\b)\Big\}\, v(\check{p}_2,h_2) + \mathcal{O}(\rm NNLP)
\end{align}
At a common transverse position, the color structure simplifies using the composition property of Wilson lines and the relation between the fundamental and adjoint representations. For the longitudinal ordering $v'^+>v^+$, we obtain
\begin{align}
\label{eq:simp_color_struc_dec_on_q}
&
U_F(+\infty, v'^+;\b)\, t^a\, \Psi(v'^+,\b)\, {U_A(v'^+,v^+;\b)}_{ab}\, \bar\Psi(v^+,\b)\, t^b\, U_F(v^+, -\infty;\b)\, U^\dagger_F(\b)
\nonumber\\
& \hspace{5cm}
= 
U_F(+\infty, v'^+;\b)\, t^a\, \Psi(v'^+,\b)\,\bar\Psi(v^+,\b)\, U^\dagger_F(v'^+,v^+;\b)\, t^a\, U^\dagger_F(+\infty, v'^+;\b)
\nonumber\\
& \hspace{5cm}
= t^a\, U_F(+\infty, v'^+;\b)\, \Psi(v'^+,\b)\, \bar\Psi(v^+,\b)\, U^\dagger_F(+\infty, v^+;\b)\, t^a
\end{align} 
In the last equality, we have used the invariance of the sum over the contracted color generators under a common color rotation. Substituting Eq.~\eqref{eq:simp_color_struc_dec_on_q} into the amplitude and performing the \(\K\) integration, we find
\begin{align}
i{\mathcal M}_{\gamma^*_\lambda\to q_1\bar q_2}^{\rm NEik, \Psi\bar\Psi\, on\, q}
&
=
\frac{e\, e_f}{2q^+}\, \frac{g^2\, \delta^{ij}}{2p_1^+}\, \frac{1}{[\P^2+\Qb^2]}\, \int_{\b}e^{-i\b\cdot(\k-\q)}
\int_{v^+,v'^+}\theta(v'^+-v^+) \,  {\bar u}(\check{p}_1,h_1)\, \frac{\gamma^+\gamma^i\gamma^-}{2}\, 
\nonumber\\
& \hspace{-1cm}
\times 
T\Big\{ t^a\, U_F(+\infty, v'^+;\b)\, \Psi(v'^+,\b)\, \bar\Psi(v^+,\b)\, U^\dagger_F(+\infty, v^+;\b)\, t^a\Big\}
\, \frac{\gamma^-\gamma^j}{2}\, \Gamma^\lambda(\P)\, v(\check{p}_2,h_2)+\mathcal O({\rm NNLP}).
\end{align}
Here, the remainder denotes terms whose interference with the leading eikonal amplitude contributes at NNLP or beyond in the cross section.

We finally consider the NEik contribution with the double quark-background insertions on the antiquark line. In the back-to-back variables, the amplitude reads
\begin{align}
i{\mathcal M}_{\gamma^*_\lambda\to q_1\bar q_2}^{\rm NEik, \Psi\bar\Psi\, on\, \bar q}
& 
= 
-\frac{e\, e_f}{2q^+}\, \frac{g^2\, \delta^{ij}}{2p_2^+}\, \int\frac{\dd^2\K}{(2\pi)^2}\frac{1}{[\K^2+\Qb^2]}\, \int_{\b} e^{-i\b\cdot(\k-\q)}\, \int_{\r} e^{-i\r\cdot(\P-\K)}
\int_{w^+,w'^+}\theta(w^+-w'^+)\, 
\nonumber\\
& 
\times 
T\, {\bar u}(\check{p}_1,h_1)\, \Gamma^\lambda(\K)\, \frac{\gamma^i\gamma^-}{2}\, U_F\big(\b+(1-z)\r\big)\, U_F\big(-\infty, w'^+;\b-z\r \big)\, t^a\, \Psi(w'^+,\b-z\r)
\nonumber\\
& 
\times 
{U_A\big(w'^+,w^+;\b-z\r\big)}_{ab}\, \bar\Psi(w^+,\b-z\r)\, t^b\, U_F\big(w^+, +\infty;\b-z\r\big)
\, \frac{\gamma^-\gamma^j\gamma^+}{2}\, v(\check{p}_2,h_2)
\end{align}
As for the quark-line contribution, we evaluate the operator structure at $\r=0$. The transverse-separation integral then gives a momentum delta function, yielding
\begin{align}
\label{eq:Scat_amp_dec_on_barq_1_b2b}
i{\mathcal M}_{\gamma^*_\lambda\to q_1\bar q_2}^{\rm NEik, \Psi\bar\Psi\, on\, \bar q}
& 
= 
-\frac{e\, e_f}{2q^+}\, \frac{g^2\, \delta^{ij}}{2p_2^+}\, \int_{\b} e^{-i\b\cdot(\k-\q)}\, \int\frac{\dd^2\K}{(2\pi)^2}\, \frac{(2\pi)^2\, \delta^{(2)}(\P-\K)}{[\K^2+\Qb^2]}\, \int_{w^+, w'^+}\theta(w^+-w'^+)
\, {\bar u}(\check{p}_1,h_1)\, \Gamma^\lambda(\K)\, 
\nonumber\\
& 
\times \, 
\frac{\gamma^i\gamma^-}{2}
T \Big\{U_F(\b)\, U_F(-\infty, w'^+;\b)\, t^a
\, \Psi(w'^+,\b)\, {U_A(w'^+,w^+;\b)}_{ab}\, \bar\Psi(w^+,\b)\, 
\nonumber\\
& 
\times 
\, t^b\, U_F(w^+,+\infty; \b)\Big\}
 \frac{\gamma^-\gamma^j\gamma^+}{2}\, v(\check{p}_2,h_2)+{\mathcal O}(\rm NNLP)
\end{align}
Using the same Wilson-line identities as above, the color structure at the common transverse position $\b$ simplifies to
\begin{align}
&
U_F(\b)\, U_F(-\infty, w'^+;\b)\, t^a\, \Psi(w'^+,\b)\, {U_A(w'^+,w^+;\b)}_{ab}\, \bar\Psi(w^+,\b)\, t^b\, U_F(w^+,+\infty;\b)
\nonumber\\
& \hspace{6cm}
=
t^a\, U_F(+\infty, w'^+;\b)\, \Psi(w'^+,\b)\, \bar\Psi(w^+,\b)\, U^\dagger_F(+\infty, w^+;\b)\, t^a 
\end{align}
Substituting this identity into Eq.~\eqref{eq:Scat_amp_dec_on_barq_1_b2b}, performing the $\K$ integration, and relabeling $w^+\to v^+$ and $w'^+\to v'^+$, we obtain
\begin{align} 
\label{eq:Scat_amp_dec_on_barq_2_b2b}
i{\mathcal M}_{\gamma^*_\lambda\to q_1\bar q_2}^{\rm NEik, \Psi\bar\Psi\, on\, \bar q}
& 
= 
-\frac{e\, e_f}{2q^+}\, \frac{g^2\, \delta^{ij}}{2p_2^+}\, \int_{\b} \frac{e^{-i\b\cdot(\k-\q)}}{[\P^2+\Qb^2]}\,
\int_{v^+,v'^+}\theta(v^+-v'^+)\, {\bar u}(\check{p}_1,h_1)\, \Gamma^\lambda(\P)\, \frac{\gamma^i\gamma^-}{2}\, 
\nonumber
\\
& 
\times
T\Big\{ t^a\, U_F(+\infty, v'^+;\b)\, \Psi(v'^+,\b)\, \bar\Psi(v^+,\b)\, U^\dagger_F(+\infty, v^+;\b)\, t^a\Big\} \frac{\gamma^-\gamma^j\gamma^+}{2}\, v(\check{p}_2,h_2)+ {\mathcal O}(\rm NNLP)
\end{align}
The two quark-background amplitudes thus involve the same Wilson-line-dressed quark bilinear, with opposite longitudinal orderings and different Dirac structures. Together with the leading correlation-limit expression for the eikonal amplitude, they provide the ingredients needed to compute the quark-background contribution to the cross section at NEik and twist-three accuracy. We evaluate the corresponding interference terms in the next subsection.

\subsection{NEik quark-antiquark dijet production cross section in DIS at NLP}
\label{qqbar_X_sec_at_NLP}


We now use the amplitudes derived in the previous subsection to compute the quark-background contribution to the quark--antiquark dijet production cross section at NEik and NLP accuracy. We consider longitudinal and transverse photon polarizations separately, starting with the longitudinal case.
Throughout this subsection, we work in a frame in which the incoming photon has vanishing transverse momentum, $\q=\mathbf{0}$.
 
Before computing the cross section, it is convenient to combine the NEik contributions associated with the double quark-background insertions on the quark and antiquark lines,
\begin{align}
i{\mathcal M}_{\gamma^*_\lambda\to q_1\bar q_2}^{\rm NEik,\Psi\bar\Psi}
&=
i{\mathcal M}_{\gamma^*_\lambda\to q_1\bar q_2}^{\rm NEik,\Psi\bar\Psi\,on\,q}
+
i{\mathcal M}_{\gamma^*_\lambda\to q_1\bar q_2}^{\rm NEik,\Psi\bar\Psi\,on\,\bar q}.
\end{align}
For this purpose, we introduce the operator
\begin{align}
\label{eq:def_O_PsibarPsi}
{\mathcal O}_{\Psi\bar\Psi}(v'^+,v^+;\b)
\equiv
T\Big[
U_F(+\infty,v'^+;\b)\,
\Psi(v'^+,\b)\,
\bar\Psi(v^+,\b)\,
U_F^\dagger(+\infty,v^+;\b)
\Big],
\end{align}
which carries both color and Dirac indices. Using $p_1^+=zq^+$ and $p_2^+=(1-z)q^+$, the combined amplitude becomes
\begin{align}
\label{eq:qqbar_NEik_amp_combined}
i{\mathcal M}_{\gamma^*_\lambda\to q_1\bar q_2}^{\rm NEik,\Psi\bar\Psi}
&=
\frac{e\,e_f}{2q^+}\,
\frac{g^2\delta^{ij}}{2q^+}\,
\frac{1}{[\P^2+\Qb^2]}
\int_{\b}e^{-i\b\cdot\k}
\int_{v^+,v'^+}
\bar u(\check p_1,h_1)
\nonumber\\
&\quad\times
\Bigg\{
\frac{\theta(v'^+-v^+)}{z}\,
\frac{\gamma^+\gamma^i\gamma^-}{2}\,
t^c\,{\mathcal O}_{\Psi\bar\Psi}(v'^+,v^+;\b)\,t^c\,
\frac{\gamma^-\gamma^j}{2}\,
\Gamma^\lambda(\P)
\nonumber\\
&\qquad-
\frac{\theta(v^+-v'^+)}{1-z}\,
\Gamma^\lambda(\P)\,
\frac{\gamma^i\gamma^-}{2}\,
t^c\,{\mathcal O}_{\Psi\bar\Psi}(v'^+,v^+;\b)\,t^c\,
\frac{\gamma^-\gamma^j\gamma^+}{2}
\Bigg\}
v(\check p_2,h_2)
+\mathcal O({\rm NNLP}).
\end{align}
As in the previous subsection, the remainder denotes terms that contribute beyond NLP when interfering with the leading term of the eikonal amplitude.

Let us now consider a longitudinally polarized photon. In the static approximation, the quark-background contribution to the cross section is
\begin{align}
\label{eq:Long_X_sec_total}
&(2\pi)^5\,2z(1-z)\,
\left.
\frac{\dd\sigma^{\gamma^*_L\to q_1\bar q_2}}
{\dd^2\k\,\dd z\,\dd^2\P}
\right|_{\Psi\bar\Psi}
=
\sum_{h_1,h_2}\sum_{\rm col.}
\Big\langle
\big({\mathcal M}^{\rm Eik}_{\gamma^*_L\to q_1\bar q_2}\big)^\dagger
{\mathcal M}^{\rm NEik,\Psi\bar\Psi}_{\gamma^*_L\to q_1\bar q_2}
+
\big({\mathcal M}^{\rm NEik,\Psi\bar\Psi}_{\gamma^*_L\to q_1\bar q_2}\big)^\dagger
{\mathcal M}^{\rm Eik}_{\gamma^*_L\to q_1\bar q_2}
\Big\rangle
+\mathcal O({\rm NNLP}).
\end{align}
Here, the subscript $\Psi\bar\Psi$ specifies the quark-background contribution. We first evaluate the interference in which the quark-background insertions appear in the amplitude and the eikonal contribution appears in its complex conjugate. In the intermediate steps, we keep the photon polarization $\lambda$ fixed but arbitrary, and specialize to the longitudinal case when evaluating the Dirac structures. For the transverse cross section, we will average the corresponding expressions over the two incoming photon polarizations. Substituting the amplitudes, we obtain
\begin{align}
\label{eq:Long_interference_1_1}
&
\sum_{h_1,h_2}\sum_{\rm col.}
\Big\langle
\big({\mathcal M}^{\rm Eik}_{\gamma^*_\lambda\to q_1\bar q_2}\big)^\dagger
{\mathcal M}^{\rm NEik,\Psi\bar\Psi}_{\gamma^*_L\to q_1\bar q_2}
\Big\rangle=
-\frac{e^2\,e_f^2\,g^2\,\delta^{ij}}{(2q^+)^3}\,
\frac{1}{[\P^2+\Qb^2]}
\int_{\b\b'}e^{i\k\cdot(\b'-\b)}
\int_{w^+,v^+,v'^+}
\sum_{h_1,h_2}
\nonumber\\
&\quad\times
\Big\langle
g{F_{j'}^{\; -}}_a(w^+,\b')\,
\big[U_A^\dagger(+\infty,w^+;\b')\big]_{ab}\,
{\rm tr}_{D}\bigg[
{\rm tr}_{F}\big[
t^b t^c\,{\mathcal O}_{\Psi\bar\Psi}(v'^+,v^+;\b)\,t^c
\big]
\nonumber\\
&\qquad\times
\bigg\{
\frac{\theta(v'^+-v^+)}{z}\,
\frac{\gamma^-\gamma^j}{2}\,
\Gamma^\lambda(\P)\,
v(\check p_2,h_2)\bar v(\check p_2,h_2)
\left[
\partial_{\P^{j'}}
\frac{\gamma^0[\Gamma^\lambda(\P)]^\dagger\gamma^0}
{[\P^2+\Qb^2]}
\right]
u(\check p_1,h_1)\bar u(\check p_1,h_1)\,
\frac{\gamma^+\gamma^i\gamma^-}{2}
\nonumber\\
&\qquad\quad-
\frac{\theta(v^+-v'^+)}{1-z}\,
\frac{\gamma^-\gamma^j\gamma^+}{2}\,
v(\check p_2,h_2)\bar v(\check p_2,h_2)
\left[
\partial_{\P^{j'}}
\frac{\gamma^0[\Gamma^\lambda(\P)]^\dagger\gamma^0}
{[\P^2+\Qb^2]}
\right]
u(\check p_1,h_1)\bar u(\check p_1,h_1)\,
\Gamma^\lambda(\P)\,
\frac{\gamma^i\gamma^-}{2}
\bigg\}
\bigg]
\bigg\rangle
+\mathcal O({\rm NNLP}).
\end{align}
The Dirac structures in Eq.~\eqref{eq:Long_interference_1_1} project onto the good components of the outgoing quark and antiquark spinors. The sums over their helicities can therefore be performed using
\begin{align}
\label{eq:sum_helicity_1}
\sum_{h_1}
u_G(\check p_1,h_1)\bar u_G(\check p_1,h_1)\gamma^+
&=
(2p_1^+)\frac{\gamma^-\gamma^+}{2}
=
(2q^+)\,z\,\frac{\gamma^-}{2}\gamma^+,
\\
\label{eq:sum_helicity_2}
\sum_{h_2}
v_G(\check p_2,h_2)\bar v_G(\check p_2,h_2)\gamma^+
&=
(2p_2^+)\frac{\gamma^-\gamma^+}{2}
=
(2q^+)\,(1-z)\,\frac{\gamma^-}{2}\gamma^+.
\end{align}
Using these identities and the cyclicity of the color trace, the first interference contribution becomes
\begin{align}
\label{eq:Long_interference_1_2}
& \hspace{-1.2cm}
\sum_{h_1,h_2}\sum_{\rm col.}
\Big\langle
\big({\mathcal M}^{\rm Eik}_{\gamma^*_\lambda\to q_1\bar q_2}\big)^\dagger
{\mathcal M}^{\rm NEik,\Psi\bar\Psi}_{\gamma^*_\lambda\to q_1\bar q_2}
\Big\rangle
=
-\frac{e^2\,e_f^2\,g^2\,\delta^{ij}}{2q^+}\,
\frac{1}{[\P^2+\Qb^2]}
\int_{\b\b'}e^{i\k\cdot(\b'-\b)}
\int_{w^+,v^+,v'^+}
\nonumber\\
& \hspace{1.5cm}
\times
\bigg\langle
g{F_{j'}^{\; -}}_a(w^+,\b')\,
\big[U_A^\dagger(+\infty,w^+;\b')\big]_{ab}\,
{\rm tr}_{D}\Big[
{\rm tr}_{F}\big[
t^c t^b t^c\,{\mathcal O}_{\Psi\bar\Psi}(v'^+,v^+;\b)
\Big]
\nonumber\\
& \hspace{1.7cm}
\times
\bigg\{
(1-z)\,\theta(v'^+-v^+)\,
\frac{\gamma^-\gamma^j}{2}\,
\Gamma^\lambda(\P)\,
\frac{\gamma^-}{2}
\left[
\partial_{\P^{j'}}
\frac{\gamma^0[\Gamma^\lambda(\P)]^\dagger\gamma^0}
{[\P^2+\Qb^2]}
\right]
\frac{\gamma^-}{2}\,
\frac{\gamma^+\gamma^i\gamma^-}{2}
\nonumber\\
&  \hspace{2.1cm}
-z\,\theta(v^+-v'^+)\,
\frac{\gamma^-\gamma^j\gamma^+}{2}\,
\frac{\gamma^-}{2}
\left[
\partial_{\P^{j'}}
\frac{\gamma^0[\Gamma^\lambda(\P)]^\dagger\gamma^0}
{[\P^2+\Qb^2]}
\right]
\frac{\gamma^-}{2}\,
\Gamma^\lambda(\P)\,
\frac{\gamma^i\gamma^-}{2}
\bigg\}
\bigg]
\bigg\rangle
+\mathcal O({\rm NNLP}).
\end{align}
For a longitudinally polarized photon, the hard factor takes the form
\begin{align}
\Gamma^L(\P)=2Q\,z(1-z)\,\gamma^+.
\end{align}
Since $\Gamma^L$ is independent of $\P$, the transverse derivative acts only on the denominator, giving
\begin{align}
\label{eq:qqbar_derivative_long_hard}
\partial_{\P^{j'}}
\left[
\frac{\gamma^0[\Gamma^L(\P)]^\dagger\gamma^0}
{[\P^2+\Qb^2]}
\right]
=
2Q\,z(1-z)\,\gamma^+\,
\frac{-2\P^{j'}}{[\P^2+\Qb^2]^2}.
\end{align}
The Dirac structures associated with the quark-background insertions on the quark and antiquark lines then yield the same result,
\begin{align}
\label{eq:qqbar_long_Dirac_q}
&\delta^{ij}\,
\frac{\gamma^-\gamma^j}{2}\,
\Gamma^L(\P)\,
\frac{\gamma^-}{2}
\left[
\partial_{\P^{j'}}
\frac{\gamma^0[\Gamma^L(\P)]^\dagger\gamma^0}
{[\P^2+\Qb^2]}
\right]
\frac{\gamma^-}{2}\,
\frac{\gamma^+\gamma^i\gamma^-}{2}
=
4Q^2z^2(1-z)^2\,
\frac{-2\P^{j'}}{[\P^2+\Qb^2]^2}\,
\gamma^-,
\\
\label{eq:qqbar_long_Dirac_barq}
&\delta^{ij}\,
\frac{\gamma^-\gamma^j\gamma^+}{2}\,
\frac{\gamma^-}{2}
\left[
\partial_{\P^{j'}}
\frac{\gamma^0[\Gamma^L(\P)]^\dagger\gamma^0}
{[\P^2+\Qb^2]}
\right]
\frac{\gamma^-}{2}\,
\Gamma^L(\P)\,
\frac{\gamma^i\gamma^-}{2}
=
4Q^2z^2(1-z)^2\,
\frac{-2\P^{j'}}{[\P^2+\Qb^2]^2}\,
\gamma^-.
\end{align}
Substituting these results into Eq.~\eqref{eq:Long_interference_1_2} and using the color identity
\begin{align}
t^ct^bt^c=\bigg(C_F-\frac{C_A}{2}\bigg)t^b\, , 
\label{eq:color_identity_1}
\end{align}
one obtains
\begin{align}
\label{eq:long_interference_1_explicit_Fin}
\sum_{h_1,h_2}\sum_{\rm col.}
\Big\langle
\big({\mathcal M}^{\rm Eik}_{\gamma^*_L\to q_1\bar q_2}\big)^\dagger
{\mathcal M}^{\rm NEik,\Psi\bar\Psi}_{\gamma^*_L\to q_1\bar q_2}
\Big\rangle
&=
-\frac{e^2\,e_f^2\,g^2}{2q^+}\,
4Q^2z^2(1-z)^2\,
\frac{-2\P^j}{[\P^2+\Qb^2]^3}\,
\left(C_F-\frac{C_A}{2}\right)
\int_{\b\b'}e^{i\k\cdot(\b'-\b)}
\int_{v^+,v'^+,w^+}
\nonumber\\
&\hspace{-2cm}
\times
\Big[(1-z)\,\theta(v'^+-v^+)
-z\,\theta(v^+-v'^+)\Big]
\nonumber\\
&\hspace{-2cm}
\times
\Big\langle
g{F_j^{\; -}}_a(w^+,\b')\,
\big[U_A^\dagger(+\infty,w^+;\b')\big]_{ab}\,
{\rm tr}_{F,D}\Big[
t^b\,\gamma^-\,
{\mathcal O}_{\Psi\bar\Psi}(v'^+,v^+;\b)
\Big]
\Big\rangle
+\mathcal O({\rm NNLP}).
\end{align}

We next consider the interference in which the quark-background insertions appear in the complex conjugate amplitude. This contribution is the complex conjugate of the one computed above. Taking the Hermitian conjugate of Eq.~\eqref{eq:qqbar_NEik_amp_combined}, we obtain
\begin{align}
\label{eq:NEik_amp_long_dagger}
\big(
i{\mathcal M}_{\gamma^*_\lambda\to q_1\bar q_2}^{\rm NEik,\Psi\bar\Psi}
\big)^\dagger
&=
\frac{e\,e_f}{2q^+}\,
\frac{g^2\delta^{ij}}{2q^+}\,
\frac{1}{[\P^2+\Qb^2]}
\int_{\b'}e^{i\b'\cdot\k}
\int_{v^+,v'^+}
\bar v(\check p_2,h_2)
\nonumber\\
&\quad\times
\bigg\{
\frac{\theta(v'^+-v^+)}{z}\,
\big(\gamma^0[\Gamma^\lambda(\P)]^\dagger\gamma^0\big)\,
\frac{\gamma^j\gamma^-}{2}\,
t^c
\big(
\gamma^0[
{\mathcal O}_{\Psi\bar\Psi}(v'^+,v^+;\b')
]^\dagger\gamma^0
\big)
t^c\,
\frac{\gamma^-\gamma^i\gamma^+}{2}
\nonumber\\
&\qquad-
\frac{\theta(v^+-v'^+)}{1-z}\,
\frac{\gamma^+\gamma^j\gamma^-}{2}\,
t^c
\big(
\gamma^0[
{\mathcal O}_{\Psi\bar\Psi}(v'^+,v^+;\b')
]^\dagger\gamma^0
\big)
t^c\,
\frac{\gamma^-\gamma^i}{2}\,
\big(\gamma^0[\Gamma^\lambda(\P)]^\dagger\gamma^0\big)
\bigg\}
u(\check p_1,h_1)
+\mathcal O({\rm NNLP}).
\end{align}
Combining this expression with the eikonal amplitude, we obtain the second interference contribution at fixed photon polarization,
\begin{align}
\label{eq:Long_interference_2_1}
&
\sum_{h_1,h_2}\sum_{\rm col.}
\Big\langle
\big(
{\mathcal M}^{\rm NEik,\Psi\bar\Psi}_{\gamma^*_\lambda\to q_1\bar q_2}
\big)^\dagger
{\mathcal M}^{\rm Eik}_{\gamma^*_\lambda\to q_1\bar q_2}
\Big\rangle
=
-\frac{e^2\,e_f^2\,g^2\,\delta^{ij}}{(2q^+)^3}\,
\frac{1}{[\P^2+\Qb^2]}
\int_{\b\b'}e^{i\k\cdot(\b'-\b)}
\int_{v^+,v'^+,w^+}
\sum_{h_1,h_2}
\nonumber\\
& \hspace{2cm}
\times
\Big\langle
{\rm tr}_{D}\bigg[
{\rm tr}_{F}\Big[
t^c
\big(
\gamma^0[
{\mathcal O}_{\Psi\bar\Psi}(v'^+,v^+;\b')
]^\dagger\gamma^0
\big)
t^c t^b
\Big]
\nonumber\\
&\hspace{2cm}
\times
\bigg\{
\frac{\theta(v'^+-v^+)}{z}\,
\frac{\gamma^-\gamma^i\gamma^+}{2}\,
u(\check p_1,h_1)\bar u(\check p_1,h_1)
\left[
\partial_{\P^{j'}}
\frac{\Gamma^\lambda(\P)}{[\P^2+\Qb^2]}
\right]
v(\check p_2,h_2)\bar v(\check p_2,h_2)\,
\gamma^0[\Gamma^\lambda(\P)]^\dagger\gamma^0\,
\frac{\gamma^j\gamma^-}{2}
\nonumber\\
& \hspace{2.4cm}
-
\frac{\theta(v^+-v'^+)}{1-z}\,
\frac{\gamma^-\gamma^i}{2}\,
\gamma^0[\Gamma^\lambda(\P)]^\dagger\gamma^0\,
u(\check p_1,h_1)\bar u(\check p_1,h_1)
\left[
\partial_{\P^{j'}}
\frac{\Gamma^\lambda(\P)}{[\P^2+\Qb^2]}
\right]
v(\check p_2,h_2)\bar v(\check p_2,h_2)\,
\frac{\gamma^+\gamma^j\gamma^-}{2}
\bigg\}
\bigg]
\nonumber\\
& \hspace{2cm}
\times
\big[U_A(+\infty,w^+;\b)\big]_{ba}\,
g{F_{j'}^{\; -}}_a(w^+,\b)
\Big\rangle
+\mathcal O({\rm NNLP}).
\end{align}
Performing the helicity sums using Eqs.~\eqref{eq:sum_helicity_1} and \eqref{eq:sum_helicity_2}, and simplifying the color trace by using Eq. \eqref{eq:color_identity_1}, we obtain
\begin{align}
\label{eq:Long_interference_2_2}
&
\sum_{h_1,h_2}\sum_{\rm col.}
\Big\langle
\big(
{\mathcal M}^{\rm NEik,\Psi\bar\Psi}_{\gamma^*_\lambda\to q_1\bar q_2}
\big)^\dagger
{\mathcal M}^{\rm Eik}_{\gamma^*_\lambda\to q_1\bar q_2}
\Big\rangle
=
-\frac{e^2\,e_f^2\,g^2\,\delta^{ij}}{2q^+}\,
\left(C_F-\frac{C_A}{2}\right)
\frac{1}{[\P^2+\Qb^2]}
\int_{\b\b'}e^{i\k\cdot(\b'-\b)}
\int_{v^+,v'^+,w^+}
\nonumber\\
& \hspace{4cm}
\times
\Big\langle
{\rm tr}_{D}\Big[
{\rm tr}_{F}\Big[
t^b
\big(
\gamma^0[
{\mathcal O}_{\Psi\bar\Psi}(v'^+,v^+;\b')
]^\dagger\gamma^0
\big)
\Big]
\nonumber\\
& \hspace{4cm}
\times
\Big\{
(1-z)\,\theta(v'^+-v^+)\,
\frac{\gamma^-\gamma^i\gamma^+}{2}\,
\frac{\gamma^-}{2}
\left[
\partial_{\P^{j'}}
\frac{\Gamma^\lambda(\P)}{[\P^2+\Qb^2]}
\right]
\frac{\gamma^-}{2}\,
\gamma^0[\Gamma^\lambda(\P)]^\dagger\gamma^0\,
\frac{\gamma^j\gamma^-}{2}
\nonumber\\
&\hspace{4.4cm}
-
z\,\theta(v^+-v'^+)\,
\frac{\gamma^-\gamma^i}{2}\,
\gamma^0[\Gamma^\lambda(\P)]^\dagger\gamma^0\,
\frac{\gamma^-}{2}
\left[
\partial_{\P^{j'}}
\frac{\Gamma^\lambda(\P)}{[\P^2+\Qb^2]}
\right]
\frac{\gamma^-}{2}\,
\frac{\gamma^+\gamma^j\gamma^-}{2}
\Big\}
\Big]
\nonumber\\
& \hspace{4cm}
\times
\big[U_A(+\infty,w^+;\b)\big]_{ba}\,
g{F_{j'}^{\; -}}_a(w^+,\b)
\Big\rangle
+\mathcal O({\rm NNLP}).
\end{align}
For a longitudinally polarized photon, the two Dirac structures in Eq.~\eqref{eq:Long_interference_2_2} again give the same result. Using $\Gamma^L(\P)=2Qz(1-z)\gamma^+$, we obtain
\begin{align}
\label{eq:qqbar_long_conj_Dirac_q}
&\delta^{ij}\,
\frac{\gamma^-\gamma^i\gamma^+}{2}\,
\frac{\gamma^-}{2}
\left[
\partial_{\P^{j'}}
\frac{\Gamma^L(\P)}{[\P^2+\Qb^2]}
\right]
\frac{\gamma^-}{2}\,
\gamma^0[\Gamma^L(\P)]^\dagger\gamma^0\,
\frac{\gamma^j\gamma^-}{2}
=
4Q^2z^2(1-z)^2\,
\frac{-2\P^{j'}}{[\P^2+\Qb^2]^2}\,
\gamma^-,
\\
\label{eq:qqbar_long_conj_Dirac_barq}
&\delta^{ij}\,
\frac{\gamma^-\gamma^i}{2}\,
\gamma^0[\Gamma^L(\P)]^\dagger\gamma^0\,
\frac{\gamma^-}{2}
\left[
\partial_{\P^{j'}}
\frac{\Gamma^L(\P)}{[\P^2+\Qb^2]}
\right]
\frac{\gamma^-}{2}\,
\frac{\gamma^+\gamma^j\gamma^-}{2}
=
4Q^2z^2(1-z)^2\,
\frac{-2\P^{j'}}{[\P^2+\Qb^2]^2}\,
\gamma^-.
\end{align}
Substituting these results into Eq.~\eqref{eq:Long_interference_2_2}, the second interference contribution becomes
\begin{align}
\label{eq:long_interference_2_explicit_Fin}
\sum_{h_1,h_2}\sum_{\rm col.}
\Big\langle
\big(
{\mathcal M}^{\rm NEik,\Psi\bar\Psi}_{\gamma^*_L\to q_1\bar q_2}
\big)^\dagger
{\mathcal M}^{\rm Eik}_{\gamma^*_L\to q_1\bar q_2}
\Big\rangle
&=
-\frac{e^2\,e_f^2\,g^2}{2q^+}\,
4Q^2z^2(1-z)^2\,
\frac{-2\P^j}{[\P^2+\Qb^2]^3}\,
\left(C_F-\frac{C_A}{2}\right)
\int_{\b\b'}e^{i\k\cdot(\b'-\b)}
\int_{v^+,v'^+,w^+}
\nonumber\\
& \hspace{-3cm}
\times
\Big[(1-z)\,\theta(v'^+-v^+)
-z\,\theta(v^+-v'^+)\Big]
\nonumber\\
& \hspace{-3cm}
\times 
\Big\langle
{\rm tr}_{F,D}\Big[
t^b
\Big(
\gamma^0[
{\mathcal O}_{\Psi\bar\Psi}(v'^+,v^+;\b')
]^\dagger\gamma^0
\Big)
\gamma^-
\Big]\,
\big[U_A(+\infty,w^+;\b)\big]_{ba}\,
g{F_j^{\; -}}_a(w^+,\b)
\Big\rangle
+\mathcal O({\rm NNLP}).
\end{align}
The quark-background contribution to the longitudinal cross section follows by adding Eqs.~\eqref{eq:long_interference_1_explicit_Fin} and \eqref{eq:long_interference_2_explicit_Fin}, as prescribed in Eq.~\eqref{eq:Long_X_sec_total}. The two terms are complex conjugates and have the same longitudinal hard coefficient, with the field-strength insertion appearing on opposite sides of the interference. \\

Let us now consider a transversely polarized photon. The quark-background contribution again arises from the interference of the eikonal and NEik quark-background amplitudes. Averaging over the two transverse polarizations of the incoming photon, the cross section reads
\begin{align}
\label{eq:Trans_X_sec_total}
&(2\pi)^5\,2z(1-z)\,
\left.
\frac{\dd\sigma^{\gamma^*_T\to q_1\bar q_2}}
{\dd^2\k\,\dd z\,\dd^2\P}
\right|_{\Psi\bar\Psi}
=
\frac{1}{2}\sum_{\lambda=\pm1}
\sum_{h_1,h_2}\sum_{\rm col.}
\Big\langle
\big(
{\mathcal M}^{\rm Eik}_{\gamma^*_\lambda\to q_1\bar q_2}
\big)^\dagger
{\mathcal M}^{\rm NEik,\Psi\bar\Psi}_{\gamma^*_\lambda\to q_1\bar q_2}
+
\big(
{\mathcal M}^{\rm NEik,\Psi\bar\Psi}_{\gamma^*_\lambda\to q_1\bar q_2}
\big)^\dagger
{\mathcal M}^{\rm Eik}_{\gamma^*_\lambda\to q_1\bar q_2}
\Big\rangle
+\mathcal O({\rm NNLP}).
\end{align}
The two interference contributions follow from Eqs.~\eqref{eq:Long_interference_1_2} and \eqref{eq:Long_interference_2_2} by using the transverse photon hard factor and performing the polarization average. In terms of the longitudinal momentum fraction $z$, this hard factor is
\begin{align}
\label{eq:gamma_T_qbarq_in_z}
\Gamma_T^\lambda(\P)
=
-\varepsilon_\lambda^l
\left[
(1-2z)\P^l\gamma^+
+\P^m\gamma^+\frac{[\gamma^l,\gamma^m]}{2}
+m\gamma^+\gamma^l
\right],
\end{align}
We first consider the contribution in which the quark-background insertions appear in the amplitude. The Dirac algebra is presented in Appendix~\ref{sec:app_Dirac_qbarq_T}. After averaging over the transverse photon polarizations, the structures associated with the insertions on the quark and antiquark lines give the same result (see Eqs. \eqref{app-eq:eik_dag_Neik_dec_on_q} and \eqref{app-eq:eik_dag_Neik_dec_on_barq}),
\begin{align}
\label{eq:qqbar_trans_Dirac_first}
&
\frac{1}{2}\sum_{\lambda}\delta^{ij}\,
\frac{\gamma^-\gamma^j}{2}\,
\Gamma_T^\lambda(\P)\,
\frac{\gamma^-}{2}
\left[
\partial_{\P^{j'}}
\frac{\gamma^0[\Gamma^\lambda(\P)]^\dagger\gamma^0}
{[\P^2+\Qb^2]}
\right]
\frac{\gamma^-}{2}\,
\frac{\gamma^+\gamma^i\gamma^-}{2}
\nonumber\\
& \hspace{3cm}
=
\frac{1}{2}\sum_{\lambda}\delta^{ij}\,
\frac{\gamma^-\gamma^j\gamma^+}{2}\,
\frac{\gamma^-}{2}
\left[
\partial_{\P^{j'}}
\frac{\gamma^0[\Gamma^\lambda(\P)]^\dagger\gamma^0}
{[\P^2+\Qb^2]}
\right]
\frac{\gamma^-}{2}\,
\Gamma^\lambda(\P)\,
\frac{\gamma^i\gamma^-}{2}
\nonumber\\
& \hspace{3cm}
=
\gamma^-
\bigg\{
\frac{\P^{j'}}{[\P^2+\Qb^2]^2}
\Big[
\big(z^2+(1-z)^2\big)(\Qb^2-\P^2)-2m^2
\Big]
+
(1-2z)\,
\frac{\P^l}{[\P^2+\Qb^2]}\,
\frac{[\gamma^l,\gamma^{j'}]}{2}
\bigg\}.
\end{align}
Substituting Eq.~\eqref{eq:qqbar_trans_Dirac_first} into Eq.~\eqref{eq:Long_interference_1_2} and using $t^ct^bt^c=(C_F-C_A/2)t^b$, we obtain
\begin{align}
\label{eq:Trans_interference_1_fin}
& \hspace{-0.5cm}
\frac{1}{2}\sum_{\lambda}
\sum_{h_1,h_2}\sum_{\rm col.}
\Big\langle
\big(
{\mathcal M}^{\rm Eik}_{\gamma^*_\lambda\to q_1\bar q_2}
\big)^\dagger
{\mathcal M}^{\rm NEik,\Psi\bar\Psi}_{\gamma^*_\lambda\to q_1\bar q_2}
\Big\rangle
=
-\frac{e^2\,e_f^2\,g^2}{2q^+}\,
\left(C_F-\frac{C_A}{2}\right)
\int_{\b\b'}e^{i\k\cdot(\b'-\b)}
\int_{w^+,v^+,v'^+}
\nonumber\\
& 
\times
\big[(1-z)\theta(v'^+-v^+)-z\theta(v^+-v'^+)\big]
\nonumber\\
& 
\times 
\bigg\{
\frac{\P^j}{[\P^2+\Qb^2]^3}
\big[
\big(z^2+(1-z)^2\big)(\Qb^2-\P^2)-2m^2
\big]
\nonumber\\
& \hspace{4cm}
\times
\Big\langle
g{F_j^{\; -}}_a(w^+,\b')\,
\big[U_A^\dagger(+\infty,w^+;\b')\big]_{ab}\,
{\rm tr}_{F,D}\Big[
t^b\,\gamma^-\,
{\mathcal O}_{\Psi\bar\Psi}(v'^+,v^+;\b)
\Big]
\Big\rangle
\nonumber\\
&
+
(1-2z)\frac{\P^l}{[\P^2+\Qb^2]^2}
\Big\langle
g{F_j^{\; -}}_a(w^+,\b')\,
\big[U_A^\dagger(+\infty,w^+;\b')\big]_{ab}\,
{\rm tr}_{F,D}\Big[
t^b\,\gamma^-\,
\frac{[\gamma^l,\gamma^j]}{2}\,
{\mathcal O}_{\Psi\bar\Psi}(v'^+,v^+;\b)
\Big]
\Big\rangle
\Bigg\}
+\mathcal O({\rm NNLP}).
\end{align}
We next evaluate the conjugate interference using Eq.~\eqref{eq:Long_interference_2_2}. After averaging over the transverse photon polarizations, the coefficient of $\gamma^-$ is unchanged, while the term proportional to $\gamma^-[\gamma^l,\gamma^j]/2$ changes sign.
\begin{align}
\label{eq:qqbar_trans_Dirac_second}
&
\frac{1}{2}\sum_{\lambda}\delta^{ij}\,
\frac{\gamma^-\gamma^i\gamma^+}{2}\,
\frac{\gamma^-}{2}
\left[
\partial_{\P^{j'}}
\frac{\Gamma^\lambda(\P)}{[\P^2+\Qb^2]}
\right]
\frac{\gamma^-}{2}\,
\gamma^0[\Gamma^\lambda(\P)]^\dagger\gamma^0\,
\frac{\gamma^j\gamma^-}{2}
\nonumber\\
& \hspace{3cm}
=
\frac{1}{2}\sum_{\lambda=\pm1}\delta^{ij}\,
\frac{\gamma^-\gamma^i}{2}\,
\gamma^0[\Gamma^\lambda(\P)]^\dagger\gamma^0\,
\frac{\gamma^-}{2}
\left[
\partial_{\P^{j'}}
\frac{\Gamma^\lambda(\P)}{[\P^2+\Qb^2]}
\right]
\frac{\gamma^-}{2}\,
\frac{\gamma^+\gamma^j\gamma^-}{2}
\nonumber\\
& \hspace{3cm}
=
\gamma^-
\Bigg\{
\frac{\P^{j'}}{[\P^2+\Qb^2]^2}
\Big[
\big(z^2+(1-z)^2\big)(\Qb^2-\P^2)-2m^2
\Big]
-
(1-2z)\frac{\P^l}{[\P^2+\Qb^2]}\,
\frac{[\gamma^l,\gamma^{j'}]}{2}
\Bigg\}.
\end{align}
Substituting Eq.~\eqref{eq:qqbar_trans_Dirac_second} into Eq.~\eqref{eq:Long_interference_2_2}, we obtain the second transverse interference contribution,
\begin{align}
\label{eq:Trans_interference_2_fin}
&
\frac{1}{2}\sum_{\lambda}
\sum_{h_1,h_2}\sum_{\rm col.}
\Big\langle
\big(
{\mathcal M}^{\rm NEik,\Psi\bar\Psi}_{\gamma^*_\lambda\to q_1\bar q_2}
\big)^\dagger
{\mathcal M}^{\rm Eik}_{\gamma^*_\lambda\to q_1\bar q_2}
\Big\rangle
=
-\frac{e^2\,e_f^2\,g^2}{2q^+}\,
\Big(C_F-\frac{C_A}{2}\Big)
\int_{\b\b'}e^{i\k\cdot(\b'-\b)}
\int_{w^+,v^+,v'^+}
\nonumber\\
& 
\times
\big[(1-z)\theta(v'^+-v^+)-z\theta(v^+-v'^+)\big]
\bigg\{
\frac{\P^j}{[\P^2+\Qb^2]^3}
\Big[
\big(z^2+(1-z)^2\big)(\Qb^2-\P^2)-2m^2
\Big]
\nonumber\\
&
\times
\Big\langle
{\rm tr}_{F,D}\Big[
t^b
\Big(
\gamma^0[
{\mathcal O}_{\Psi\bar\Psi}(v'^+,v^+;\b')
]^\dagger\gamma^0
\Big)
\gamma^-
\Big]\,
\big[U_A(+\infty,w^+;\b)\big]_{ba}\,
g{F_j^{\; -}}_a(w^+,\b)
\Big\rangle
\nonumber\\
&
-
(1-2z)\frac{\P^l}{[\P^2+\Qb^2]^2}
\Big\langle
{\rm tr}_{F,D}\Big[
t^b
\Big(
\gamma^0[
{\mathcal O}_{\Psi\bar\Psi}(v'^+,v^+;\b')
]^\dagger\gamma^0
\Big)
\gamma^-\frac{[\gamma^l,\gamma^j]}{2}
\Big]\,
\big[U_A(+\infty,w^+;\b)\big]_{ba}\,
g{F_j^{\; -}}_a(w^+,\b)
\Big\rangle
\bigg\}
+\mathcal O({\rm NNLP}).
\end{align}
Adding Eqs.~\eqref{eq:Trans_interference_1_fin} and \eqref{eq:Trans_interference_2_fin} gives the quark-background contribution to the transverse cross section in Eq.~\eqref{eq:Trans_X_sec_total}. The terms proportional to $\gamma^-$ enter with the same hard coefficient, while those involving $\gamma^-[\gamma^l,\gamma^j]/2$ enter with opposite signs. These contributions therefore involve, respectively, the sum and the difference of the corresponding target-averaged operator structures.

The longitudinal and transverse cross sections can be written in a common form by introducing target-averaged operators for the two interference contributions. We define
\begin{align}
\label{eq:T_op_for_gqbarq}
\T_{j;\,g,q\bar q}^{[\Gamma]}[\phi](\k)
&\equiv
\int_{\b\b'}e^{i\k\cdot(\b'-\b)}
\int_{w^+,v^+,v'^+}
\phi(v'^+-v^+)
\nonumber\\
&\quad\times
\Big\langle
g{F_j^{\; -}}_a(w^+,\b')\,
\big[U_A^\dagger(+\infty,w^+;\b')\big]_{ab}\,
{\rm tr}_{F,D}\Big[
\Gamma\,t^b\,
{\mathcal O}_{\Psi\bar\Psi}(v'^+,v^+;\b)
\Big]
\Big\rangle,
\\
\label{eq:T_op_for_qbarqg}
\T_{j;\,q\bar q,g}^{[\Gamma]}[\phi](\k)
&\equiv
\int_{\b\b'}e^{i\k\cdot(\b'-\b)}
\int_{w^+,v^+,v'^+}
\phi(v'^+-v^+)
\nonumber\\
&\quad\times
\Bigg\langle
{\rm tr}_{F,D}\Big[
t^b
\Big(
\gamma^0[
{\mathcal O}_{\Psi\bar\Psi}(v'^+,v^+;\b')
]^\dagger\gamma^0
\Big)\Gamma
\Big]\,
\big[U_A(+\infty,w^+;\b)\big]_{ba}\,
g{F_j^{\; -}}_a(w^+,\b)
\Bigg\rangle.
\end{align}
Here, $\Gamma$ denotes the Dirac matrix specifying the projection, and $\phi$ is a longitudinal weight function. The labels $g,q\bar q$ and $q\bar q,g$ distinguish the contributions with the field-strength insertion in the complex conjugate amplitude and in the amplitude, respectively. The weight appearing in the cross section is
\begin{align}
\label{eq:qqbar_longitudinal_weight}
\phi_z(s)\equiv(1-z)\theta(s)-z\theta(-s).
\end{align}
We also introduce the hard coefficients
\begin{align}
\label{eq:qqbar_hard_L_unpol}
{\mathcal C}^{[g,q\bar q],j}_{\rm L,unpol.}(z,\P)
&\equiv
-8Q^2z^2(1-z)^2
\frac{\P^j}{[\P^2+\Qb^2]^3},
\\
\label{eq:qqbar_hard_L_hpol}
{\mathcal C}^{[g,q\bar q],i}_{\rm L,h.\,pol.}(z,\P)
&\equiv 0,
\\
\label{eq:qqbar_hard_T_unpol}
{\mathcal C}^{[g,q\bar q],j}_{\rm T,unpol.}(z,\P)
&\equiv
\frac{\P^j}{[\P^2+\Qb^2]^3}
\Big[
\big(z^2+(1-z)^2\big)(\Qb^2-\P^2)-2m^2
\Big],
\\
\label{eq:qqbar_hard_T_hpol}
{\mathcal C}^{[g,q\bar q],i}_{\rm T,h.\,pol.}(z,\P)
&\equiv
(1-2z)\frac{\P^i}{[\P^2+\Qb^2]^2}.
\end{align}
Using the same transverse Levi-Civita convention as in the quark–gluon calculation, the commutator structure can be expressed as an axial projection,
\begin{align}
\label{eq:qqbar_axial_projection}
\gamma^-\frac{[\gamma^i,\gamma^j]}{2}
=
i\epsilon^{ij}\gamma^-\gamma_5.
\end{align}
Combining the two interference contributions, the quark-background cross sections for longitudinally and transversely polarized photons take the common form
\begin{align}
\label{eq:qqbar_Xsec_compact_operators}
(2\pi)^5\,2z(1-z)\,
\left.
\frac{\dd\sigma^{\gamma^*_{T,L}\to q_1\bar q_2}}
{\dd^2\k\,\dd z\,\dd^2\P}
\right|_{\Psi\bar\Psi}
&=
-\frac{e^2\,e_f^2\,g^2}{2q^+}
\left(C_F-\frac{C_A}{2}\right)
\Bigg\{
{\mathcal C}^{[g,q\bar q],j}_{\rm T,L;\,unpol.}(z,\P)
\Big[
\T_{j;\,g,q\bar q}^{[\gamma^-]}[\phi_z](\k)
+
\T_{j;\,q\bar q,g}^{[\gamma^-]}[\phi_z](\k)
\Big]
\nonumber\\
&\qquad\qquad+
i\epsilon^{ij}\,
{\mathcal C}^{[g,q\bar q],i}_{\rm T,L;\,h.\,pol.}(z,\P)
\Big[
\T_{j;\,g,q\bar q}^{[\gamma^-\gamma_5]}[\phi_z](\k)
-
\T_{j;\,q\bar q,g}^{[\gamma^-\gamma_5]}[\phi_z](\k)
\Big]
\Bigg\}
+\mathcal O({\rm NNLP}).
\end{align}
For the transverse cross section, the hard coefficients already include the average over the two incoming photon polarizations. The longitudinal result involves only the vector projection, while the transverse result also contains an axial projection. In the next subsection, we express these target-averaged operators as quantum expectation values and examine their relation to TMD correlators.

\subsection{NEik quark-antiquark dijet production cross section in terms of TMD correlators}

We now express the quark-background contribution to quark-antiquark dijet production in terms of TMD correlators. The three-body operators appearing in this channel have a different transverse arrangement of the fields from those studied in Ref.~\cite{Rodini:2022wki}. We introduce the corresponding correlators and their parametrizations following an analogous approach. For this purpose, we first express the target-averaged operators in Eq.~\eqref{eq:qqbar_Xsec_compact_operators} as quantum expectation values and use translational invariance to rewrite them in terms of relative coordinates.\\

For the operator $\T_{j;\,g,q\bar q}^{[\Gamma]}[\phi](\k)$, we introduce
\begin{align}
\Delta\b=\b'-\b,
\qquad
\Delta v^+=v^+-w^+,
\qquad
\Delta v'^+=v'^+-w^+.
\end{align}
Using the target-state normalization specified in Eq.~\eqref{norm_states}, the integrations over the common longitudinal and transverse coordinates give
\begin{align}
\label{eq:T_gqbarq_quantum_shifted}
\T_{j;\,g,q\bar q}^{[\Gamma]}[\phi](\k)
&=
\frac{1}{2P^-_{\rm tar}}
\int_{\Delta\b}e^{i\k\cdot\Delta\b}
\int_{\Delta v^+,\Delta v'^+}
\phi(\Delta v'^+-\Delta v^+)
\nonumber\\
&\quad\times
\Big\langle P_{\rm tar},s\Big|
g{F_j^{\; -}}_a(0,\Delta\b)\,
\big[U_A^\dagger(+\infty,0;\Delta\b)\big]_{ab}
{\rm tr}_{F,D}\Big[
\Gamma\,t^b\,
{\mathcal O}_{\Psi\bar\Psi}
(\Delta v'^+,\Delta v^+;\mathbf0)
\Big]
\Big|P_{\rm tar},s\Big\rangle.
\end{align}
The same shifts applied to the conjugate interference operator yield
\begin{align}
\label{eq:T_qbarqg_quantum_shifted}
\T_{j;\,q\bar q,g}^{[\Gamma]}[\phi](\k)
&=
\frac{1}{2P^-_{\rm tar}}
\int_{\Delta\b}e^{i\k\cdot\Delta\b}
\int_{\Delta v^+,\Delta v'^+}
\phi(\Delta v'^+-\Delta v^+)
\nonumber\\
&\quad\times
\Big\langle P_{\rm tar},s\Big|
{\rm tr}_{F,D}\Big[
t^b
\Big(
\gamma^0\big[
{\mathcal O}_{\Psi\bar\Psi}
(\Delta v'^+,\Delta v^+;\Delta\b)
\big]^\dagger\gamma^0
\Big)\Gamma
\Big]
\big[U_A(+\infty,0;\mathbf0)\big]_{ba}\,
g{F_j^{\; -}}_a(0,\mathbf0)
\Big|P_{\rm tar},s\Big\rangle.
\end{align}
In both expressions, the longitudinal weight is unchanged, since it depends only on the separation between the two quark fields. The time ordering contained in ${\mathcal O}_{\Psi\bar\Psi}$, and the corresponding conjugate ordering, are kept explicit through the operator definitions.

Following the notation of Ref.~\cite{Rodini:2022wki}, we introduce a three-body correlator adapted to the operator structure appearing in quark–antiquark dijet production. In this case, the quark and antiquark fields are located at the same transverse position, while the field-strength insertion is separated from them by a transverse displacement. We define
\begin{align}
\label{eq:def_TMD_2prime1prime}
\widetilde\Phi^{[\Gamma]}_{\mu,2'1'}
(z_1^+,z_2^+,z_3^+;\b)
&\equiv
\Big\langle P_{\rm tar},s\Big|
T\Big\{
\bar\Psi(z_1^+,\b)\,
U_F^\dagger(+\infty,z_1^+;\b)\,
t^c\,\frac{\Gamma}{2}\,
U_F(+\infty,z_2^+;\b)\,
\Psi(z_2^+,\b)
\nonumber\\
&\hspace{4cm}\times
\big[U_A(+\infty,z_3^+;\mathbf0)\big]_{ca}\,
g{F_\mu^{\; -}}_a(z_3^+,\mathbf0)
\Big\}
\Big|P_{\rm tar},s\Big\rangle.
\end{align}
The label $2'1'$ distinguishes this arrangement from the correlators with a quark–gluon pair at one transverse position and the remaining quark field at the other. As above, an average over the target spin is understood.

To express the operator traces in terms of quark bilinears, we first examine the ordering of the two fermion fields. Reversing their order generates an anticommutator contribution in addition to the fermionic minus sign. The reduction to a bilinear therefore requires this contribution to vanish after the color and Dirac projections entering the cross section.

Before identifying the target matrix elements with TMD correlators, we examine the ordering of the quark fields. The projections $\Gamma=\gamma^-$ and $\gamma^-\gamma_5$ involve only the good components of the target quark field. At equal $x^-$, their canonical anticommutator generates a contact term proportional to the identity in color space. Evaluating its color projection in the gauge $A^-=0$, this term vanishes upon contraction with $t^b$, before taking coincident transverse coordinates.
Restoring the Wilson lines, the projected operator traces can consequently be written as
\begin{align}
\label{eq:qqbar_trace_to_bilinear}
&-
{\rm tr}_{F,D}\Big[
\frac{\Gamma}{2}\,t^b\,
{\mathcal O}_{\Psi\bar\Psi}(v'^+,v^+;\b)
\Big]
=
\bar\Psi(v^+,\b)\,
U_F^\dagger(+\infty,v^+;\b)\,
\frac{\Gamma}{2}\,t^b\,
U_F(+\infty,v'^+;\b)\,
\Psi(v'^+,\b),
\\
\label{eq:qqbar_conj_trace_to_bilinear}
&-
{\rm tr}_{F,D}\Big[
\frac{\Gamma}{2}\,t^b\,
\gamma^0
\big[
{\mathcal O}_{\Psi\bar\Psi}(v'^+,v^+;\b)
\big]^\dagger
\gamma^0
\Big]
=
\bar\Psi(v'^+,\b)\,
U_F^\dagger(+\infty,v'^+;\b)\,
\frac{\Gamma}{2}\,t^b\,
U_F(+\infty,v^+;\b)\,
\Psi(v^+,\b),
\end{align}
with $\Gamma$ being either $\gamma^-$ or $\gamma^-\gamma_5$. These relations hold for the projected traces entering the cross section, where the contact term vanishes. At nonzero transverse separation, the gluonic insertion and the quark bilinear are spacelike separated and can be reordered. At coincident transverse coordinates, the same reordering follows in the gauge specified above from the equal-$x^-$ commutation of $F_i^{\; -}$ with the canonical good quark fields. We can therefore place the gluonic insertion in the ordering used in the correlator definition.

%

Using Eqs.~\eqref{eq:qqbar_trace_to_bilinear} and \eqref{eq:qqbar_conj_trace_to_bilinear}, we rewrite the target-averaged operators in terms of the correlator defined in Eq.~\eqref{eq:def_TMD_2prime1prime}. For the first interference contribution, an additional transverse translation places the field-strength insertion at the origin and the quark bilinear at $-\Delta\b$. For the conjugate contribution, the field-strength insertion is already at the origin, while the quark bilinear is at $\Delta\b$. The cross sections therefore take the form

\begin{align}
\label{eq:qqbar_Xsec_coordinate_TMD}
&(2\pi)^5\,2z(1-z)\,
\left.
\frac{\dd\sigma^{\gamma^*_{T,L}\to q_1\bar q_2}}
{\dd^2\k\,\dd z\,\dd^2\P}
\right|_{\Psi\bar\Psi}
=
\frac{e^2\,e_f^2\,g^2}{2q^+P^-_{\rm tar}}\,
\left(C_F-\frac{C_A}{2}\right)
\int_{\Delta\b}e^{i\k\cdot\Delta\b}
\int_{\Delta v^+,\Delta v'^+}
\phi_z(\Delta v'^+-\Delta v^+)
\nonumber\\
&\quad\times
\Bigg\{
{\mathcal C}^{[g,q\bar q],j}_{\rm T,L;\,unpol.}(z,\P)
\Big[
\widetilde\Phi^{[\gamma^-]}_{j,2'1'}
(\Delta v^+,\Delta v'^+,0;-\Delta\b)
+
\widetilde\Phi^{[\gamma^-]}_{j,2'1'}
(\Delta v'^+,\Delta v^+,0;\Delta\b)
\Big]
\nonumber\\
&\qquad+
i\epsilon^{ij}\,
{\mathcal C}^{[g,q\bar q],i}_{\rm T,L;\,h.\,pol.}(z,\P)
\Big[
\widetilde\Phi^{[\gamma^-\gamma_5]}_{j,2'1'}
(\Delta v^+,\Delta v'^+,0;-\Delta\b)
-
\widetilde\Phi^{[\gamma^-\gamma_5]}_{j,2'1'}
(\Delta v'^+,\Delta v^+,0;\Delta\b)
\Big]
\Bigg\}
+\mathcal O({\rm NNLP}).
\end{align}
Here, $\phi_z$ is the longitudinal weight defined in Eq.~\eqref{eq:qqbar_longitudinal_weight}, and the hard coefficients are unchanged. The factor of two associated with the $\Gamma/2$ normalization of the correlator combines with the target-state normalization $1/(2P^-_{\rm tar})$ to give $1/P^-_{\rm tar}$. The transverse result includes the average over the incoming photon polarizations.

With the ordering prescription discussed above, Hermitian conjugation exchanges the longitudinal coordinates of the quark and antiquark fields. The correlator therefore satisfies
\begin{align}
\label{eq:qqbar_TMD_coordinate_conjugation}
\left[
\widetilde\Phi^{[\Gamma]}_{\mu,2'1'}
(z_1^+,z_2^+,z_3^+;\b)
\right]^*
=
\widetilde\Phi^{[\gamma^0\Gamma^\dagger\gamma^0]}_{\mu,2'1'}
(z_2^+,z_1^+,z_3^+;\b).
\end{align}
For the two projections entering the cross section,
\begin{align}
\gamma^0(\gamma^-)^\dagger\gamma^0
&=\gamma^-,
&
\gamma^0(\gamma^-\gamma_5)^\dagger\gamma^0
&=\gamma^-\gamma_5.
\end{align}
Thus, complex conjugation leaves the Dirac projection unchanged and interchanges $z_1^+$ and $z_2^+$.

We next introduce the momentum-fraction representation, using the same Fourier-phase convention as for the three-body correlators in the quark–gluon channel,

\begin{align}
\label{eq:qqbar_TMD_longitudinal_Fourier}
\widetilde\Phi^{[\Gamma]}_{\mu,2'1'}
(z_1^+,z_2^+,z_3^+;\b)
&=
(P^-_{\rm tar})^2
\int[\dd\rmx]\,
e^{-iP^-_{\rm tar}
(\rmx_1z_1^++\rmx_2z_2^++\rmx_3z_3^+)}
\Phi^{[\Gamma]}_{\mu,2'1'}
(\rmx_1,\rmx_2,\rmx_3;\b),
\end{align}
where the integration measure is defined in Eq. \eqref{eq:twist_3_fraction_measure}. 
The constraint $\rmx_1+\rmx_2+\rmx_3=0$ follows from invariance under a common longitudinal translation. Unlike the unweighted longitudinal integrals encountered in the quark–gluon calculation, the integrals here contain the step-function weight $\phi_z$. Their evaluation produces both delta-function and principal-value contributions.

We first consider the correlator with longitudinal arguments $(\Delta v^+,\Delta v'^+,0)$ and transverse separation $-\Delta\b$. Substituting Eq.~\eqref{eq:qqbar_TMD_longitudinal_Fourier} and introducing $\Delta^+=\Delta v'^+-\Delta v^+$, we obtain
\begin{align}
\label{eq:qqbar_weighted_integral_first}
&
\int_{\Delta v^+,\Delta v'^+}
\phi_z(\Delta v'^+-\Delta v^+)\,
\widetilde\Phi^{[\Gamma]}_{\mu,2'1'}
(\Delta v^+,\Delta v'^+,0;-\Delta\b)
\nonumber\\
&
=
(P^-_{\rm tar})^2
\int[\dd\rmx]\,
\Phi^{[\Gamma]}_{\mu,2'1'}
(\rmx_1,\rmx_2,\rmx_3;-\Delta\b)
\int_{\Delta v^+,\Delta^+}
\Big[(1-z)\theta(\Delta^+)-z\theta(-\Delta^+)\Big]
e^{-iP^-_{\rm tar}
[(\rmx_1+\rmx_2)\Delta v^++\rmx_2\Delta^+]}.
\end{align}
The integration over $\Delta v^+$ gives $2\pi\delta((\rmx_1+\rmx_2)P^-_{\rm tar})$. Together with the constraint in $[\dd\rmx]$, this sets $\rmx_3=0$. The remaining half-line integrals are defined with an infinitesimal convergence factor,
\begin{align}
\int_0^\infty\dd\Delta^+\,
e^{-i\rmx_2P^-_{\rm tar}\Delta^+-\epsilon\Delta^+}
&=
\frac{-i}{\rmx_2P^-_{\rm tar}-i\epsilon},
\\
\int_{-\infty}^0\dd\Delta^+\,
e^{-i\rmx_2P^-_{\rm tar}\Delta^++\epsilon\Delta^+}
&=
\frac{i}{\rmx_2P^-_{\rm tar}+i\epsilon},
\qquad 
\end{align}
for $\epsilon>0$. Taking $\epsilon\to0^+$ and using $P^-_{\rm tar}>0$, the weighted integral becomes
\begin{align}
\label{eq:qqbar_weighted_integral_poles}
\int_{\Delta v^+,\Delta v'^+}
\phi_z(\Delta v'^+-\Delta v^+)\,
\widetilde\Phi^{[\Gamma]}_{\mu,2'1'}
(\Delta v^+,\Delta v'^+,0;-\Delta\b)
&
=
2\pi\int[\dd\rmx]\,
\delta(\rmx_3)
\left[
-\frac{i(1-z)}{\rmx_2-i0}
-\frac{iz}{\rmx_2+i0}
\right]
\nonumber\\
& \hspace{1cm}
\times 
\Phi^{[\Gamma]}_{\mu,2'1'}
(\rmx_1,\rmx_2,\rmx_3;-\Delta\b).
\end{align}
Using the distributional identity
\begin{align}
\frac{1}{\rmx\pm i0}
=
{\rm P.V.}\bigg(\frac{1}{\rmx}\bigg)
\mp i\pi\delta(\rmx),
\end{align}
the longitudinal kernel can be written as
\begin{align}
-\frac{i(1-z)}{\rmx_2-i0}
-\frac{iz}{\rmx_2+i0}
=
-i\,{\rm P.V.}\bigg(\frac{1}{\rmx_2}\bigg)
+\pi(1-2z)\delta(\rmx_2).
\end{align}
Finally, setting $\rmx_1=\rmx$ and $\rmx_2=-\rmx$, we find
\begin{align}
\label{eq:qqbar_weighted_integral_first_fin}
& \hspace{-2.5cm}
\int_{\Delta v^+,\Delta v'^+}
\phi_z(\Delta v'^+-\Delta v^+)\,
\widetilde\Phi^{[\Gamma]}_{\mu,2'1'}
(\Delta v^+,\Delta v'^+,0;-\Delta\b)
\nonumber\\
&=
2\pi^2(1-2z)\,
\Phi^{[\Gamma]}_{\mu,2'1'}(0,0,0;-\Delta\b)
+
2\pi i\,\!\int_{-1}^{1}\, \dd\rmx\; 
{\rm P.V.}\bigg(\frac{1}{\rmx}\bigg)\,
\Phi^{[\Gamma]}_{\mu,2'1'}
(\rmx,-\rmx,0;-\Delta\b).
\end{align}
For the second correlator, the longitudinal arguments are interchanged. With the same change of variables, $\Delta^+=\Delta v'^+-\Delta v^+$, its Fourier phase becomes
\begin{align}
e^{-iP^-_{\rm tar}
(\rmx_1\Delta v'^++\rmx_2\Delta v^+)}
=
e^{-iP^-_{\rm tar}
[(\rmx_1+\rmx_2)\Delta v^++\rmx_1\Delta^+]}.
\end{align}
The half-line integrals therefore generate the same pole prescriptions as before, now in $\rmx_1$. Performing the longitudinal integrations gives

\begin{align}
\label{eq:qqbar_weighted_integral_second_poles}
&\hspace{-2.5cm}
\int_{\Delta v^+,\Delta v'^+}
\phi_z(\Delta v'^+-\Delta v^+)\,
\widetilde\Phi^{[\Gamma]}_{\mu,2'1'}
(\Delta v'^+,\Delta v^+,0;\Delta\b)
\nonumber\\
&=
2\pi\int[\dd\rmx]\,\delta(\rmx_3)
\left[
-\frac{i(1-z)}{\rmx_1-i0}
-\frac{iz}{\rmx_1+i0}
\right]
\Phi^{[\Gamma]}_{\mu,2'1'}
(\rmx_1,\rmx_2,\rmx_3;\Delta\b)
\nonumber\\
&=
2\pi^2(1-2z)\,
\Phi^{[\Gamma]}_{\mu,2'1'}(0,0,0;\Delta\b)
-
2\pi i\,\!\int_{-1}^{1}\, \dd\rmx\, \, 
{\rm P.V.}\bigg(\frac{1}{\rmx}\bigg)\,
\Phi^{[\Gamma]}_{\mu,2'1'}
(\rmx,-\rmx,0;\Delta\b).
\end{align}
The delta-function contribution is unchanged, while the principal-value term has the opposite sign. Combining the two results, the sum entering the vector contribution is
\begin{align}
\label{eq:qqbar_weighted_sum_coordinate}
& \hspace{-2.5cm}
\int_{\Delta v^+,\Delta v'^+}
\phi_z(\Delta v'^+-\Delta v^+)
\Big[
\widetilde\Phi^{[\Gamma]}_{j,2'1'}
(\Delta v^+,\Delta v'^+,0;-\Delta\b)
+
\widetilde\Phi^{[\Gamma]}_{j,2'1'}
(\Delta v'^+,\Delta v^+,0;\Delta\b)
\Big]
\nonumber\\
&=
2\pi^2(1-2z)
\Big[
\Phi^{[\Gamma]}_{j,2'1'}(0,0,0;-\Delta\b)
+
\Phi^{[\Gamma]}_{j,2'1'}(0,0,0;\Delta\b)
\Big]
\nonumber\\
&
-
2\pi i\,\!\int_{-1}^{1}\, \dd\rmx\, \, 
{\rm P.V.}\bigg(\frac{1}{\rmx}\bigg)
\Big[
\Phi^{[\Gamma]}_{j,2'1'}(\rmx,-\rmx,0;\Delta\b)
-
\Phi^{[\Gamma]}_{j,2'1'}(\rmx,-\rmx,0;-\Delta\b)
\Big].
\end{align}
Similarly, the difference entering the axial contribution reads
\begin{align}
\label{eq:qqbar_weighted_difference_coordinate}
& \hspace{-2.5cm}
\int_{\Delta v^+,\Delta v'^+}
\phi_z(\Delta v'^+-\Delta v^+)
\Big[
\widetilde\Phi^{[\Gamma]}_{j,2'1'}
(\Delta v^+,\Delta v'^+,0;-\Delta\b)
-
\widetilde\Phi^{[\Gamma]}_{j,2'1'}
(\Delta v'^+,\Delta v^+,0;\Delta\b)
\Big]
\nonumber\\
&=
2\pi^2(1-2z)
\Big[
\Phi^{[\Gamma]}_{j,2'1'}(0,0,0;-\Delta\b)
-
\Phi^{[\Gamma]}_{j,2'1'}(0,0,0;\Delta\b)
\Big]
\nonumber\\
&
+
2\pi i\,\!\int_{-1}^{1}\, \dd\rmx\, \, 
{\rm PV}\bigg( \frac{1}{\rmx}\bigg)
\Big[
\Phi^{[\Gamma]}_{j,2'1'}(\rmx,-\rmx,0;-\Delta\b)
+
\Phi^{[\Gamma]}_{j,2'1'}(\rmx,-\rmx,0;\Delta\b)
\Big].
\end{align}
We define the transverse-momentum-space correlator by
\begin{align}
\label{eq:qqbar_TMD_transverse_Fourier}
\Phi^{[\Gamma]}_{j,2'1'}
(\rmx_1,\rmx_2,\rmx_3;\k)
\equiv
\int\frac{\dd^2\b}{(2\pi)^2}\,
e^{i\k\cdot\b}\,
\Phi^{[\Gamma]}_{j,2'1'}
(\rmx_1,\rmx_2,\rmx_3;\b).
\end{align}
With this convention, a correlator evaluated at $-\Delta\b$ transforms into a momentum-space correlator at $-\k$. The transverse Fourier transform of Eq.~\eqref{eq:qqbar_weighted_sum_coordinate} therefore gives
\begin{align}
\label{eq:qqbar_weighted_sum_momentum}
&\hspace{-2.5cm}
\int_{\Delta\b}e^{i\k\cdot\Delta\b}
\int_{\Delta v^+,\Delta v'^+}
\phi_z(\Delta v'^+-\Delta v^+)
\Big[
\widetilde\Phi^{[\Gamma]}_{j,2'1'}
(\Delta v^+,\Delta v'^+,0;-\Delta\b)
+
\widetilde\Phi^{[\Gamma]}_{j,2'1'}
(\Delta v'^+,\Delta v^+,0;\Delta\b)
\Big]
\nonumber\\
&=
\frac{(2\pi)^4}{2}(1-2z)
\Big[
\Phi^{[\Gamma]}_{j,2'1'}(0,0,0;\k)
+
\Phi^{[\Gamma]}_{j,2'1'}(0,0,0;-\k)
\Big]
\nonumber\\
&
-
(2\pi)^3i\,\!\int_{-1}^{1}\, \dd\rmx\, \, 
{\rm P.V.}\bigg(\frac{1}{\rmx}\bigg)
\Big[
\Phi^{[\Gamma]}_{j,2'1'}(\rmx,-\rmx,0;\k)
-
\Phi^{[\Gamma]}_{j,2'1'}(\rmx,-\rmx,0;-\k)
\Big].
\end{align}
Likewise, the Fourier transform of Eq.~\eqref{eq:qqbar_weighted_difference_coordinate} reads
\begin{align}
\label{eq:qqbar_weighted_difference_momentum}
& \hspace{-2.5cm}
\int_{\Delta\b}e^{i\k\cdot\Delta\b}
\int_{\Delta v^+,\Delta v'^+}
\phi_z(\Delta v'^+-\Delta v^+)
\Big[
\widetilde\Phi^{[\Gamma]}_{j,2'1'}
(\Delta v^+,\Delta v'^+,0;-\Delta\b)
-
\widetilde\Phi^{[\Gamma]}_{j,2'1'}
(\Delta v'^+,\Delta v^+,0;\Delta\b)
\Big]
\nonumber\\
&=
\frac{(2\pi)^4}{2}(1-2z)
\Big[
\Phi^{[\Gamma]}_{j,2'1'}(0,0,0;-\k)
-
\Phi^{[\Gamma]}_{j,2'1'}(0,0,0;\k)
\Big]
\nonumber\\
&
+
(2\pi)^3i\,\!\int_{-1}^{1}\, \dd\rmx\, \, 
{\rm P.V.}\bigg(\frac{1}{\rmx}\bigg)
\Big[
\Phi^{[\Gamma]}_{j,2'1'}(\rmx,-\rmx,0;-\k)
+
\Phi^{[\Gamma]}_{j,2'1'}(\rmx,-\rmx,0;\k)
\Big].
\end{align}
The vector contribution thus involves the transverse-momentum-even combination of the zero-fraction correlator and the transverse-momentum-odd combination of its principal-value integral. The axial contribution involves the opposite combinations. Substituting these expressions into Eq.~\eqref{eq:qqbar_Xsec_coordinate_TMD}, with $\Gamma=\gamma^-$ for the sum and $\Gamma=\gamma^-\gamma_5$ for the difference, gives the cross sections in terms of momentum-space correlators. \\

To organize the result in a form parallel to the quark–gluon channel, we introduce the correlator corresponding to the opposite arrangement of the gluonic insertion and the quark pair. While $\widetilde\Phi_{\mu,2'1'}^{[\Gamma]}$ places the quark pair at $\b$ and the field-strength insertion at the transverse origin, its companion is defined by
\begin{align}
\label{eq:def_TMD_1prime2prime}
\widetilde\Phi^{[\Gamma]}_{\mu,1'2'}
(z_1^+,z_2^+,z_3^+;\b)
&\equiv
\Big\langle P_{\rm tar},s\Big|
T\Big\{
\big[U_A(+\infty,z_3^+;\b)\big]_{ca}\,
g{F_\mu^{\; -}}_a(z_3^+,\b)
\nonumber\\
& \hspace{3cm}
\times
\bar\Psi(z_1^+,\mathbf0)\,
U_F^\dagger(+\infty,z_1^+;\mathbf0)\,
t^c\,\frac{\Gamma}{2}\,
U_F(+\infty,z_2^+;\mathbf0)\,
\Psi(z_2^+,\mathbf0)
\Big\}
\Big|P_{\rm tar},s\Big\rangle.
\end{align}

In both correlators, the longitudinal coordinates $z_1^+$, $z_2^+$ and $z_3^+$ label the antiquark field, quark field and field-strength insertion, respectively. With the ordering prescription discussed above, a common transverse translation gives
\begin{align}
\label{eq:qqbar_companion_coordinate_relation}
\widetilde\Phi^{[\Gamma]}_{\mu,1'2'}
(z_1^+,z_2^+,z_3^+;\b)
=
\widetilde\Phi^{[\Gamma]}_{\mu,2'1'}
(z_1^+,z_2^+,z_3^+;-\b).
\end{align}
We use the same longitudinal Fourier convention for both correlators. Their momentum-fraction representations therefore satisfy
\begin{align}
\label{eq:qqbar_companion_fraction_relation}
\Phi^{[\Gamma]}_{\mu,1'2'}
(\rmx_1,\rmx_2,\rmx_3;\b)
=
\Phi^{[\Gamma]}_{\mu,2'1'}
(\rmx_1,\rmx_2,\rmx_3;-\b).
\end{align}
Following the same construction as in the quark–gluon channel, we introduce
\begin{align}
\label{eq:qqbar_Phi_plus_def}
\Phi_{\mu,\oplus}^{\prime[\Gamma]}
(\rmx_1,\rmx_2,\rmx_3;\b)
&\equiv
\frac{1}{2}
\Big[
\Phi_{\mu,2'1'}^{[\Gamma]}
(\rmx_1,\rmx_2,\rmx_3;\b)
+
\Phi_{\mu,1'2'}^{[\Gamma]}
(-\rmx_2,-\rmx_1,-\rmx_3;\b)
\Big],
\\
\label{eq:qqbar_Phi_minus_def}
\Phi_{\mu,\ominus}^{\prime[\Gamma]}
(\rmx_1,\rmx_2,\rmx_3;\b)
&\equiv
\frac{i}{2}
\Big[
\Phi_{\mu,2'1'}^{[\Gamma]}
(\rmx_1,\rmx_2,\rmx_3;\b)
-
\Phi_{\mu,1'2'}^{[\Gamma]}
(-\rmx_2,-\rmx_1,-\rmx_3;\b)
\Big].
\end{align}
The prime distinguishes these combinations from those introduced for quark–gluon production. The transformation of the longitudinal arguments follows from our field-label convention: conjugation exchanges the quark and antiquark coordinates, while complex conjugation of the Fourier phase reverses the signs of all three momentum fractions.
Using Eq.~\eqref{eq:qqbar_TMD_coordinate_conjugation} together with Eq.~\eqref{eq:qqbar_companion_fraction_relation}, these combinations satisfy
\begin{align}
\label{eq:qqbar_Phi_pm_conjugation}
\left[
\Phi_{\mu,\oplus}^{\prime[\Gamma]}
(\rmx_1,\rmx_2,\rmx_3;\b)
\right]^*
&=
\Phi_{\mu,\oplus}^{\prime[\gamma^0\Gamma^\dagger\gamma^0]}
(\rmx_1,\rmx_2,\rmx_3;-\b),
\\
\left[
\Phi_{\mu,\ominus}^{\prime[\Gamma]}
(\rmx_1,\rmx_2,\rmx_3;\b)
\right]^*
&=
\Phi_{\mu,\ominus}^{\prime[\gamma^0\Gamma^\dagger\gamma^0]}
(\rmx_1,\rmx_2,\rmx_3;-\b).
\end{align}
For $\Gamma=\gamma^-$ and $\gamma^-\gamma_5$, the Dirac projection is unchanged. The corresponding transverse-momentum-space combinations are therefore real. The original correlator is recovered through
\begin{align}
\label{eq:qqbar_Phi_pm_inverse}
\Phi_{\mu,2'1'}^{[\Gamma]}
(\rmx_1,\rmx_2,\rmx_3;\b)
=
\Phi_{\mu,\oplus}^{\prime[\Gamma]}
(\rmx_1,\rmx_2,\rmx_3;\b)
-i\,
\Phi_{\mu,\ominus}^{\prime[\Gamma]}
(\rmx_1,\rmx_2,\rmx_3;\b).
\end{align}
On the momentum-fraction line selected by the longitudinal integrations, $(\rmx_1,\rmx_2,\rmx_3)=(\rmx,-\rmx,0)$, the transformed arguments coincide with the original ones. The combinations then reduce to
\begin{align}
\label{eq:qqbar_Phi_plus_softline}
\Phi_{\mu,\oplus}^{\prime[\Gamma]}
(\rmx,-\rmx,0;\b)
&=
\frac{1}{2}
\Big[
\Phi_{\mu,2'1'}^{[\Gamma]}(\rmx,-\rmx,0;\b)
+
\Phi_{\mu,2'1'}^{[\Gamma]}(\rmx,-\rmx,0;-\b)
\Big],
\\
\label{eq:qqbar_Phi_minus_softline}
\Phi_{\mu,\ominus}^{\prime[\Gamma]}
(\rmx,-\rmx,0;\b)
&=
\frac{i}{2}
\Big[
\Phi_{\mu,2'1'}^{[\Gamma]}(\rmx,-\rmx,0;\b)
-
\Phi_{\mu,2'1'}^{[\Gamma]}(\rmx,-\rmx,0;-\b)
\Big].
\end{align}
On the line $(\rmx,-\rmx,0)$, the transverse-momentum-space combinations satisfy
\begin{align}
\Phi_{j,2'1'}^{[\Gamma]}(\rmx,-\rmx,0;\k)
+
\Phi_{j,2'1'}^{[\Gamma]}(\rmx,-\rmx,0;-\k)
&=
2\Phi_{j,\oplus}^{\prime[\Gamma]}(\rmx,-\rmx,0;\k),
\\
\Phi_{j,2'1'}^{[\Gamma]}(\rmx,-\rmx,0;\k)
-
\Phi_{j,2'1'}^{[\Gamma]}(\rmx,-\rmx,0;-\k)
&=
-2i\Phi_{j,\ominus}^{\prime[\Gamma]}(\rmx,-\rmx,0;\k).
\end{align}
The weighted sum in Eq.~\eqref{eq:qqbar_weighted_sum_momentum} therefore becomes
\begin{align}
\label{eq:qqbar_weighted_sum_pm}
\int_{\Delta\b}e^{i\k\cdot\Delta\b}
\int_{\Delta v^+,\Delta v'^+}
\phi_z(\Delta v'^+-\Delta v^+)
&\Big[
\widetilde\Phi^{[\Gamma]}_{j,2'1'}
(\Delta v^+,\Delta v'^+,0;-\Delta\b)
+
\widetilde\Phi^{[\Gamma]}_{j,2'1'}
(\Delta v'^+,\Delta v^+,0;\Delta\b)
\Big]
\nonumber\\
& \hspace{-0.9cm}
=
(2\pi)^4(1-2z)\,
\Phi_{j,\oplus}^{\prime[\Gamma]}(0,0,0;\k)
-
2(2\pi)^3\,\!\int_{-1}^{1}\, \dd\rmx\, \, 
{\rm P.V.}\bigg(\frac{1}{\rmx}\bigg)\,
\Phi_{j,\ominus}^{\prime[\Gamma]}(\rmx,-\rmx,0;\k).
\end{align}
Similarly, the weighted difference in Eq.~\eqref{eq:qqbar_weighted_difference_momentum} reads
\begin{align}
\label{eq:qqbar_weighted_difference_pm}
\int_{\Delta\b}e^{i\k\cdot\Delta\b}
\int_{\Delta v^+,\Delta v'^+}
\phi_z(\Delta v'^+-\Delta v^+)
&
\Big[
\widetilde\Phi^{[\Gamma]}_{j,2'1'}
(\Delta v^+,\Delta v'^+,0;-\Delta\b)
-
\widetilde\Phi^{[\Gamma]}_{j,2'1'}
(\Delta v'^+,\Delta v^+,0;\Delta\b)
\Big]
\nonumber\\
& \hspace{-0.9cm}
=
i(2\pi)^4(1-2z)\,
\Phi_{j,\ominus}^{\prime[\Gamma]}(0,0,0;\k)
+
2i(2\pi)^3\,\!\int_{-1}^{1}\, \dd\rmx\, \, 
{\rm P.V.}\bigg(\frac{1}{\rmx}\bigg)\,
\Phi_{j,\oplus}^{\prime[\Gamma]}(\rmx,-\rmx,0;\k).
\end{align}
Substituting these expressions into Eq.~\eqref{eq:qqbar_Xsec_coordinate_TMD}, the longitudinal and transverse cross sections take the common form
\begin{align}
\label{eq:qqbar_Xsec_pm_correlators}
&(2\pi)^5\,2z(1-z)\,
\frac{\dd\sigma^{\gamma^*_{T,L}\to q_1\bar q_2}}
{\dd^2\k\,\dd z\,\dd^2\P}
\bigg|_{\Psi\bar\Psi}
=
\frac{e^2\,e_f^2\,g^2}{2q^+P^-_{\rm tar}}\,
\bigg(C_F-\frac{C_A}{2}\bigg)
\nonumber\\
&
\times 
\bigg\{
\mathcal C^{[g,q\bar q],j}_{\rm T,L;\,unpol.}(z,\P)
\Bigg[
(2\pi)^4(1-2z)\,
\Phi_{j,\oplus}^{\prime[\gamma^-]}(0,0,0;\k)
-
2(2\pi)^3\,\!\int_{-1}^{1}\, \dd\rmx\, \, 
{\rm P.V.}\bigg(\frac{1}{\rmx}\bigg)\,
\Phi_{j,\ominus}^{\prime[\gamma^-]}(\rmx,-\rmx,0;\k)
\Bigg]
\nonumber\\
&-
\epsilon^{ij}\,
\mathcal C^{[g,q\bar q],i}_{\rm T,L;\,h.\,pol.}(z,\P)
\Bigg[
(2\pi)^4(1-2z)\,
\Phi_{j,\ominus}^{\prime[\gamma^-\gamma_5]}(0,0,0;\k)
+
2(2\pi)^3\,\!\int_{-1}^{1}\, \dd\rmx\, \, 
{\rm P.V.}\bigg(\frac{1}{\rmx}\bigg)\,
\Phi_{j,\oplus}^{\prime[\gamma^-\gamma_5]}
(\rmx,-\rmx,0;\k)
\Bigg]
\Bigg\}
+\mathcal O({\rm NNLP}).
\end{align}
At this point, for simplicity we consider an unpolarized target. In that case, we parametrize the $\oplus$ and $\ominus$ correlators in terms of scalar distributions. Using the same first-Bessel-moment convention as in the quark–gluon channel, we write
\begin{align}
\label{eq:qqbar_pm_vector_param}
\Phi_{j,\oplus}^{\prime[\gamma^-]}
(\rmx_1,\rmx_2,\rmx_3;\b)
&=
-i\,\b^j M^2\,
\widetilde f_{\oplus}^{\prime\perp(1)}
(\rmx_1,\rmx_2,\rmx_3;b),
\\
\Phi_{j,\ominus}^{\prime[\gamma^-]}
(\rmx_1,\rmx_2,\rmx_3;\b)
&=
-i\,\b^j M^2\,
\widetilde f_{\ominus}^{\prime\perp(1)}
(\rmx_1,\rmx_2,\rmx_3;b),
\\
\label{eq:qqbar_pm_axial_param}
\Phi_{j,\oplus}^{\prime[\gamma^-\gamma_5]}
(\rmx_1,\rmx_2,\rmx_3;\b)
&=
-i\,\epsilon^{jl}\b^l M^2\,
\widetilde g_{\oplus}^{\prime\perp(1)}
(\rmx_1,\rmx_2,\rmx_3;b),
\\
\Phi_{j,\ominus}^{\prime[\gamma^-\gamma_5]}
(\rmx_1,\rmx_2,\rmx_3;\b)
&=
-i\,\epsilon^{jl}\b^l M^2\,
\widetilde g_{\ominus}^{\prime\perp(1)}
(\rmx_1,\rmx_2,\rmx_3;b).
\end{align}
Here, $(1)$ denotes the first Bessel moment, and the prime distinguishes these distributions from those introduced in the quark–gluon channel. 
The corresponding momentum-space parametrizations define the scalar distributions $f_{\oplus}^{\prime\perp}$, $f_{\ominus}^{\prime\perp}$, $g_{\oplus}^{\prime\perp}$ and $g_{\ominus}^{\prime\perp}$, whose transverse-momentum dependence is expressed in terms of $k=|\k|$, as in the quark--gluon channel:
\begin{align}
\label{eq:qqbar_pm_momentum_param}
\Phi_{j,\oplus}^{\prime[\gamma^-]}
(\rmx_1,\rmx_2,\rmx_3;\k)
&=
\k^j\,f_{\oplus}^{\prime\perp}
(\rmx_1,\rmx_2,\rmx_3;k),
\\
\Phi_{j,\ominus}^{\prime[\gamma^-]}
(\rmx_1,\rmx_2,\rmx_3;\k)
&=
\k^j\,f_{\ominus}^{\prime\perp}
(\rmx_1,\rmx_2,\rmx_3;k),
\\
\Phi_{j,\oplus}^{\prime[\gamma^-\gamma_5]}
(\rmx_1,\rmx_2,\rmx_3;\k)
&=
\epsilon^{jl}\k^l\,g_{\oplus}^{\prime\perp}
(\rmx_1,\rmx_2,\rmx_3;k),
\\
\Phi_{j,\ominus}^{\prime[\gamma^-\gamma_5]}
(\rmx_1,\rmx_2,\rmx_3;\k)
&=
\epsilon^{jl}\k^l\,g_{\ominus}^{\prime\perp}
(\rmx_1,\rmx_2,\rmx_3;k).
\end{align}
The conjugation properties established above imply that these scalar distributions are real. Using $\epsilon^{ij}\epsilon^{jl}=-\delta^{il}$, the axial contractions entering the cross section become
\begin{align}
\label{eq:qqbar_pm_axial_contracted}
\epsilon^{ij}
\Phi_{j,\oplus}^{\prime[\gamma^-\gamma_5]}
(\rmx_1,\rmx_2,\rmx_3;\k)
&=
-\k^i\,g_{\oplus}^{\prime\perp}
(\rmx_1,\rmx_2,\rmx_3;k),
\\
\epsilon^{ij}
\Phi_{j,\ominus}^{\prime[\gamma^-\gamma_5]}
(\rmx_1,\rmx_2,\rmx_3;\k)
&=
-\k^i\,g_{\ominus}^{\prime\perp}
(\rmx_1,\rmx_2,\rmx_3;k).
\end{align}

On the line $(\rmx,-\rmx,0)$, the $\oplus$ combination is even under transverse-momentum reversal. The unpolarized-target vector and axial parametrizations are odd, so both $\oplus$ projections vanish on this line. Only the $\ominus$ distributions contribute, and Eq.~\eqref{eq:qqbar_Xsec_pm_correlators} reduces to
\begin{align}
\label{eq:qqbar_Xsec_final_scalar_pm}
& \hspace{-2cm}
(2\pi)^5\,2z(1-z)\,
\frac{\dd\sigma^{\gamma^*_{T,L}\to q_1\bar q_2}}
{\dd^2\k\,\dd z\,\dd^2\P}
\bigg|_{\Psi\bar\Psi}
=
\frac{e^2\,e_f^2\,g^2}{2q^+P^-_{\rm tar}}\,
\bigg(C_F-\frac{C_A}{2}\bigg)
\nonumber\\
& \hspace{3cm}
\times 
\bigg\{
-2(2\pi)^3\,
\mathcal C^{[g,q\bar q],j}_{\rm T,L;\,unpol.}(z,\P)\,
\k^j\, \int_{-1}^{1}\, \dd\rmx\, \, {\rm P.V.}\bigg(\frac{1}{\rmx}\bigg)\,
f_{\ominus}^{\prime\perp}
(\rmx,-\rmx,0;k)
\nonumber\\
& \hspace{3.8cm}
+
(2\pi)^4(1-2z)\,
\mathcal C^{[g,q\bar q],i}_{\rm T,L;\,h.\,pol.}(z,\P)\,
\k^i\,
g_{\ominus}^{\prime\perp}
(0,0,0;k)
\bigg\}
+\mathcal O({\rm NNLP}).
\end{align}
Both photon polarizations probe the principal-value integral of $f_{\ominus}^{\prime\perp}$ along $(\rmx,-\rmx,0)$. The transverse cross section also probes $g_{\ominus}^{\prime\perp}$ at vanishing longitudinal momentum fractions, whereas this contribution is absent for longitudinal polarization because $\mathcal C^{[g,q\bar q],i}_{\rm L;\,h.\,pol.}=0$.

\section{Summary and outlook}
\label{sec:summary}
In this work, we have studied back-to-back dijet production in DIS at NEik accuracy within the CGC framework, focusing on the quark-background contributions. We have considered both quark–gluon and quark–antiquark production and computed the corresponding cross sections through NLP in the correlation limit for longitudinally and transversely polarized photons. Our analysis connects the quark-background operators arising beyond the eikonal approximation to the three-body correlators relevant at twist-three accuracy.

For the quark–gluon channel, we start from the scattering amplitudes derived previously in general kinematics and extend the correlation-limit calculation beyond leading power. The expansion gives two distinct sources of NLP contributions. The kinematic corrections modify the hard coefficients accompanying the twist-two quark distribution, while the genuine twist-three contributions involve quark–gluon–quark operators with a field-strength insertion. We express these operators in terms of the TMD correlators discussed in Ref.~\cite{Rodini:2022wki} and provide the corresponding hard coefficients explicitly. For an unpolarized target, both photon polarizations probe $f_{\oplus}^{\perp}$, while the transverse cross section also probes $g_{\ominus}^{\perp}$. In addition, the transverse result contains a quark-mass-dependent helicity-flip contribution, whose tensor correlator is left explicit, and which can contribute at NLP accuracy in the heavy quark case, when the quark mass is considered of the same order as the hard scale. Note that our results, including full quark mass dependence, are valid both in this heavy quark regime and for small or vanishing quark mass. We only assume that the quark mass is not parametrically larger than the hard scale, given by the typical transverse momentum of the jets.  

For quark–antiquark production, we first derive the scattering amplitudes associated with two quark-background insertions on either the quark or antiquark line. At NEik accuracy, the quark-background contribution to the cross section arises from their interference with the eikonal amplitude. In the correlation limit, the leading term of the eikonal amplitude contains a field-strength insertion, whereas the quark-background amplitudes can be evaluated at zero transverse separation. Their interference therefore gives three-body operators in which the quark pair is located at one transverse position and the gluonic insertion at the other. We introduce correlators adapted to this arrangement and distinguish them from those entering the quark–gluon channel.

The longitudinal ordering of the two quark-background insertions in the quark--antiquark channel leads to a further difference between the two processes. In quark--gluon production, the unrestricted longitudinal integrations give formal zero-momentum-fraction projections of the TMD correlators. In quark--antiquark production, the ordering weights generate both delta-function and principal-value contributions. For an unpolarized target, both photon polarizations probe a principal-value integral of $f_{\ominus}^{\prime\perp}$ along $(x_1,x_2,x_3)=(x,-x,0)$, while the transverse cross section also probes $g_{\ominus}^{\prime\perp}$ at vanishing longitudinal momentum fractions. The two dijet channels therefore probe different operator arrangements and longitudinal projections of three-body quark correlations.

The results presented here extend the connection between the NEik CGC description and quark TMD correlators beyond leading power in the correlation limit. An important continuation of this work is to investigate the renormalization and evolution of the three-body operators identified in the two channels. In particular, the relation between their small-$x$ evolution and the evolution of twist-three TMD correlators deserves a dedicated study. Such an analysis should also address the regulated interpretation of the zero-fraction projections and principal-value integrals appearing in the cross sections.

\acknowledgements
TA is supported in part by the National Science Centre (Poland) under the research Grant No. 2023/50/E/ST2/00133
(SONATA BIS 13). GB is supported in part by the National Science Centre (Poland) under the research
Grant No. 2020/38/E/ST2/00122 (SONATA BIS 10). AC is supported in part by the National Science Centre (Poland) under the research Grant No. 2021/43/D/ST2/01154 (SONATA17).

\appendix

\section{Details of the Dirac algebra for quark-gluon dijet production}
\label{sec:app_1}
\subsection{Dirac algebra for the inside--inside contribution to the transverse cross section}
\label{app:inside-inside_cont}
Here, we present the Dirac algebra entering the square of the ``inside'' amplitude for quark--gluon dijet production with a transversely polarized incoming photon. The relevant structure is
\begin{align}
{\mathcal D}\big|^{\rm T}_{\rm in-in}
&\equiv
\frac{1}{2}\sum_{\lambda}\sum_{h_1,\lambda_2}
(1-z)^4\,
\gamma^0\big(\Gamma_T^{\rm in}\big)^\dagger\gamma^0\,
\frac{\gamma^-\gamma^+}{2}\,
u(1)\bar u(1)\,
\frac{\gamma^+\gamma^-}{2}\,
\Gamma_T^{\rm in}.
\end{align}
The factor of $1/2$ accounts for the average over the incoming photon polarizations. Summing over the outgoing quark helicity and using $k_1^+=zq^+$, we obtain
\begin{align}
\sum_{h_1}
\frac{\gamma^-\gamma^+}{2}\,
u(1)\bar u(1)\,
\frac{\gamma^+\gamma^-}{2}
&=
(2q^+)\,z\,\frac{\gamma^-}{2}.
\end{align}
Substituting the hard factor given in Eq.~\eqref{gamma_in_new}, the Dirac structure becomes
\begin{align}
{\mathcal D}\big|^{\rm T}_{\rm in-in}
&=
\frac{(2q^+)\,z\,(1-z)^4}{4}
\sum_{\lambda,\lambda_2}
\varepsilon_{\lambda}^{l}
\varepsilon_{\lambda}^{l'*}
\varepsilon_{\lambda_2}^{j*}
\varepsilon_{\lambda_2}^{j'}\,
\gamma^{l'}
\nonumber\\
&\quad\times
\bigg\{
\frac{\P^{k'}}{(1-z)^2}
\bigg[
-(1+z)\delta^{j'k'}
+(1-z)\frac{[\gamma^{j'},\gamma^{k'}]}{2}
\bigg]
+m\gamma^{j'}
\bigg\}
\nonumber\\
&\quad\times
\gamma^-
\bigg\{
\frac{\P^k}{(1-z)^2}
\bigg[
-(1+z)\delta^{jk}
+(1-z)\frac{[\gamma^k,\gamma^j]}{2}
\bigg]
+m\gamma^j
\bigg\}\gamma^l.
\end{align}
We work in four spacetime dimensions, so the transverse indices take two values. Performing the polarization sums and using
$\{\gamma^i,\gamma^j\}=-2\delta^{ij}$, we find
\begin{align}
{\mathcal D}\big|^{\rm T}_{\rm in-in}
&=
\frac{(2q^+)\,z}{4}\,
\delta^{ll'}\delta^{jj'}\,
\gamma^{l'}\gamma^-
\Big\{
\P^{k'}
\big[-(1-z)\gamma^{k'}\gamma^{j'}
-2\delta^{j'k'}\big]
-(1-z)^2m\gamma^{j'}
\Big\}
\nonumber\\
&\quad\times
\Big\{
\P^k
\big[-(1-z)\gamma^j\gamma^k
-2\delta^{jk}\big]
+(1-z)^2m\gamma^j
\Big\}\gamma^l.
\end{align}
Here, moving $\gamma^-$ to the left of the first brace changes the sign of its mass term. Expanding the product and contracting the gluon polarization indices gives
\begin{align}
{\mathcal D}\big|^{\rm T}_{\rm in-in}
&=
-\frac{(2q^+)\,z}{4}\,
\gamma^-\,\delta^{ll'}\gamma^{l'}
\Bigg\{
\P^{k'}\P^k
\Big[
\big(-2(1-z)^2+4(1-z)\big)
\gamma^{k'}\gamma^k
+4\delta^{k'k}
\Big]
\nonumber\\
&\qquad
+2(1-z)^4m^2
-(1-z)^2m\P^k
\big[2(1-z)\gamma^k-2\gamma^k\big]
\nonumber\\
&\qquad
+(1-z)^2m\P^k
\big[2(1-z)\gamma^k-2\gamma^k\big]
\Bigg\}\gamma^l.
\end{align}
The terms linear in $m$ cancel. Since $\P^{k'}\P^k$ is symmetric, only the anticommutator contributes to its contraction with $\gamma^{k'}\gamma^k$, yielding
\begin{align}
\P^{k'}\P^k\gamma^{k'}\gamma^k
&=
\P^{k'}\P^k
\frac{\{\gamma^{k'},\gamma^k\}}{2}
=-\P^2.
\end{align}
The remaining expression therefore reduces to
\begin{align}
\label{app:DT_in_in}
{\mathcal D}\big|^{\rm T}_{\rm in-in}
&=
-\frac{(2q^+)\,z}{4}\,
\gamma^-\,\delta^{ll'}\gamma^{l'}
\Big\{
\P^2\big[2(1-z)^2-4(1-z)+4\big]
+2(1-z)^4m^2
\Big\}\gamma^l
\nonumber\\
&=
-\frac{(2q^+)\,z}{2}\,
\Big[(1+z^2)\P^2+(1-z)^4m^2\Big]\,
\gamma^-\,\delta^{ll'}
\frac{\{\gamma^{l'},\gamma^l\}}{2}
\nonumber\\
&=
(2q^+)\,z\,
\Big[(1+z^2)\P^2+(1-z)^4m^2\Big]\gamma^-.
\end{align}
%

\subsection{Dirac algebra for the before--before contribution to the transverse cross section}
\label{app:before-before_dirac}

We next evaluate the Dirac structures entering the square of the ``before'' amplitude. Through NLP accuracy, we need the product of the LP hard factors and their interference with the NLP hard factors. The quark helicity sum has already been performed, while the factor of $1/2$ accounts for the average over the incoming photon polarizations.

We begin with the LP contribution,
\begin{align}
{\mathcal D}\big|_{\rm bef-bef}^{\rm T,\,(LP)^\dagger(LP)}
&\equiv
\frac{1}{2}\sum_{\lambda,\lambda_2}
\frac{(2q^+)\,z}{[\P^2+\Qb^2]^2}\,
\gamma^0
\big[\Gamma_T^{\rm bef}(\P)\big]^\dagger
\gamma^0\,
\frac{\gamma^-}{2}\,
\Gamma_T^{\rm bef}(\P).
\end{align}
Using the hard factor given in Eq.~\eqref{gamma_bef_new}, its Dirac conjugate can be written as
\begin{align}
\gamma^0
\big[\Gamma_T^{\rm bef}(\P)\big]^\dagger
\gamma^0
&=
\varepsilon_\lambda^{l'*}
\varepsilon_{\lambda_2}^{j'}\,
\gamma^{j'}
\Big\{
\P^{k'}
\big[-\gamma^{k'}\gamma^{l'}
-2(1-z)\delta^{k'l'}\big]
+m\gamma^{l'}
\Big\}.
\end{align}
After performing the polarization sums and moving $\gamma^-$ to the left, we obtain
\begin{align}
{\mathcal D}\big|_{\rm bef-bef}^{\rm T,\,(LP)^\dagger(LP)}
&=
-\frac{(2q^+)\,z}{4[\P^2+\Qb^2]^2}\,
\gamma^-\,\delta^{l'l}\delta^{j'j}\,
\gamma^{j'}
\Big\{
\P^{k'}
\big[-\gamma^{k'}\gamma^{l'}
-2(1-z)\delta^{k'l'}\big]
-m\gamma^{l'}
\Big\}
\nonumber\\
&\quad\times
\Big\{
\P^k
\big[-\gamma^l\gamma^k
-2(1-z)\delta^{kl}\big]
+m\gamma^l
\Big\}\gamma^j.
\end{align}
For the transverse contractions in four spacetime dimensions, we use
\begin{align}
\delta^{ij}\gamma^i\gamma^j
&=-2,
&
\delta^{ij}\gamma^i\gamma^l\gamma^j
&=0,
&
\delta^{ij}\gamma^i\gamma^l\gamma^k\gamma^j
&=-2\gamma^k\gamma^l.
\end{align}
Contracting the photon polarization indices gives
\begin{align}
{\mathcal D}\big|_{\rm bef-bef}^{\rm T,\,(LP)^\dagger(LP)}
&=
-\frac{(2q^+)\,z}{4[\P^2+\Qb^2]^2}\,
\gamma^-\,\delta^{j'j}\gamma^{j'}
\Big\{
\P^k\P^{k'}
\big[
-2\gamma^{k'}\gamma^k
+4(1-z)\gamma^{k'}\gamma^k
+4(1-z)^2\delta^{k'k}
\big]
\nonumber\\
&\qquad
-m\P^k
\big[2\gamma^k-2(1-z)\gamma^k\big]
+m\P^{k'}
\big[2\gamma^{k'}-2(1-z)\gamma^{k'}\big]
+2m^2
\Big\}\gamma^j.
\end{align}
The terms linear in $m$ cancel after relabeling the dummy indices. Using $\P^{k'}\P^k\gamma^{k'}\gamma^k=-\P^2$, the remaining contractions yield
\begin{align}
{\mathcal D}\big|_{\rm bef-bef}^{\rm T,\,(LP)^\dagger(LP)}
&=
-\frac{(2q^+)\,z}{4[\P^2+\Qb^2]^2}\,
\gamma^-\,\delta^{j'j}\gamma^{j'}
\Big\{
\P^2\big[2-4(1-z)+4(1-z)^2\big]+2m^2
\Big\}\gamma^j
\nonumber\\
&=
\frac{(2q^+)\,z}{[\P^2+\Qb^2]^2}
\Big\{
\big[z^2+(1-z)^2\big]\P^2+m^2
\Big\}\gamma^-.
\label{D_bef_LP_dag_LP}
\end{align}

We next consider the interference of the conjugate LP hard factor with the NLP hard factor,
\begin{align}
{\mathcal D}\big|_{\rm bef-bef}^{\rm T,\,(LP)^\dagger(NLP)}
&\equiv
\frac{1}{2}\sum_{\lambda,\lambda_2}
(2q^+)\,z\,
\gamma^0
\frac{\big[\Gamma_T^{\rm bef}(\P)\big]^\dagger}
{[\P^2+\Qb^2]}
\gamma^0\,
\frac{\gamma^-}{2}\,
\Gamma_T^{{\rm bef,NLP},i}.
\end{align}
Substituting the explicit hard factors gives
\begin{align}
{\mathcal D}\big|_{\rm bef-bef}^{\rm T,\,(LP)^\dagger(NLP)}
&=
\frac{(2q^+)\,z}{4[\P^2+\Qb^2]^2}\,
\delta^{j'j}\delta^{l'l}\,
\gamma^{j'}
\Big\{
\P^{k'}
\big[-\gamma^{k'}\gamma^{l'}
-2(1-z)\delta^{k'l'}\big]
+m\gamma^{l'}
\Big\}\gamma^-
\nonumber\\
&\quad\times
\bigg\{
\bigg[
\delta^{ki}
-\frac{2\P^i\P^k}{[\P^2+\Qb^2]}
\bigg]
\big[-\gamma^l\gamma^k-2(1-z)\delta^{kl}\big]
-\frac{2m\P^i}{[\P^2+\Qb^2]}\gamma^l
\bigg\}\gamma^j.
\end{align}
Contracting the photon polarization indices, we obtain
\begin{align}
{\mathcal D}\big|_{\rm bef-bef}^{\rm T,\,(LP)^\dagger(NLP)}
&=
-\frac{(2q^+)\,z}{4[\P^2+\Qb^2]^2}\,
\gamma^-\,\delta^{j'j}\gamma^{j'}
\bigg\{
\P^{k'}
\bigg[
\delta^{ki}
-\frac{2\P^i\P^k}{[\P^2+\Qb^2]}
\bigg]
\nonumber\\
&\qquad\times
\big[
-2\gamma^{k'}\gamma^k
+4(1-z)\gamma^{k'}\gamma^k
+4(1-z)^2\delta^{k'k}
\big]
-\frac{4m^2\P^i}{[\P^2+\Qb^2]}
\nonumber\\
&\qquad
-\frac{2m\P^i\P^{k'}}{[\P^2+\Qb^2]}
\big[2\gamma^{k'}-2(1-z)\gamma^{k'}\big]
\nonumber\\
&\qquad
-m\bigg[
\delta^{ki}
-\frac{2\P^i\P^k}{[\P^2+\Qb^2]}
\bigg]
\big[2\gamma^k-2(1-z)\gamma^k\big]
\bigg\}\gamma^j.
\end{align}
The terms linear in $m$ vanish upon contracting the remaining gluon polarization indices, since $\delta^{j'j}\gamma^{j'}\gamma^k\gamma^j=0$. The same contraction reverses the order of the two transverse gamma matrices in the momentum-dependent term, giving
\begin{align}
{\mathcal D}\big|_{\rm bef-bef}^{\rm T,\,(LP)^\dagger(NLP)}
&=
\frac{(2q^+)\,z}{[\P^2+\Qb^2]^2}\,
\gamma^-
\bigg\{
\P^{k'}
\bigg[
\delta^{ki}
-\frac{2\P^i\P^k}{[\P^2+\Qb^2]}
\bigg]
\nonumber\\
&\qquad\times
\bigg[
(1-2z)
\bigg(
\frac{[\gamma^k,\gamma^{k'}]}{2}
-\delta^{kk'}
\bigg)
+2(1-z)^2\delta^{kk'}
\bigg]
-\frac{2m^2\P^i}{[\P^2+\Qb^2]}
\bigg\}.
\end{align}
The contraction of the commutator with $\P^k\P^{k'}$ vanishes by antisymmetry. Using $2(1-z)^2+2z-1=z^2+(1-z)^2$, we then find
\begin{align}
{\mathcal D}\big|_{\rm bef-bef}^{\rm T,\,(LP)^\dagger(NLP)}
&=
\frac{(2q^+)\,z}{[\P^2+\Qb^2]^2}\,
\gamma^-
\bigg\{
(1-2z)\P^j\frac{[\gamma^i,\gamma^j]}{2}
\nonumber\\
&\qquad
+\big[z^2+(1-z)^2\big]
\bigg[
\P^i-\frac{2\P^i\P^2}{[\P^2+\Qb^2]}
\bigg]
-\frac{2m^2\P^i}{[\P^2+\Qb^2]}
\bigg\}
\nonumber\\
&=
\frac{(2q^+)\,z}{[\P^2+\Qb^2]^2}\,
\gamma^-
\bigg\{
\Big[
\big[z^2+(1-z)^2\big](\Qb^2-\P^2)-2m^2
\Big]
\frac{\P^i}{[\P^2+\Qb^2]}
+(1-2z)\P^j\frac{[\gamma^i,\gamma^j]}{2}
\bigg\}.
\label{D_bef_LP_dag_NLP}
\end{align}

Finally, the interference of the conjugate NLP hard factor with the LP hard factor is
\begin{align}
{\mathcal D}\big|_{\rm bef-bef}^{\rm T,\,(NLP)^\dagger(LP)}
&\equiv
\frac{1}{2}\sum_{\lambda,\lambda_2}
(2q^+)\,z\,
\gamma^0
\big[\Gamma_T^{{\rm bef,NLP},i}\big]^\dagger
\gamma^0\,
\frac{\gamma^-}{2}\,
\frac{\Gamma_T^{\rm bef}(\P)}{[\P^2+\Qb^2]}.
\end{align}
This structure follows by Dirac conjugation of the preceding result,
\begin{align}
{\mathcal D}\big|_{\rm bef-bef}^{\rm T,\,(NLP)^\dagger(LP)}
&=
\gamma^0
\left[
{\mathcal D}\big|_{\rm bef-bef}^{\rm T,\,(LP)^\dagger(NLP)}
\right]^\dagger
\gamma^0.
\end{align}
The scalar term is unchanged, while the commutator term changes sign. We therefore obtain
\begin{align}
{\mathcal D}\big|_{\rm bef-bef}^{\rm T,\,(NLP)^\dagger(LP)}
&=
\frac{(2q^+)\,z}{[\P^2+\Qb^2]^2}\,
\gamma^-
\bigg\{
\Big[
\big[z^2+(1-z)^2\big](\Qb^2-\P^2)-2m^2
\Big]
\frac{\P^i}{[\P^2+\Qb^2]}
-(1-2z)\P^j\frac{[\gamma^i,\gamma^j]}{2}
\bigg\}.
\label{D_bef_NLP_dag_LP}
\end{align}
%

\subsection{Dirac algebra for the before--inside interference contributions to the transverse cross section}
\label{sec:app_bef_in_interfrence}

We finally evaluate the Dirac structures entering the interference between the ``inside'' and ``before'' amplitudes. Since the ``inside'' amplitude has no NLP correction in the transverse-separation expansion, we consider its interference with the LP and NLP terms of the ``before'' amplitude.

We begin with the product of the conjugate ``inside'' hard factor and the LP ``before'' hard factor,
\begin{align}
{\mathcal D}\big|_{\rm in-bef}^{\rm T,\,(LP)^\dagger(LP)}
&\equiv
\frac{(1-z)^2}{[\P^2+\Qb^2]}\,
\frac{1}{2}\sum_{\lambda,\lambda_2}
\gamma^0\big(\Gamma_T^{\rm in}\big)^\dagger\gamma^0\,
\frac{\gamma^-}{2}\,
\Gamma_T^{\rm bef}(\P).
\end{align}
Substituting the explicit hard factors and performing the polarization sums gives
\begin{align}
{\mathcal D}\big|_{\rm in-bef}^{\rm T,\,(LP)^\dagger(LP)}
&=
\frac{(1-z)^2}{4[\P^2+\Qb^2]}\,
\delta^{ll'}\delta^{jj'}\,
\gamma^{l'}
\bigg\{
\frac{\P^{k'}}{(1-z)^2}
\big[(1-z)\gamma^{j'}\gamma^{k'}
-2z\delta^{j'k'}\big]
+m\gamma^{j'}
\bigg\}
\nonumber\\
&\quad\times
\gamma^-
\Big\{
\P^k
\big[-\gamma^l\gamma^k
-2(1-z)\delta^{kl}\big]
+m\gamma^l
\Big\}\gamma^j.
\end{align}
Using the transverse algebra, we rewrite
\begin{align}
(1-z)\gamma^{j'}\gamma^{k'}
-2z\delta^{j'k'}
&=
-(1-z)\gamma^{k'}\gamma^{j'}
-2\delta^{j'k'}.
\end{align}
Moving $\gamma^-$ to the left then yields
\begin{align}
{\mathcal D}\big|_{\rm in-bef}^{\rm T,\,(LP)^\dagger(LP)}
&=
-\frac{(1-z)^2}{4[\P^2+\Qb^2]}\,
\gamma^-\,\delta^{ll'}\delta^{jj'}\,
\gamma^{l'}
\bigg\{
\frac{\P^{k'}}{(1-z)^2}
\big[-(1-z)\gamma^{k'}\gamma^{j'}
-2\delta^{j'k'}\big]
-m\gamma^{j'}
\bigg\}
\nonumber\\
&\quad\times
\Big\{
\P^k
\big[-\gamma^l\gamma^k
-2(1-z)\delta^{kl}\big]
+m\gamma^l
\Big\}\gamma^j.
\end{align}
Applying the transverse gamma-matrix identities used in the previous subsection, the term proportional to $m^2$ vanishes. The remaining contractions give
\begin{align}
{\mathcal D}\big|_{\rm in-bef}^{\rm T,\,(LP)^\dagger(LP)}
&=
-\frac{(1-z)^2}{4[\P^2+\Qb^2]}\,
\gamma^-
\bigg\{
4m\P^j\gamma^j
\bigg[\frac{1}{(1-z)^2}-(1-z)\bigg]
+\frac{4z^2\P^2}{(1-z)^2}
\bigg\}
\nonumber\\
&=
\frac{\gamma^-}{[\P^2+\Qb^2]}
\Big\{
-z^2\P^2
+m\big[(1-z)^3-1\big]\P^j\gamma^j
\Big\}.
\end{align}

We next consider the interference with the NLP ``before'' hard factor,
\begin{align}
{\mathcal D}\big|_{\rm in-bef}^{\rm T,\,(LP)^\dagger(NLP)}
&\equiv
(1-z)^2\,
\frac{1}{2}\sum_{\lambda,\lambda_2}
\gamma^0\big(\Gamma_T^{\rm in}\big)^\dagger\gamma^0\,
\frac{\gamma^-}{2}\,
\Gamma_T^{{\rm bef,NLP},i}.
\end{align}
The NLP hard factor contains the derivative of the denominator and the derivative of the numerator,
\begin{align}
\Gamma_T^{{\rm bef,NLP},i}
&=
-\frac{2\P^i}{[\P^2+\Qb^2]^2}\,
\Gamma_T^{\rm bef}(\P)
+\frac{1}{[\P^2+\Qb^2]}\,
\partial_{\P^i}\Gamma_T^{\rm bef}(\P).
\end{align}
The first term is proportional to the LP structure computed above. Substituting the explicit numerator derivative in the second term, we obtain
\begin{align}
{\mathcal D}\big|_{\rm in-bef}^{\rm T,\,(LP)^\dagger(NLP)}
&=
-\frac{2\P^i}{[\P^2+\Qb^2]}\,
{\mathcal D}\big|_{\rm in-bef}^{\rm T,\,(LP)^\dagger(LP)}
\nonumber\\
&\quad+
\frac{(1-z)^2}{4[\P^2+\Qb^2]}\,
\delta^{ll'}\delta^{jj'}\,
\gamma^{l'}
\bigg\{
\frac{\P^{k'}}{(1-z)^2}
\big[(1-z)\gamma^{j'}\gamma^{k'}
-2z\delta^{j'k'}\big]
+m\gamma^{j'}
\bigg\}
\nonumber\\
&\qquad\times
\gamma^-
\big[-\gamma^l\gamma^i
-2(1-z)\delta^{li}\big]\gamma^j.
\end{align}
Contracting the transverse indices in the second term gives
\begin{align}
{\mathcal D}\big|_{\rm in-bef}^{\rm T,\,(LP)^\dagger(NLP)}
&=
-\frac{2\P^i}{[\P^2+\Qb^2]}\,
{\mathcal D}\big|_{\rm in-bef}^{\rm T,\,(LP)^\dagger(LP)}
\nonumber\\
&\quad+
\frac{\gamma^-}{[\P^2+\Qb^2]}
\bigg\{
-z^2\P^i
+z^2\P^j\frac{[\gamma^i,\gamma^j]}{2}
+m(1-z)^3\gamma^i
\bigg\}.
\end{align}
Combining the two contributions, we find
\begin{align}
{\mathcal D}\big|_{\rm in-bef}^{\rm T,\,(LP)^\dagger(NLP)}
&=
\frac{\gamma^-}{[\P^2+\Qb^2]}
\bigg\{
-z^2\frac{\P^i(\Qb^2-\P^2)}{[\P^2+\Qb^2]}
+z^2\P^j\frac{[\gamma^i,\gamma^j]}{2}
\nonumber\\
&\qquad
+2m\big[1-(1-z)^3\big]
\frac{\P^i\P^j}{[\P^2+\Qb^2]}\gamma^j
+m(1-z)^3\gamma^i
\bigg\}.
\end{align}
In this calculation, the derivative acts only on the ``before'' hard factor. The momentum dependence of $\Gamma_T^{\rm in}$ is not differentiated.

The remaining interference structures follow by Dirac conjugation. For the LP contribution, we define
\begin{align}
{\mathcal D}\big|_{\rm bef-in}^{\rm T,\,(LP)^\dagger(LP)}
&\equiv
\frac{(1-z)^2}{[\P^2+\Qb^2]}\,
\frac{1}{2}\sum_{\lambda,\lambda_2}
\gamma^0
\big[\Gamma_T^{\rm bef}(\P)\big]^\dagger
\gamma^0\,
\frac{\gamma^-}{2}\,
\Gamma_T^{\rm in}.
\end{align}
The scalar term is unchanged under Dirac conjugation, whereas the structure $\gamma^-\gamma^j$ changes sign. Consequently,
\begin{align}
{\mathcal D}\big|_{\rm bef-in}^{\rm T,\,(LP)^\dagger(LP)}
&=
\frac{\gamma^-}{[\P^2+\Qb^2]}
\Big\{
-z^2\P^2
-m\big[(1-z)^3-1\big]\P^j\gamma^j
\Big\}.
\end{align}
Similarly, the NLP interference structure is
\begin{align}
{\mathcal D}\big|_{\rm bef-in}^{\rm T,\,(NLP)^\dagger(LP)}
&\equiv
(1-z)^2\,
\frac{1}{2}\sum_{\lambda,\lambda_2}
\gamma^0
\big[\Gamma_T^{{\rm bef,NLP},i}\big]^\dagger
\gamma^0\,
\frac{\gamma^-}{2}\,
\Gamma_T^{\rm in}.
\end{align}
Both the commutator structure and the terms proportional to $\gamma^-\gamma^j$ change sign, giving
\begin{align}
{\mathcal D}\big|_{\rm bef-in}^{\rm T,\,(NLP)^\dagger(LP)}
&=
\frac{\gamma^-}{[\P^2+\Qb^2]}
\bigg\{
-z^2\frac{\P^i(\Qb^2-\P^2)}{[\P^2+\Qb^2]}
-z^2\P^j\frac{[\gamma^i,\gamma^j]}{2}
\nonumber\\
&\qquad
+2m\big[(1-z)^3-1\big]
\frac{\P^i\P^j}{[\P^2+\Qb^2]}\gamma^j
-m(1-z)^3\gamma^i
\bigg\}.
\end{align}
%

\section{Details of the Dirac algebra for quark-antiquark dijet production}
\label{sec:app_2}

In this appendix, we present the details of the Dirac algebra used in the quark-antiquark dijet calculation. We first derive the photon-splitting hard factors for longitudinal and transverse photon polarizations. We then evaluate the Dirac structures entering the transverse-photon interference between the eikonal amplitude and the quark-background amplitudes, with the background insertions on either the quark or the antiquark line.

\subsection{Computation of the photon-splitting hard factor}
\label{subsec:appphoton-splitting_hard_factor}

In this subsection, we evaluate the photon-splitting Dirac structure for longitudinal and transverse photon polarizations. With the normalization used in the main text, the contracted photon vertex is $2q^+\Gamma^\lambda(\K)$. Before specifying the polarization, it reads
\begin{align}
2q^+\Gamma^\lambda(\K)
&=
\epsilon^\lambda_\mu(q)\gamma^+
\Big[
p_1^+\gamma^-
-\gamma^m\Big(\K^m+\frac{p_1^+}{q^+}\q^m\Big)
+m
\Big]\gamma^\mu
\nonumber\\
&\quad\times
\Big[
p_2^+\gamma^-
-\gamma^l\Big(-\K^l+\frac{p_2^+}{q^+}\q^l\Big)
-m
\Big]\gamma^+,
\end{align}
where $p_1^++p_2^+=q^+$. Anticommuting the transverse gamma matrices with $\gamma^+$, we obtain
\begin{align}
2q^+\Gamma^\lambda(\K)
&=
\epsilon^\lambda_\mu(q)
\bigg[
p_1^+\gamma^+\gamma^-
+\bigg(
\gamma^m\Big(\K^m+\frac{p_1^+}{q^+}\q^m\Big)+m
\bigg)\gamma^+
\bigg]\gamma^\mu
\nonumber\\
&\quad\times
\bigg[
p_2^+\gamma^-\gamma^+
+\gamma^+\bigg(
\gamma^l\Big(-\K^l+\frac{p_2^+}{q^+}\q^l\Big)-m
\bigg)
\bigg].
\end{align}
Expanding the product and using $(\gamma^+)^2=(\gamma^-)^2=0$ gives
\begin{align}
2q^+\Gamma^\lambda(\K)
&=
\epsilon^\lambda_\mu(q)
\bigg\{
p_1^+p_2^+\,
\gamma^+\gamma^-\{\gamma^\mu,\gamma^-\}\gamma^+
\nonumber\\
&\quad+
p_2^+
\bigg(
\gamma^m\Big(\K^m+\frac{p_1^+}{q^+}\q^m\Big)+m
\bigg)
\gamma^+\gamma^\mu\gamma^-\gamma^+
\nonumber\\
&\quad+
p_1^+\gamma^+\gamma^-\gamma^\mu\gamma^+
\bigg(
\gamma^l\Big(-\K^l+\frac{p_2^+}{q^+}\q^l\Big)-m
\bigg)
\nonumber\\
&\quad+
\bigg(
\gamma^m\Big(\K^m+\frac{p_1^+}{q^+}\q^m\Big)+m
\bigg)
\gamma^+\{\gamma^\mu,\gamma^+\}
\bigg(
\gamma^l\Big(-\K^l+\frac{p_2^+}{q^+}\q^l\Big)-m
\bigg)
\bigg\}.
\end{align}
Applying the Clifford algebra, this expression becomes
\begin{align}
\label{eq:app_gen_photon_vertex}
2q^+\Gamma^\lambda(\K)
&=
\epsilon^\lambda_\mu(q)
\bigg\{
(2p_1^+)(2p_2^+)g^{\mu-}\gamma^+
\nonumber\\
&\quad+
2g^{\mu+}\gamma^+
\bigg(
-\gamma^m\Big(\K^m+\frac{p_1^+}{q^+}\q^m\Big)+m
\bigg)
\bigg(
\gamma^l\Big(-\K^l+\frac{p_2^+}{q^+}\q^l\Big)-m
\bigg)
\nonumber\\
&\quad+
(2p_2^+)g^\mu_{\;l}
\bigg(
\gamma^m\Big(\K^m+\frac{p_1^+}{q^+}\q^m\Big)+m
\bigg)\gamma^+\gamma^l
\nonumber\\
&\quad+
(2p_1^+)g^\mu_{\;l}\gamma^l\gamma^+
\bigg(
\gamma^m\Big(-\K^m+\frac{p_2^+}{q^+}\q^m\Big)-m
\bigg)
\bigg\}.
\end{align}

For a longitudinally polarized photon, we use the effective polarization vector obtained after applying current conservation, as defined in Eq.~\eqref{eq:def_epsilon_L},
\begin{align}
\epsilon^L_\mu(q)
&=
\frac{Q}{q^+}g_\mu^{\;+}.
\end{align}
Only the first term in Eq.~\eqref{eq:app_gen_photon_vertex} contributes, yielding
\begin{align}
2q^+\Gamma_L(\K)
&=
\frac{Q}{q^+}(2p_1^+)(2p_2^+)\gamma^+.
\end{align}
Dividing by $2q^+$, we recover the longitudinal hard factor,
\begin{align}
\Gamma_L(\K)
&=
2Q\,\frac{p_1^+p_2^+}{(q^+)^2}\gamma^+
=
2Qz(1-z)\gamma^+.
\end{align}

For a transversely polarized photon, we use the polarization vector defined in Eq.~\eqref{eq:def_epsilon_T}. Substituting it into Eq.~\eqref{eq:app_gen_photon_vertex}, we obtain
\begin{align}
2q^+\Gamma_T^\lambda(\K)
&=
-\varepsilon_\lambda^l
\bigg\{
-\frac{(2p_1^+)(2p_2^+)}{q^+}\q^l\gamma^+
\nonumber\\
&\quad-
(2p_2^+)\gamma^+
\bigg(
\gamma^m\Big(\K^m+\frac{p_1^+}{q^+}\q^m\Big)-m
\bigg)\gamma^l
\nonumber\\
&\quad-
(2p_1^+)\gamma^+\gamma^l
\bigg(
\gamma^m\Big(-\K^m+\frac{p_2^+}{q^+}\q^m\Big)-m
\bigg)
\bigg\}.
\end{align}
Separating the terms proportional to $\q$, $\K$ and $m$, and decomposing the transverse gamma-matrix products into commutators and anticommutators, gives
\begin{align}
2q^+\Gamma_T^\lambda(\K)
&=
-\varepsilon_\lambda^l
\bigg\{
-\frac{(2p_1^+)(2p_2^+)}{q^+}\q^l\gamma^+
+(2q^+)m\gamma^+\gamma^l
\nonumber\\
&\quad-
\frac{(2p_1^+)(2p_2^+)}{q^+}\q^m\gamma^+
\frac{\{\gamma^m,\gamma^l\}}{2}
\nonumber\\
&\quad-
(2p_2^+)\K^m\gamma^+
\bigg(
\frac{\{\gamma^m,\gamma^l\}}{2}
-\frac{[\gamma^l,\gamma^m]}{2}
\bigg)
\nonumber\\
&\quad+
(2p_1^+)\K^m\gamma^+
\bigg(
\frac{\{\gamma^m,\gamma^l\}}{2}
+\frac{[\gamma^l,\gamma^m]}{2}
\bigg)
\bigg\}.
\end{align}
Using $\{\gamma^m,\gamma^l\}=-2\delta^{ml}$, the two terms proportional to $\q$ cancel. The remaining expression reads
\begin{align}
2q^+\Gamma_T^\lambda(\K)
&=
-\varepsilon_\lambda^l
\Big\{
2(p_2^+-p_1^+)\K^l\gamma^+
+(2q^+)\K^m\gamma^+
\frac{[\gamma^l,\gamma^m]}{2}
+(2q^+)m\gamma^+\gamma^l
\Big\}.
\end{align}
The normalized transverse hard factor is therefore
\begin{align}
\Gamma_T^\lambda(\K)
&=
-\varepsilon_\lambda^l
\bigg[
\frac{p_2^+-p_1^+}{q^+}\K^l\gamma^+
+\K^m\gamma^+\frac{[\gamma^l,\gamma^m]}{2}
+m\gamma^+\gamma^l
\bigg]
\nonumber\\
&=
-\varepsilon_\lambda^l
\bigg[
(1-2z)\K^l\gamma^+
+\K^m\gamma^+\frac{[\gamma^l,\gamma^m]}{2}
+m\gamma^+\gamma^l
\bigg].
\end{align}

\subsection{Dirac algebra for quark-antiquark dijet production with a transversely polarized photon}
\label{sec:app_Dirac_qbarq_T}
In this subsection, we evaluate the Dirac structures entering the interference of the eikonal amplitude with the quark-background amplitudes for a transversely polarized photon. We use the normalized photon-splitting hard factor derived in the preceding subsection. Its transverse-momentum derivative is
\begin{align}
\partial_{\P^{j'}}\Gamma_T^\lambda(\P)
&=
-\varepsilon_\lambda^l
\bigg[
(1-2z)\delta^{lj'}\gamma^+
+\gamma^+\frac{[\gamma^l,\gamma^{j'}]}{2}
\bigg].
\end{align}
The Dirac-conjugate hard factor reads
\begin{align}
\gamma^0
\big[\Gamma_T^\lambda(\P)\big]^\dagger
\gamma^0
&=
-\varepsilon_\lambda^{l'*}
\bigg[
(1-2z)\P^{l'}\gamma^+
+\P^{m'}\gamma^+
\frac{[\gamma^{m'},\gamma^{l'}]}{2}
-m\gamma^+\gamma^{l'}
\bigg].
\end{align}
Consequently, the derivative entering the conjugate eikonal amplitude is
\begin{align}
\partial_{\P^{j'}}
\bigg(
\gamma^0
\frac{\big[\Gamma_T^\lambda(\P)\big]^\dagger}
{[\P^2+\Qb^2]}
\gamma^0
\bigg)
&=
\varepsilon_\lambda^{l'*}
\bigg\{
\bigg[
\frac{2\P^{j'}\P^{m'}}{[\P^2+\Qb^2]^2}
-\frac{\delta^{j'm'}}{[\P^2+\Qb^2]}
\bigg]
\nonumber\\
&\quad\times
\bigg[
\gamma^+\frac{[\gamma^{m'},\gamma^{l'}]}{2}
+(1-2z)\delta^{l'm'}\gamma^+
\bigg]
-\frac{2m\P^{j'}}{[\P^2+\Qb^2]^2}
\gamma^+\gamma^{l'}
\bigg\}.
\end{align}

We first consider the interference in which the quark-background insertions lie on the quark line in the amplitude, while the complex-conjugate amplitude is eikonal. The corresponding Dirac structure is
\begin{align}
{\mathcal D}_{\rm dec.\,on\,q}^{\rm (Eik)^\dagger(NEik)}
&\equiv
\frac{1}{2}\sum_\lambda\delta^{ij}\,
\frac{\gamma^-\gamma^j}{2}\,
\Gamma_T^\lambda(\P)\,
\frac{\gamma^-}{2}
\bigg[
\partial_{\P^{j'}}
\bigg(
\gamma^0
\frac{\big[\Gamma_T^\lambda(\P)\big]^\dagger}
{[\P^2+\Qb^2]}
\gamma^0
\bigg)
\bigg]
\frac{\gamma^-}{2}\,
\frac{\gamma^+\gamma^i\gamma^-}{2}.
\end{align}
The factor of $1/2$ accounts for the average over the incoming photon polarizations. Substituting the hard factors and performing this polarization sum, we obtain
\begin{align}
{\mathcal D}_{\rm dec.\,on\,q}^{\rm (Eik)^\dagger(NEik)}
&=
-\frac{\delta^{ij}\delta^{ll'}}{2}\,
\frac{\gamma^-\gamma^j}{2}
\bigg[
(1-2z)\P^l\gamma^+
+\P^m\gamma^+\frac{[\gamma^l,\gamma^m]}{2}
+m\gamma^+\gamma^l
\bigg]\frac{\gamma^-}{2}
\nonumber\\
&\quad\times
\bigg\{
\bigg[
\frac{2\P^{j'}\P^{m'}}{[\P^2+\Qb^2]^2}
-\frac{\delta^{j'm'}}{[\P^2+\Qb^2]}
\bigg]
\bigg[
\gamma^+\frac{[\gamma^{m'},\gamma^{l'}]}{2}
+(1-2z)\delta^{l'm'}\gamma^+
\bigg]
\nonumber\\
&\qquad
-\frac{2m\P^{j'}}{[\P^2+\Qb^2]^2}
\gamma^+\gamma^{l'}
\bigg\}
\frac{\gamma^-}{2}\,
\frac{\gamma^+\gamma^i\gamma^-}{2}.
\end{align}
Using the light-cone Clifford algebra and the transverse gamma-matrix identities given above, the terms linear in $m$ vanish. The remaining contractions give
\begin{align}
{\mathcal D}_{\rm dec.\,on\,q}^{\rm (Eik)^\dagger(NEik)}
&=
-\bigg\{
\bigg[
\frac{2\P^{j'}\P^{m'}}{[\P^2+\Qb^2]^2}
-\frac{\delta^{j'm'}}{[\P^2+\Qb^2]}
\bigg]
\Big[
2(2z-1)\P^m\gamma^{m'}\gamma^m
+4z^2\P^{m'}
\Big]
+\frac{4m^2\P^{j'}}{[\P^2+\Qb^2]^2}
\bigg\}\frac{\gamma^-}{2}.
\end{align}
Decomposing the product of transverse gamma matrices into its commutator and anticommutator parts, this becomes
\begin{align}
{\mathcal D}_{\rm dec.\,on\,q}^{\rm (Eik)^\dagger(NEik)}
&=
-\gamma^-
\bigg\{
\bigg[
\frac{2\P^{j'}\P^{m'}}{[\P^2+\Qb^2]^2}
-\frac{\delta^{j'm'}}{[\P^2+\Qb^2]}
\bigg]
\nonumber\\
&\qquad\times
\bigg[
(2z-1)\P^m
\frac{[\gamma^{m'},\gamma^m]}{2}
+\big[z^2+(1-z)^2\big]\P^{m'}
\bigg]
+\frac{2m^2\P^{j'}}{[\P^2+\Qb^2]^2}
\bigg\}.
\end{align}
The contraction of the commutator with $\P^{m'}\P^m$ vanishes by antisymmetry. We therefore find
\begin{align}
\label{app-eq:eik_dag_Neik_dec_on_q}
{\mathcal D}_{\rm dec.\,on\,q}^{\rm (Eik)^\dagger(NEik)}
&=
\gamma^-
\bigg\{
\frac{\P^{j'}}{[\P^2+\Qb^2]^2}
\Big[
\big[z^2+(1-z)^2\big](\Qb^2-\P^2)-2m^2
\Big]
\nonumber\\
&\qquad
+(1-2z)\frac{\P^l}{[\P^2+\Qb^2]}
\frac{[\gamma^l,\gamma^{j'}]}{2}
\bigg\}.
\end{align}

We next consider the contribution with the quark-background insertions on the antiquark line. Its Dirac structure is
\begin{align}
{\mathcal D}_{\rm dec.\,on\,\bar q}^{\rm (Eik)^\dagger(NEik)}
&\equiv
\frac{1}{2}\sum_\lambda\delta^{ij}\,
\frac{\gamma^-\gamma^j\gamma^+}{2}\,
\frac{\gamma^-}{2}
\bigg[
\partial_{\P^{j'}}
\bigg(
\gamma^0
\frac{\big[\Gamma_T^\lambda(\P)\big]^\dagger}
{[\P^2+\Qb^2]}
\gamma^0
\bigg)
\bigg]
\frac{\gamma^-}{2}\,
\Gamma_T^\lambda(\P)\,
\frac{\gamma^i\gamma^-}{2}.
\end{align}
Substituting the hard factors, performing the polarization sum and simplifying the light-cone gamma matrices gives
\begin{align}
{\mathcal D}_{\rm dec.\,on\,\bar q}^{\rm (Eik)^\dagger(NEik)}
&=
\frac{\delta^{ij}\delta^{ll'}}{2}\,
\gamma^j
\bigg\{
\bigg[
\frac{2\P^{j'}\P^{m'}}{[\P^2+\Qb^2]^2}
-\frac{\delta^{j'm'}}{[\P^2+\Qb^2]}
\bigg]
\Big[
\gamma^{m'}\gamma^{l'}
+2(1-z)\delta^{l'm'}
\Big]
\nonumber\\
&\qquad
-\frac{2m\P^{j'}}{[\P^2+\Qb^2]^2}\gamma^{l'}
\bigg\}
\Big[
\P^m\gamma^l\gamma^m
+2(1-z)\P^l
+m\gamma^l
\Big]\frac{\gamma^i\gamma^-}{2}.
\end{align}
After contracting the transverse indices, the terms linear in $m$ again vanish, leaving
\begin{align}
{\mathcal D}_{\rm dec.\,on\,\bar q}^{\rm (Eik)^\dagger(NEik)}
&=
-\bigg\{
\bigg[
\frac{2\P^{j'}\P^{m'}}{[\P^2+\Qb^2]^2}
-\frac{\delta^{j'm'}}{[\P^2+\Qb^2]}
\bigg]
\Big[
2(1-2z)\P^m\gamma^m\gamma^{m'}
+4(1-z)^2\P^{m'}
\Big]
+\frac{4m^2\P^{j'}}{[\P^2+\Qb^2]^2}
\bigg\}\frac{\gamma^-}{2}.
\end{align}
Using $\{\gamma^m,\gamma^{m'}\}=-2\delta^{mm'}$, we obtain
\begin{align}
\label{app-eq:eik_dag_Neik_dec_on_barq}
{\mathcal D}_{\rm dec.\,on\,\bar q}^{\rm (Eik)^\dagger(NEik)}
&=
\gamma^-
\bigg\{
\frac{\P^{j'}}{[\P^2+\Qb^2]^2}
\Big[
\big[z^2+(1-z)^2\big](\Qb^2-\P^2)-2m^2
\Big]
+(1-2z)\frac{\P^l}{[\P^2+\Qb^2]}
\frac{[\gamma^l,\gamma^{j'}]}{2}
\bigg\}.
\end{align}
Thus, the quark- and antiquark-line contributions have the same Dirac structure for this interference.

The remaining interference contributions, in which the quark-background insertions appear in the complex-conjugate amplitude, follow by Dirac conjugation. The scalar term is unchanged, whereas the commutator term changes sign. Consequently,
\begin{align}
{\mathcal D}_{\rm dec.\,on\,q}^{\rm (NEik)^\dagger(Eik)}
&=
{\mathcal D}_{\rm dec.\,on\,\bar q}^{\rm (NEik)^\dagger(Eik)}
\nonumber\\
&=
\gamma^-
\bigg\{
\frac{\P^{j'}}{[\P^2+\Qb^2]^2}
\Big[
\big[z^2+(1-z)^2\big](\Qb^2-\P^2)-2m^2
\Big]
-(1-2z)\frac{\P^l}{[\P^2+\Qb^2]}
\frac{[\gamma^l,\gamma^{j'}]}{2}
\bigg\}.
\end{align}

\vspace{0.3in}
\bibliography{mybib_New}

\end{document}